\documentclass[conference]{IEEEtran}
\ifCLASSINFOpdf
\else
\fi
\usepackage{tikz}

\usepackage[noadjust]{cite}
\usepackage{multirow}
\usepackage[hidelinks]{hyperref}
\usepackage{multicol}
\usepackage{longtable}
\usepackage{multicol}
\usepackage{caption}
\usepackage{subcaption,graphicx}
\usepackage{dblfloatfix} 
\usepackage{tabularx}
\usepackage{tabulary}
\usepackage{threeparttable}
\usepackage{booktabs}
\usepackage{pifont}

\usepackage{amsmath,amssymb}
\usepackage{xcolor, soul}
\usepackage[linewidth=0.5pt]{mdframed}
\newcommand{\rpm}{\sbox0{$1$}\sbox2{$\scriptstyle\pm$}\raise\dimexpr(\ht0-\ht2)/2\relax\box2 }

\newcommand{\new}[1]{\textcolor[rgb]{0,0,0}{#1}}

\newcommand{\hlc}[2][yellow]{{\sethlcolor{#1}\hl{#2}}}
\definecolor{light_red}{rgb}{0.96, 0.76, 0.76}
\definecolor{light_green}{rgb}{0.56, 0.93, 0.56}
\definecolor{light_yellow}{rgb}{0.99, 0.97, 0.37}
\definecolor{dark_green}{rgb}{0.0, 0.5, 0.0}
\newcommand{\model}{\textit{AWT}}

\usepackage{xcolor}

\begin{document}
%
\title{Adversarial Watermarking Transformer: Towards Tracing Text Provenance with Data Hiding}



%
\author{\IEEEauthorblockN{Sahar Abdelnabi and Mario Fritz}
\IEEEauthorblockA{CISPA Helmholtz Center for Information Security}}


\maketitle

\begin{abstract}
Recent advances in natural language generation have introduced powerful language models with high-quality output text. However, this raises concerns about the potential misuse of such models for malicious purposes. In this paper, we study natural language watermarking as a defense to help better mark and trace the provenance of text. We introduce the Adversarial Watermarking Transformer (\model{}) with a jointly trained encoder-decoder and adversarial training that, given an input text and a binary message, generates an output text that is unobtrusively encoded with the given message. We further study different training and inference strategies to achieve minimal changes to the semantics and correctness of the input text. 

\model{} is the first end-to-end model to hide data in text by automatically learning -without ground truth- word substitutions along with their locations in order to encode the message. We empirically show that our model is effective in largely preserving text utility and decoding the watermark while hiding its presence against adversaries. Additionally, we demonstrate that our method is robust against a range of attacks. 

\end{abstract}


%
\IEEEpeerreviewmaketitle

\section{Introduction} \label{intro}
Recent years have witnessed major achievements in natural language processing (NLP), generation, and understanding. This is in part driven by the introduction of attention-based models (i.e., transformers~\cite{vaswani2017attention}) that outperformed recurrent or convolutional neural networks in many language tasks such as machine translation~\cite{vaswani2017attention,conneau2019cross}, language understanding~\cite{yang2019xlnet,devlin2018bert}, and language generation~\cite{zellers2019defending}. In addition, model pre-training further fueled these advances and it is now a common practice in NLP~\cite{peters2018deep, howard2018universal}; many large-scale models are now pre-trained on large datasets with either denoising auto-encoding or language modelling objectives and then fine-tuned on other NLP downstream tasks~\cite{yang2019xlnet,devlin2018bert,radford2019language,radfordimproving,wang2019denoising,brown2020language}.

On the other hand, this raises concerns about the potential misuse of such powerful models for malicious purposes such as spreading neural-generated fake news and misinformation. For example, OpenAI used a staged release to publicize their GPT-2 language model in order to evaluate the impact and potential risks~\cite{solaiman2019release}. Moreover, Zellers et al.~\cite{zellers2019defending} proposed a generative model called Grover demonstrating that a language model such as GPT-2 can be trained on news articles and can consequently generate realistically looking fake news.

These models can generate highly fluent text which sometimes had even higher ratings than human-written text and fooled human detectors~\cite{zellers2019defending,ippolito2020automatic,adelani2020generating}. While it is now possible to perform automatic detection, it is subject to recent advances in text generation (e.g., architecture, model size, and decoding strategies)~\cite{zellers2019defending,ippolito2020automatic}, which could hinder the automatic detection in the long run. Hence, we seek a more sustainable solution that can disambiguate between real and fake text.

To this end, we aim to perform automatic and unobstructive data hiding within language 
towards eventually watermarking the output of text generation models. Specifically, we envision black-box access scenarios to the language model APIs~\cite{openaigpt3} or to services such as text generation and editing-assistance that could be misused to create misinformation. 
Watermarking can then be used to introduce detectable fingerprints in the output that enable provenance tracing and detection. As deep learning models are widely deployed in the wild as services, they are subject to many attacks that only require black-box access (e.g.,~\cite{krishna2019thieves,orekondy2019knockoff,tramer2016stealing,papernot2017practical}). Thus, it is important to proactively provide solutions for such potential attacks before their prevalence. 
\begin{figure}[!t]
\centering
\includegraphics[width=0.9\linewidth]{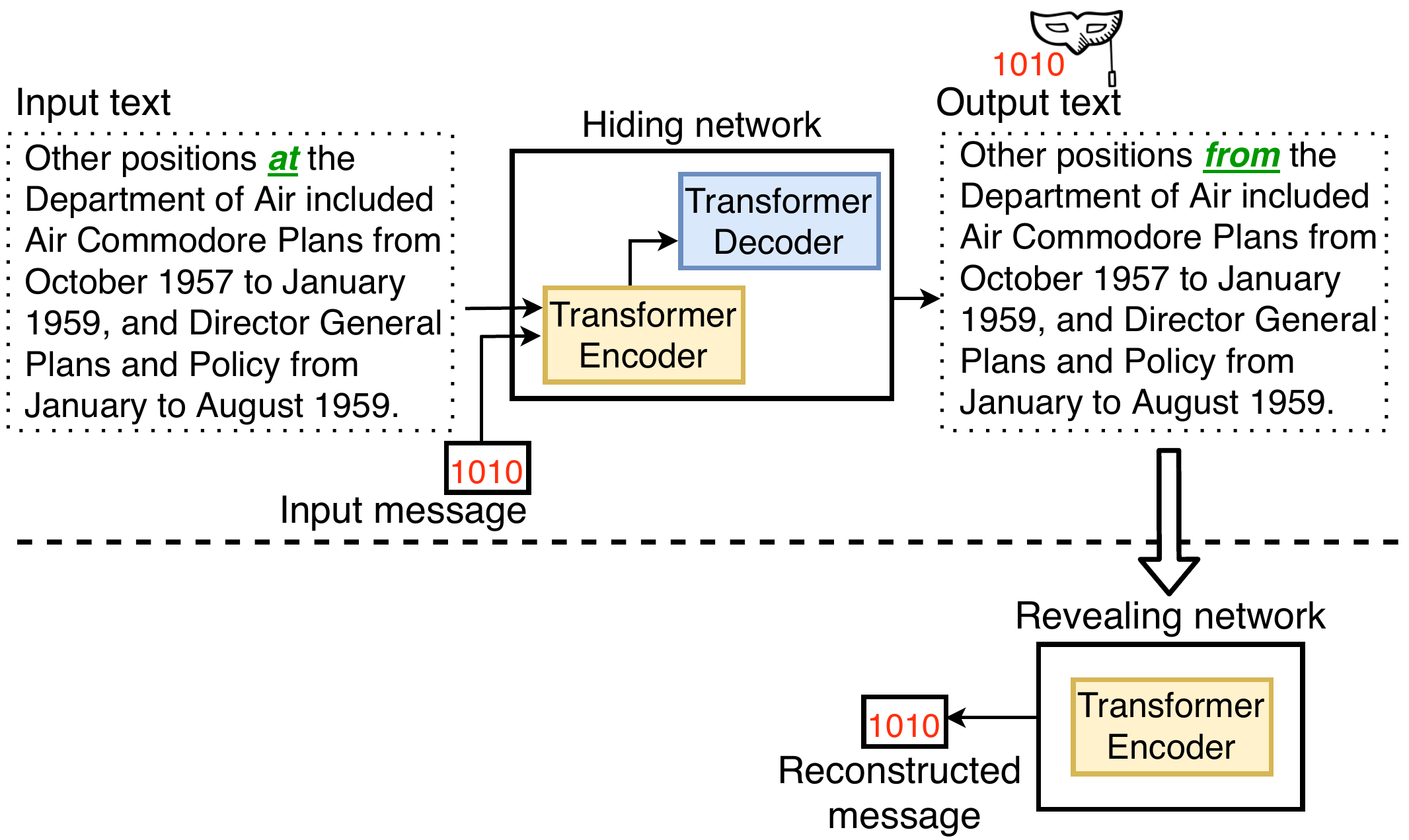}
\vspace{-4pt}
\caption{An overview of our text watermarking solution at inference time.} 
\vspace{-3mm}
\label{fig:teaser}
\end{figure}

\paragraph{Language watermarking} There have been several attempts to create watermarking methods for natural language, such as synonym substitutions~\cite{topkara2006hiding,chang2010practical}, syntactic tools (e.g., structural transformation
~\cite{topkara2006words}), 
in addition to language-specific changes~\cite{meral2007syntactic,chiang2003natural,halvani2013natural}. However, these previous methods used fixed rule-based substitutions that required extensive engineering efforts to design, in addition to human input and annotations, which hinders the automatic transformation. 
Also, the designed rules are limited as they might not apply to all sentences (e.g., no syntactic transformations can be applied~\cite{topkara2006words}). Additionally, they introduce large lexical or style changes to the original text, which is not preferred when keeping the original state is required (such as the output of an already well-trained language model). Besides, rule-based methods could impose restrictions on the use of the language (e.g., by word masking). Finally, using fixed substitutions can systematically change the text statistics which, in turn, undermines the secrecy of the watermark and enables adversaries to automatically detect and remove the watermark. 

\paragraph{Data hiding with neural networks} Data hiding can be done in other mediums as well such as images~\cite{Ingemar2007digital}. Several end-to-end methods have been proposed to substitute hand-crafted features and automatically hide and reveal data (e.g., bit strings) in images. This can be done using a jointly trained encoder and decoder architecture that is sometimes coupled with adversarial training to enforce secrecy~\cite{zhu2018hidden,baluja2017hiding,hayes2017generating,vukotic2018deep,zhang2019invisible}. However, automatic hiding approaches for language are still lacking, which could be attributed to the relatively harder discrete nature of language and having less redundancy compared to images.  

\paragraph{Our approach} We introduce the Adversarial Watermarking Transformer (\model{}); a solution for automatically hiding data in natural language without having paired training data or designing rule-based encoding. Similar to sequence-to-sequence machine translation models~\cite{sutskever2014sequence}, \model{} consists of a transformer encoder-decoder component that takes an input sentence and a binary message and produces an output text. This component works as a \textit{hiding network}, which is jointly trained with a transformer encoder that takes the output text only and works as a \textit{message decoder} to reconstruct the binary message. We utilize adversarial training~\cite{goodfellow2014generative} and train these two components against an \textit{adversary} that performs a classification between the input and modified text. The model is jointly trained to encode the message using the least amount of changes, successfully decode the message, and at the same time, fool the adversary. 
An example of using the data hiding and revealing networks at test time is shown in~\autoref{fig:teaser}.   

\paragraph{Evaluation axes} We evaluate the performance of our model on different axes inspired by the desired requirements: 1) The \textbf{effectiveness} denoted by message decoding accuracy and preserving text utility (by introducing the least amount of changes and preserving semantic similarity and grammatical correctness), 2) The \textbf{secrecy} of data encoding against adversaries. 3) The \textbf{robustness} to removing attempts. These requirements can be competing and reaching a trade-off between them is needed. For example, having a perfectly and easily decoded message can be done by changing the text substantially, which affects the text preserving, or by inserting less likely tokens, which affects the secrecy.       

\paragraph{\textbf{Contributions}} We formalize our contributions as follows: 1) We present \textbf{\model{}}; a novel approach that is the first to use a \textit{learned end-to-end framework} for data hiding in natural language that can be used for watermarking. 2) We study different variants of the model and inference strategies in order to improve the text utility, secrecy, and robustness. We measure the text utility with quantitative, qualitative, and human evaluations. To evaluate the secrecy, we analyze and visualize the modified text statistics and we evaluate the performance of different adversaries. Besides, we study the robustness under different attacks. 
3) We show that our model achieves a better trade-off between the evaluation axes compared to a rule-based synonym substitution baseline.
\section{Related Work}
We summarize previous work related to ours, such as language watermarking and steganography, model watermarking, and neural text detection. 

\subsection{Language Watermarking} \label{sec:lang_watermarking_rw}
Watermarking for multimedia documents has many applications such as identifying and protecting authorship~\cite{kamaruddin2018review,Ingemar2007digital,podilchuk2001digital,singh2013survey}. It consists of an embedding stage where the hidden information (i.e., watermark) is encoded in the cover signal, 
followed by a decoding stage where the watermark is recovered from the signal. 
Initial text watermarking attempts aimed to watermark documents, rather than language, by altering documents' characteristics such as characters' appearance, fonts, or spacing, by specific patterns depending on the codeword~\cite{brassil1995electronic}. However, these methods are prone to scanning and re-formatting attacks (e.g., copying and pasting)~\cite{kamaruddin2018review,topkara2005natural}. 

The other category of methods relies on linguistic characteristics of the natural language such as making syntactic or semantic changes to the cover text~\cite{topkara2005natural}. An example of such is the synonym substitution method in~\cite{topkara2006hiding} in which WordNet was used to find synonyms of words that are then divided into two groups to represent `0' or `1'. The authors relied on ambiguity by encoding the message with ambiguous words or homographs (i.e., a word that has multiple meanings). This was used to provide resilience as attackers would find it hard to perform automatic disambiguation to return to the original sentence. \new{However, words in the dataset were annotated/tagged by meanings from the WordNet database. These annotations were then used to select suitable synonyms,} which does not allow automatic methods with no human input. 
Generally, synonym substitution methods are vulnerable to an adversary who performs random counter synonym substitutions. In addition, they perform fixed pairwise substitutions which makes them not flexible and also vulnerable to detection. 

Additionally, sentence structure can be altered to encode the codeword according to a defined encoding~\cite{topkara2006words,topkara2006natural}. These methods introduce changes such as passivization, clefting, extraposition, and preposing~\cite{topkara2005natural,meral2009natural}. However, these transformations might not be applicable to all sentences, also, they change the sentence to a large extent. 

In contrast, we perform an end-to-end data hiding approach that is data-driven and does not require efforts to design rules and unique dictionary lookups. 

\subsection{Linguistic Steganography} Steganography hides information in text for mainly secret communication. However, it might have different requirements from watermarking~\cite{zhu2018hidden,topkara2006hiding}; while both of them target stealthiness to avoid detection, steganography does not assume an active warden. Thus, watermarking should have robustness to local changes. In our case, it should also preserve the underlying cover text and utility.

Translation by modifying a cover text was used in steganography such as the work in~\cite{wilson2014linguistic,wilson2015detection,wilson2016avoiding} that used a set of rule-based transformations to convert tweets to possible translations. The encoding and decoding were done with a keyed hash function
; the translations that map to the desired hash values were selected. Therefore, the decoding is not robust to local changes to the sentence. Another synonym-based method was proposed in~\cite{shirali2008new} based on assigning different bits to American and British words which makes it not applicable to a large number of sentences. Another direction is to generate text according to a shared key, instead of using translation. For example, the work in~\cite{fang2017generating} used a trained LSTM language model that generates sentences according to a masked vocabulary and a binary stream; the vocabulary was partitioned into different segments where each segment was assigned a sequence of bits. However, this imposes a large constraint on the usage of the language model since it needs to abide by the masking. Therefore, these steganography solutions are not suitable for our scenario as they specifically prioritize secret communication over flexibility or watermarking requirements.

\subsection{Model Watermarking} \label{sec:model_watermarking}
To protect the intellectual property of deep learning models, several approaches have been proposed to watermark models~\cite{li2019prove,lukas2019deep,adi2018turning,le2020adversarial}. This could be done by embedding the watermark into the model's weights, which requires white-box access for verification~\cite{uchida2017embedding,chen2018deepmarks,darvish2019deepsigns}, or by assigning specific labels for a trigger set (i.e., backdoors~\cite{gu2017badnets}), which only requires black-box access~\cite{adi2018turning,li2019prove,zhang2018protecting}. 

These methods were mainly addressing image classification networks; there is no previous work that attempted to watermark language models. We also differentiate our approach from model watermarking; instead of watermarking a model, we study data/language watermarking using a deep learning method that could eventually be used to watermark the language model's output. 

Our task shares some similarities in requirements with model watermarking (e.g., preserving model utility, authentication, and robustness against removal attempts), but they are different in the objective and assumptions about attacks. While the main purpose of model watermarking is to prove ownership and protect against model stealing or extraction~\cite{jia2020entangled}, our language watermarking scheme is designed to trace provenance and to prevent misuse. Thus, it should be consistently present in the output, not only a response to a trigger set. Moreover, while the adversary might aim to falsely claim or dispute ownership in model watermarking/stealing~\cite{li2019piracy}, we assume in our task that the adversary's goal is not to get detected or traced by the watermark. We elaborate on this difference in Section~\ref{sec:robust_piracy}. Finally, \new{model stealing can be done with white-box or black-box access to the victim model~\cite{jia2020entangled}, while we assume black-box access only to the language and watermarking model.}

\subsection{Neural Text Detection} Similar to the arms race in image deepfakes detection~\cite{yu2019attributing,wang2020cnn,carlini2020evading}, recent approaches were proposed to detect machine-generated text. For example, the Grover language model~\cite{zellers2019defending} was fine-tuned as a classifier to discriminate between human-written news and Grover generations. The authors reported that the model size played an important factor in the arms race; if a larger generator is used, the detection accuracy drops.  
Another limitation was observed in~\cite{ippolito2020automatic} in which the authors fine-tuned BERT to classify between human and GPT-2 generated text. The classifier was sensitive to the decoding strategy used in generation (top-\textit{k}, top-\textit{p}, and sampling from the untruncated distribution). It also had poor transferability when trained with a certain strategy and tested with another one. Therefore, while detecting machine-generated text is an interesting problem, it largely depends on the language model and decoding strategy. 

Analogous to image deepfake classifiers' limitations~\cite{zhang2020not}, this suggests that the success of classifiers might drop based on future progress in language modelling~\cite{zellers2019defending} (e.g., larger models~\cite{brown2020language}, arbitrary order generation~\cite{stern2019insertion}, and reducing exposure bias~\cite{caccia2018language}), in addition to decoding strategies that could reduce statistical abnormalities without introducing semantic artifacts~\cite{ippolito2020automatic}. 
Thus, it now becomes important to provide more sustainable solutions.

\section{Problem Statement and Threat Model} \label{sec:threat_model}
In this section, we discuss our usage scenario, requirements, assumptions about the adversary, and attacks. 
\paragraph{Watermarking as a defense against models' abuse} We study watermarking as a sustainable solution towards provenance tracing of machine-generated text in the case of models' abuse. An example of that scenario is a commercial black-box language model API~\cite{openaigpt3} or a text generation service that has legitimate usages such as editing assistance. The service is offered by the language model's owner or creator. However, it can be used in an unintended way by an adversary to automatically generate entire fake articles or misinformation at scale, aiming to achieve financial gains or serve a political agenda~\cite{zellers2019defending}. The owner can then proactively and in a responsible manner provide a way to identify and detect the model's generations by watermarking its output~\cite{zhang2020not}. 

News platforms can cooperate with the model owner, by having a copy of the watermark decoder, in order to identify the watermarks in the news articles and, thus, detect machine-generated articles. That is similar to~\cite{zellers2019defending} that suggests that news platforms can use the Grover classifier to detect Grover's articles. This is also in line with video-sharing platforms such as YouTube that uses deep networks to detect pornographic content~\cite{hosseini2017attacking}, and~\cite{mariconti2019you} which suggests using machine learning classifiers to flag videos that could be targeted by hate attacks. 

\paragraph{Watermarking using \model{}} The hiding network (message encoder) of \model{} is used by the owner to embed a watermark ($m$) into the text. The same message encoder can be used to encode different watermarks ($m_1$, $m_2$, ..., $m_n$) if needed (e.g., if the service is offered to different parties).
The multi-bit watermarking framework (as opposed to zero-bit) helps to trace
provenance to different parties. The revealing network (message decoder) of \model{} can, in turn, be used to reveal a watermark $m'$ which is then matched to the set of watermarks ($m_1$, $m_2$, ..., $m_n$).
\paragraph{Requirements}
We draw insights from digital watermarking studies in images to define the requirements. For example, the main requirements defined in~\cite{Ingemar2007digital} include: successful watermark embedding and verification, perceptual similarity (imperceptibility), robustness to removal attempts and edits (e.g., cropping, compression), and security to unauthorized detection. We adapt these requirements to our task and define the problem as a trade-off between the following:

\begin{itemize}
    \item \textbf{Effectiveness:} The watermark should be successfully embedded and verified. At the same time, it should keep the text utility; it should introduce the least amount of changes to the cover text, and ideally produce natural, grammatically and semantically correct changes, to preserve the perceptual similarity. 
    
    \item \textbf{Secrecy:} The watermark should achieve stealthiness by not introducing evident changes that can be easily detectable by automated classifiers. Ideally, it should be indistinguishable from non-watermarked text. This, in part, contributes to the text utility and naturalness preserving factor, and it helps to avoid suspicion and hinders the adversary's efforts to tamper with the watermark by identifying it. Therefore, we study the watermark secrecy and consider a range of possible discriminators.

    \item \textbf{Robustness:} The watermark should be resilient and not easily removable by simple changes. Ideally, to remove the watermark, one has to introduce heavy modifications that render the text `unusable'. Satisfying the previous two requirements (text utility and secrecy) can, in part, contribute to the robustness, since the adversary would not be able to distinguish the watermark.
\end{itemize}
\paragraph{Assumptions about the adversary and attacks} We consider a black-box API and assume that the attacker has no white-box access to the language model or the watermarking model (the watermark encoder and decoder), and also no access to the input watermark or the cover text before watermarking. We assume that the adversary aims to use the service without getting detected, thus, to \textit{tamper with (remove) the watermark while largely preserving the service's output (i.e., utility).} We consider the following robustness attacks: 1) Random changes and denoising, where the attacker has knowledge about using a translation-based watermarking scheme but not the model details. 2) Re-watermarking and de-watermarking, where the attacker has full knowledge about \model{} details and training data but no access to the model itself.

\section{Adversarial Watermarking Transformer} \label{sec:method}
\begin{figure*}[!t]
\centering
\includegraphics[width=0.9\linewidth, keepaspectratio]{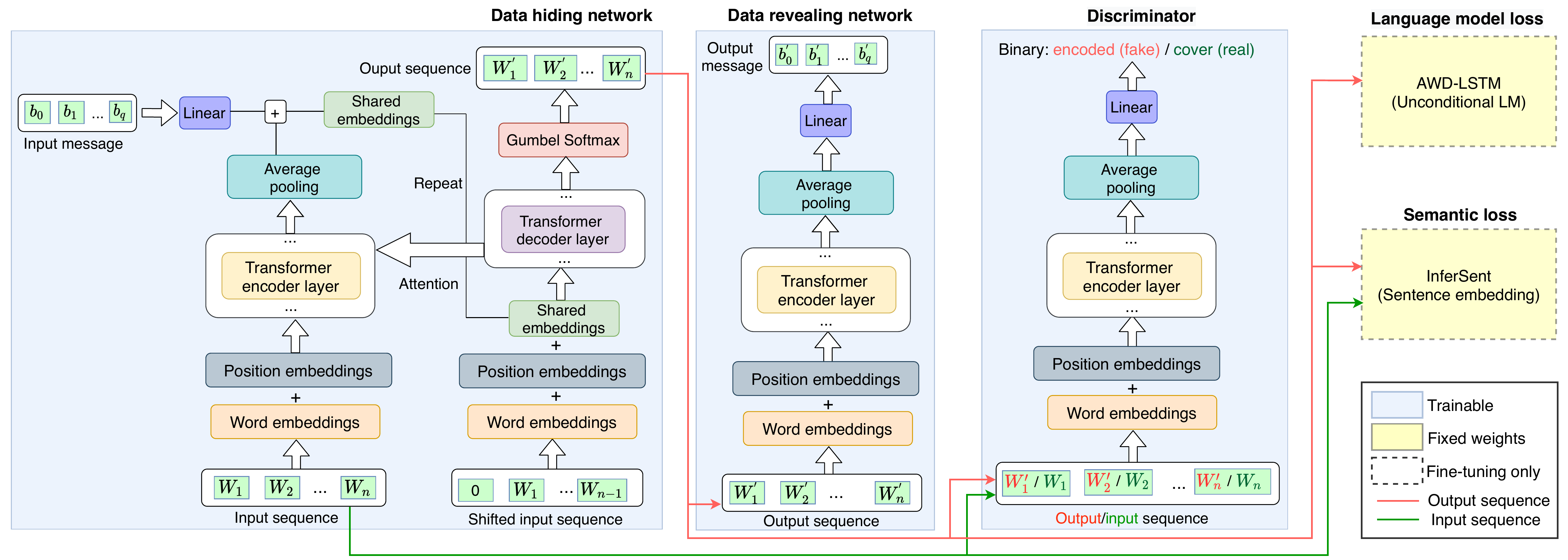}
\caption{The architecture of \model{}. The model consists of a data hiding network (sequence-to-sequence model), a data revealing network to decode the message, and a discriminator, in addition to the auxiliary components used at the fine-tuning step.}
\label{fig:awt}
\vspace{-3mm}
\end{figure*}

We propose the Adversarial Watermarking Transformer (\model{}) as an end-to-end framework for language watermarking. As shown in~\autoref{fig:awt}, the proposed solution includes a hiding network, a revealing network, and they are both trained against a discriminator. In this section, we discuss the details of these components and the training procedures. 

\subsection{Hiding Network (Message Encoder)}
This component is responsible for translating the input text to the watermarked text. Similar to sequence-to-sequence machine translation models~\cite{bahdanau2014neural,vaswani2017attention,shetty2017speaking}, it consists of an encoder and a decoder. 

\paragraph{Encoder} The encoder ($E$) is a transformer-encoder block consisting of several transformer encoder layers. Each layer consists of a self-attention block followed by a fully-connected layer. The encoder takes an input sentence $S = \{W_0, W_1, ..., W_n\}$, consisting of one-hot encoded words that are then projected to the embedding space using the word-embedding layer. As transformers are position-invariant, position embeddings (sinusoidal embeddings~\cite{vaswani2017attention}) are then added to the word embeddings. The encoder produces a fixed-length vector which is an average pooling across the time dimension of the last encoder layer~\cite{choi2019encoding}.

\paragraph{Message} The input message: $M = \{b_0, b_1, ..., b_q\}$ ($q$ binary bits sampled randomly), is first fed to a fully connected layer in order to match the embeddings' dimension and is then added to the sentence encoding produced by the encoder, producing a shared embedding between the sentence and the message, which is then passed to the autoregressive decoder and added to its input at each time-step.

\paragraph{Decoder} The decoder ($D$) has a similar architecture as the encoder, in addition to having an attention layer over the encoder's output, following the transformer architecture~\cite{vaswani2017attention}. In paired machine translation, the decoder usually takes the ground-truth target sequence (shifted right) and is trained to predict the next word at each time step. Since our problem does not have paired training data, the model is trained as an autoencoder~\cite{choi2019encoding}; the decoder takes the shifted input sentence and is trained to reconstruct the sequence given to the encoder, producing an output sentence $S^{'} = \{W^{'}_0, W^{'}_1, ..., W^{'}_n\}$. This serves as the reconstruction component in similar image data hiding methods~\cite{zhu2018hidden}, and it helps to largely preserve the input. In order to train the whole network jointly and allow back-propagation from the other components, we use Gumbel-Softmax approximation~\cite{jang2016categorical,kusner2016gans} with one-hot encoding in the forward pass (Straight-Through  Gumbel Estimator using $\textit{argmax}$~\cite{jang2016categorical}), and differentiable soft samples in the backward pass (softmax is used to approximate the $\textit{argmax}$ operation~\cite{jang2016categorical}). 
The reconstruction loss is the cross-entropy loss:  
$$L_{\textit{rec}} = \mathbb{E}_{p_{\textit{data}}(S)}[-\log p_D(S)] $$

\subsection{Revealing Network (Message Decoder)}
This part of the network is responsible for reconstructing the input message. It takes the one-hot samples produced by the autoencoder, multiplied by the embedding matrix, and with adding position embeddings. 
The message decoder ($M$) is a transformer-encoder block since it is typically used in text classification applications~\cite{devlin2018bert,ippolito2020automatic}. The output of the last transformer encoder layer is averaged across the time dimension and fed to a fully connected layer with an output size that is equivalent to the message length $q$. The message reconstruction loss is the binary cross-entropy over all bits: $$L_\textit{m} = -\sum_{i=1}^{q} b_i \log(p_M(b_i)) + (1-b_i) \log(1-p_M(b_i))$$

\textit{Weight tying:} To reduce the number of parameters in the network, we share the embedding weights across the whole network~\cite{vaswani2017attention} (i.e., text autoencoder including the encoder and decoder, message decoder, and discriminator), and also with the pre-softmax layer that maps from the embedding space to tokens in the text decoder~\cite{merity2017regularizing,vaswani2017attention,inan2016tying}. We found it beneficial in terms of the model size and faster convergence to also share the weights between the encoder part of the text autoencoder and the message decoder. 

\subsection{Discriminator}
In order to have a subtle message encoding that does not alter the language statistics, we utilize adversarial training and train the previous two components against a discriminator. The discriminator ($A$) is a transformer-encoder with a similar structure to the message decoder. It takes the non-watermarked sentences $S$ 
and the watermarked sentences $S^{'}$, multiplies the one-hot samples with the shared embeddings, and adds the position embeddings. It produces an average over the time steps of the last transformer encoder layer, which is used for the binary classification using the binary cross-entropy loss: $$L_{\textit{disc}} = -\log(A(S)) - \log(1-A(S^{'}))$$ while the adversarial loss is: $L_\textit{A} = - \log(A(S^{'}))$.
As we show later, we found this component essential in supporting the watermark secrecy against adversaries. 

\subsection{Training and Fine-tuning}
The model is first trained jointly with the above three losses with weighted averaging: $$L_\textit{1} = w_A L_\textit{A} + w_{\textit{rec}} L_{\textit{rec}} + w_\textit{m} L_\textit{m}$$
These losses are competing; e.g., a perfect sentence reconstruction would fail to encode the message. Therefore, we tuned the losses' weights on the validation set to achieve a good trade-off; e.g., it was helpful to assign a relatively higher weight to the message loss, otherwise, the reconstruction dominates. We did not need to anneal the message weight after the start. The other losses had comparable weights to each other.

The previous loss function aims to preserve the input sentence and encode the message with the least amount of changes while not changing the text statistics. However, we still do not have an explicit constraint on the type of changes done by the network to encode the message. Therefore, after training the network with $L_1$, we further fine-tune the network to achieve semantic consistency and grammatical correctness.

\paragraph{Preserving semantics} One way to force the output to be semantically similar to the input sentence is to embed both sentences into a semantic embedding space and compute the distance between the two encodings. We follow~\cite{shetty2018a4nt} and use the pre-trained Facebook sentence embedding model~\cite{conneau2017supervised} that was trained to produce a sentence representation based on the natural language inference (NLI) task. The model was trained on the Stanford Natural Language Inference (SNLI) dataset~\cite{bowman2015large}. We fix the sentence encoder $(F)$ weights and use it to compute the semantic loss between $S$ and $S^{'}$ as follows:
$$L_{\textit{sem}} = || F(S) - F(S^{'}) ||$$

\paragraph{Sentence correctness} To explicitly enforce correct grammar and structure, we fine-tune the model with a language model loss~\cite{shetty2018a4nt}. We independently trained the AWD-LSTM (ASGD Weight-Dropped LSTM)~\cite{merity2017regularizing} on the used dataset, as a medium-scale, but widely used and effective language model~\cite{howard2018universal,dai2019transformer,carlini2019secret}. We then use the trained AWD-LSTM model ($LM$) with fixing its weight to compute the likelihood of the output sentence $S^{'}$. Sentences with higher likelihood are more likely to be syntactically similar to the original text used in training. The language model loss is defined as: 
$$L_{\textit{LM}} = - \sum_{i} \log p_{\textit{LM}}(W^{'}_i | W^{'}_{<i}) $$

These previous two components take the one-hot samples and map them to their respective embedding space. We fine-tune the network using these two losses in addition to the previous ones as follows: $L_2 = w_\textit{A} L_\textit{A} + w_{\textit{rec}} L_{\textit{rec}} + w_\textit{m} L_\textit{m} + w_{\textit{sem}} L_{\textit{sem}} + w_{\textit{LM}} L_{\textit{LM}}$. 

As we later show, fine-tuning with these auxiliary losses helps to produce more realistically looking and natural samples compared to only training with reconstructing the sentence. Introducing these new losses after the first training stage was mainly to speed-up convergence and training time since the model at first has not yet learned to reconstruct the input. So after the model learns the basic function, we use this stage as a warm start for further optimization. This is similar to pre-training as an autoencoder for other translation tasks~\cite{shetty2018a4nt}.

\section{Experimental Results}
In this section, we first discuss our setup. Then, we evaluate the different aspects of our model: \textbf{effectiveness}, \textbf{secrecy}, and \textbf{robustness}. We compare \model{} to baselines and present a user study to evaluate the output's quality. 
\subsection{Setup} \label{sec:setup}
\paragraph{Dataset}
We used the word-level WikiText-2 (WT2) that is curated from Wikipedia articles with light processing and was introduced in~\cite{merity2016pointer}. We used the same tokenization, processing, and split setup as~\cite{merity2016pointer,merity2017regularizing,merity2018analysis}. The dataset is approximately twice the size of the Penn Treebank (PTB) benchmark dataset for language modelling~\cite{marcus1994penn}, besides, the WikiText-2 keeps the capitalization, punctuation, and numbers. It contains over 30,000 unique vocabulary words and has a size of 2 million words in the training set and 0.2 million in validation and test sets. Since our watermarking framework can be applied independently as a post-processing step, we experiment on human-written data to objectively judge the proposed watermarking scheme correctness and to use a benchmark pre-processed dataset.

\paragraph{Implementation Details} \label{sec:implem_details}
We used a dimension size ($d_{model}$) of 512 for all transformers blocks and embeddings. The encoder and decoder transformer blocks are composed of 3 identical layers and 4 attention heads per layer, the decoder has a masked (on future input) self-attention. The rest of the transformer hyperparameters follows~\cite{vaswani2017attention} (e.g., a dropout probability of 0.1, a dimension of 2048 for the feed-forward layers, ReLU activations, and sinusoidal position embeddings). We optimize the network with Adam optimizer~\cite{kingma2014adam} with a varying learning rate~\cite{vaswani2017attention}: 
$$ \textit{lrate}_{\textit{gen}} = d_{\textit{model}}^{-0.8} * \text{min} ({\textit{step}}^{-0.5}, {\textit{step}*{\textit{warmup}}}^{-1.5}) $$
$$ \textit{lrate}_{\textit{disc}} = d_{\textit{model}}^{-1.1} * \text{min} ({\textit{step}}^{-0.5}, {\textit{step}*{\textit{warmup}}}^{-1.5}) $$

where $\textit{step}$ is the batch counter, $\textit{lrate}_{\textit{gen}}$ is the learning rate of the autoencoder and message decoder, and $\textit{lrate}_{\textit{disc}}$ is the learning rate of the discriminator, trained alternatively. We use 6000 warmup steps and a batch size of 80. We use a Gumbel temperature of 0.5~\cite{shetty2018a4nt,shetty2017speaking}. We trained the network for 200 epochs for each stage. For training the AWD-LSTM language model, we used the authors' implementation\footnote{https://github.com/salesforce/awd-lstm-lm}. We used the trained sentence embedding model\footnote{https://github.com/facebookresearch/InferSent}. A good trade-off between losses was found when setting the message loss's weight to a relatively higher value than the others (e.g., 5x). Otherwise, the other losses dominate and the training fails to optimize the message loss. The training was not sensitive to the exact weights.  
\new{Our code and models are publicly available: \url{https://github.com/S-Abdelnabi/awt/}.}

\paragraph{Input length during training and test} The dataset is a continuous text corpus. During training, we encode a randomly sampled 4-bit message (similar to~\cite{wilson2014linguistic}) into a text segment/sentence (varying length: $\mathcal{N}(80,5)$). We test the network on fixed-length segments of 80 words per segment, which can be adapted if needed, small changes to this number (\rpm5 words) did not significantly affect the results. As our objective is to watermark machine-generated articles, this segment-level setup can be extended to a longer text or a document-level input by successively encoding and decoding concatenated segments. Thus, a longer watermark can be composed of multiple 4-bits messages with a certain pre-defined order. Using longer watermarks allows verification using null-hypothesis testing. We base the watermark verification decision on the matching accuracy of all decoded messages from the concatenated segments. In section~\ref{agg}, we evaluate the verification with respect to the total segments' length.

\subsection{Effectiveness Evaluation} \label{sec:eff}
In this section, we evaluate the \textbf{effectiveness} of the model in terms of \textbf{text utility} and \textbf{bit accuracy}. We discuss our evaluation metrics and we compare different model's variants. 
We examine two different inference strategies to improve the utility. We discuss how to verify the watermark by sentence aggregation and show the trade-off between utility and verification confidence at different input lengths. We show how to improve the bit accuracy by averaging multiple encoded segments. We then perform a qualitative analysis to visualize and assess the changes produced by the model. 

\subsubsection{Metrics} To measure the message decoding, we use the bitwise message accuracy (random chance: 50\%) of all sentences in the test set. To measure utility preserving, we use the meteor score~\cite{denkowski2014meteor} that is used in machine-translation tasks to compare the output sentence against ground-truth references. Meteor performs n-gram alignments between the candidate and output text with synonym lookups from WordNet~\cite{miller1998wordnet}. It ranges from 0 to 1 (`no' to `identical' similarity).  

However, we found the meteor score not enough to evaluate the text semantics; two output sentences can have the same number of changed words compared to the input sentence and thus a similar meteor score (assuming there is no synonym overlapping), however, one of them could be closer to the input sentence. Therefore, to approximate the semantic difference between the input and output text, we used SBERT~\cite{reimers2019sentence}, a pre-trained sentence embedding model based on fine-tuning BERT as a siamese network on the NLI task. We compute the input and output embeddings and calculate the $L_2$ difference between them (lower is better). We discuss more details about the importance of using this additional metric in Section~\ref{qual} and Appendix~\ref{appendix_metrics}. We average the meteor scores and SBERT distances for all sentences in the test set. 
\begin{table} [!b]
\centering
\resizebox{0.8\linewidth}{!}{%
\begin{tabular}{l|ccc}
\toprule
\textbf{Model} & \textbf{Bit accuracy} & \textbf{Meteor} & \textbf{SBERT distance} \\   \midrule
\model{} & 97.04\%\rpm0.16 & 0.962\rpm0.0003 & 1.26\rpm0.008 \\ \hline
$-$ fine-tuning & 95.13\%\rpm0.21 & 0.943\rpm0.0005 & 1.73\rpm0.015 \\
$-$ discriminator & 96.15\%\rpm0.22 & 0.938\rpm0.0006 & 2.29\rpm0.016\\\bottomrule
\end{tabular}}
\caption{Model's variants quantitative analysis. The first row is the full model, the second row is without fine-tuning, the third row is without fine-tuning or a discriminator.} \label{tab:ablation}
\end{table}

\subsubsection{Model ablation}
We show in~\autoref{tab:ablation} three variants of our model. We ran each one 10 times with random sampling of messages and we found the results very comparable, we report the average and standard deviation of the metrics across these runs. The first row shows the full \model{} with the fine-tuning step, the second one shows the model without fine-tuning, and the last row shows the model without discriminator and fine-tuning (trained only with text and message reconstruction). This shows that the fine-tuning step helps to improve the text preserving and semantics as suggested by the increase in the meteor score and the decrease in the SBERT distance, at the same time, it maintains a high message decoding accuracy. Additionally, the model trained with a discriminator had a lower SBERT distance compared to the model that was trained with text reconstruction only, although both of them have a comparable meteor score. As we demonstrate in our qualitative and secrecy analysis shown later, this indicates that the adversarial training setup improves the output's quality, in addition to its secrecy advantages\footnote{Unless mentioned otherwise, all the following experiments are performed on the fine-tuned model, and \model{} stands for the full model.}.

\subsubsection{Inference strategies} \label{sec:inference}
To further maintain the text utility and improve the output sequence's quality, 
we study two inference strategies. First, we sample a set of samples for each sentence and then select the best sample, based on possible quality metrics. 
Second, we deliberately leave some sentences non-watermarked. Preserving utility has a trade-off relationship with verification confidence and bit accuracy, which we discuss in Sections~\ref{agg} and~\ref{sec:avg}.
\paragraph{Best-of-many encoding} We here sample $n$ sentences for each input sentence using the Gumbel sampler in the autoencoder network. We then use the trained language model (AWD-LSTM) to compute the likelihood for each output sample. Then, we pick the sample with the highest likelihood (excluding samples with no changes to the input) and feed it to the message decoder. An alternative quality metric is to pick the sample with the lowest SBERT distance to the input sentence, we found that these two metrics give comparable results, however, using the language model gives slightly better samples in terms of grammatical and syntactic correctness (discussed in Section~\ref{qual} and Appendix~\ref{appendix_metrics}).

\begin{figure}[!t]
\centering
\includegraphics[width=0.9\linewidth]{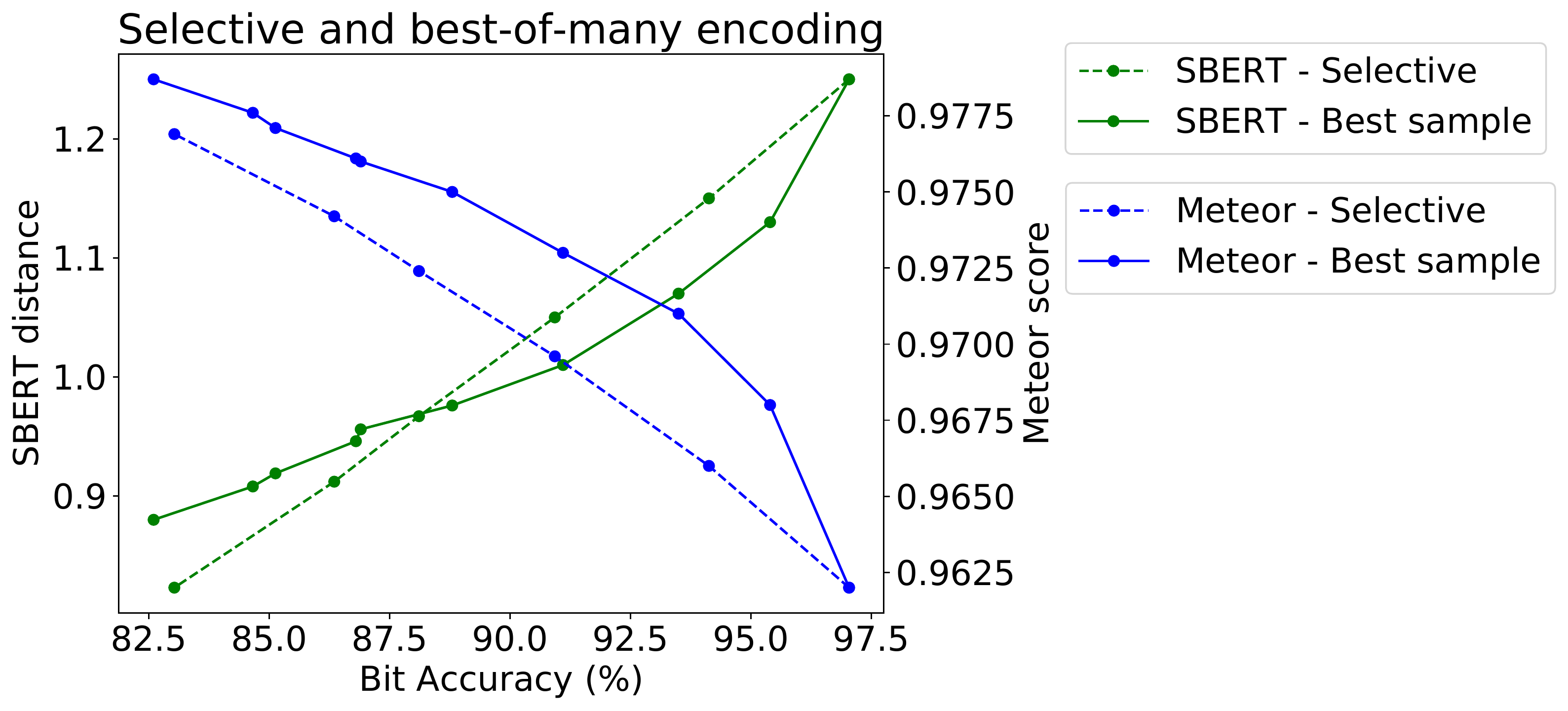}
\caption{Different operating points from selective and best-of-many sampling encoding.}
\label{fig:tradeoff_sampling_selective}
\vspace{-4mm}
\end{figure}

\begin{figure}[!b]
\centering
\begin{subfigure}{0.45\columnwidth}
  \centering
  \includegraphics[width=\linewidth]{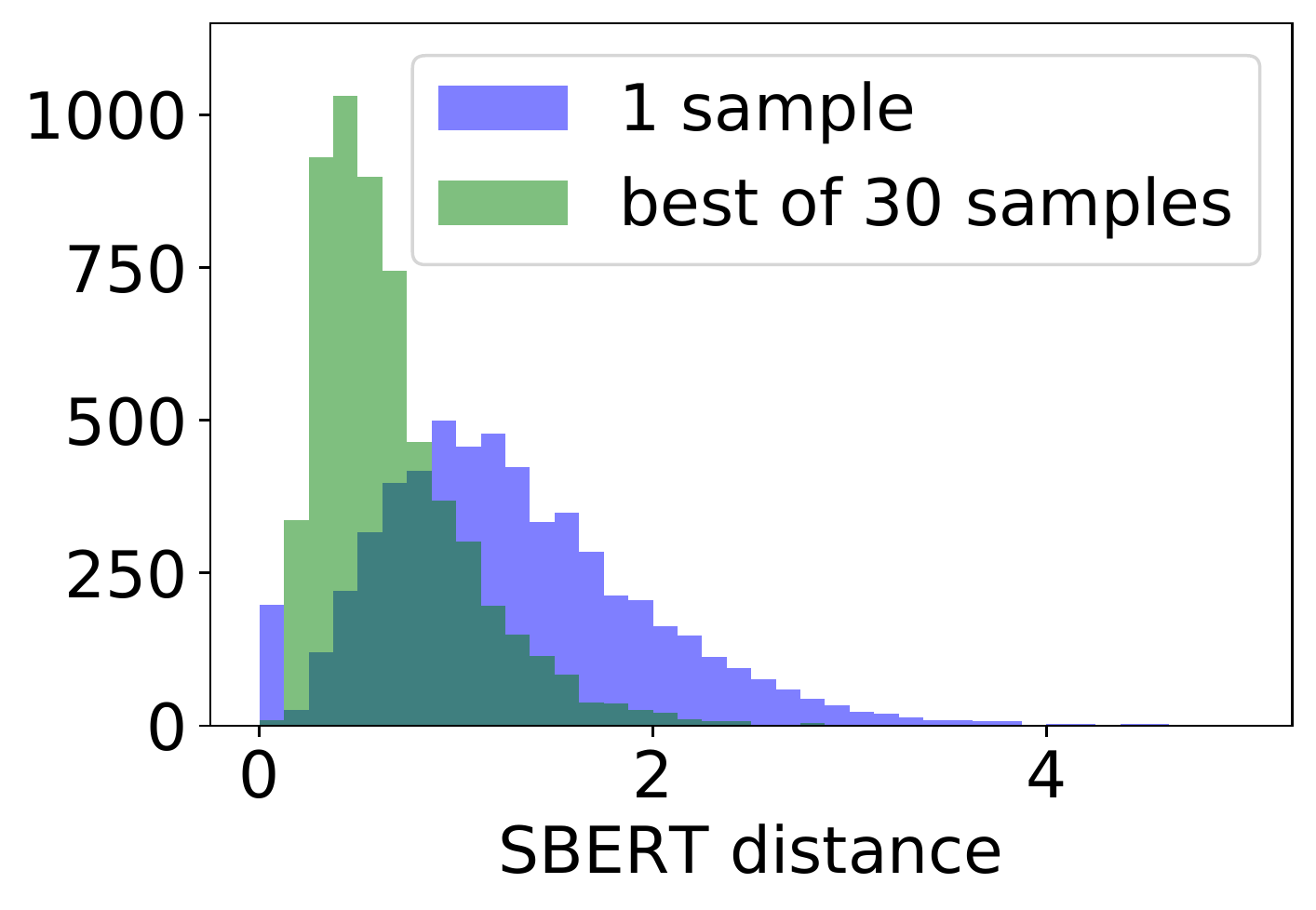}  
  \caption{SBERT distance}
  \label{fig:sampling_sbert}
\end{subfigure}
\begin{subfigure}{0.45\columnwidth}
  \centering
  \includegraphics[width=\linewidth]{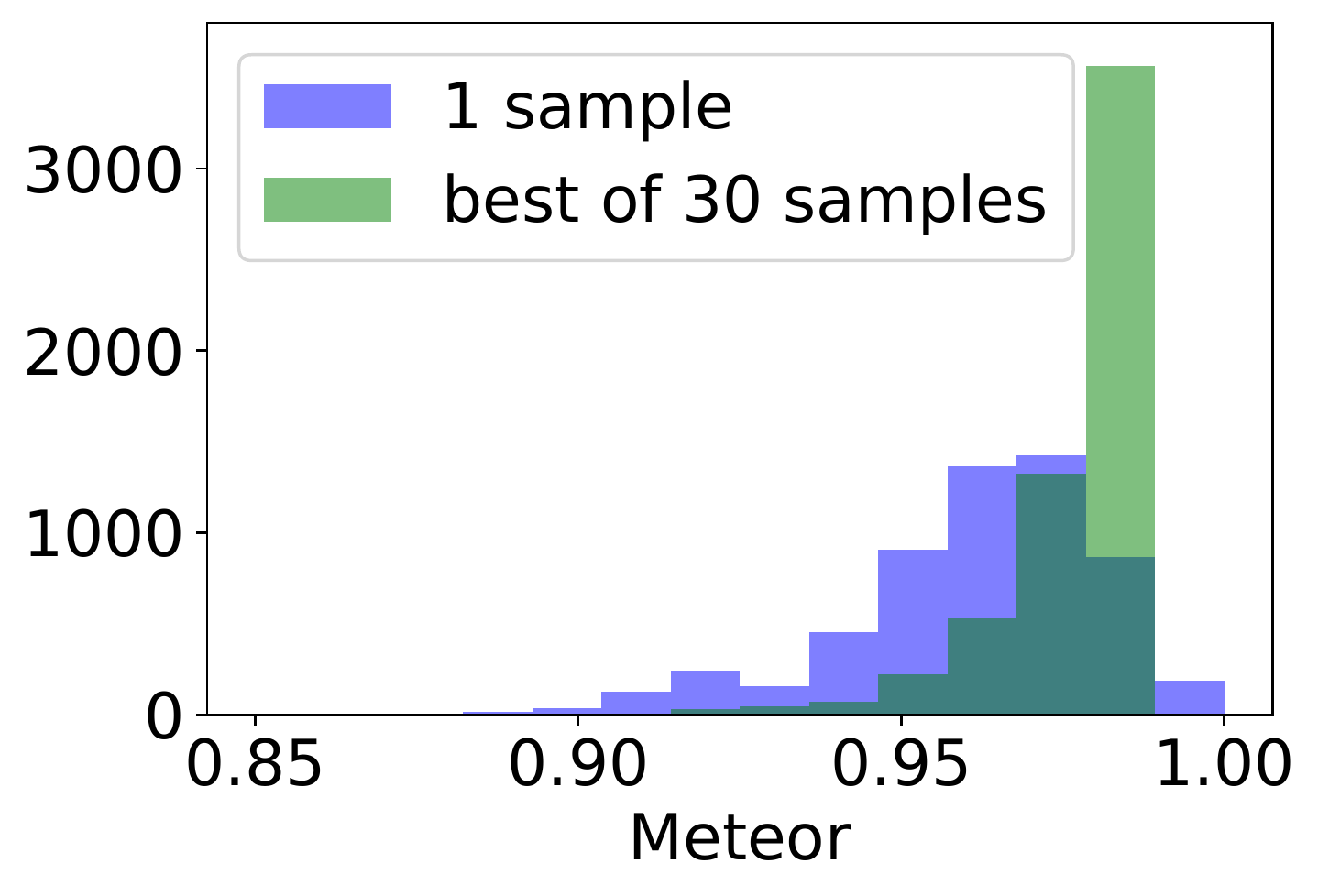}  
  \caption{Meteor}
  \label{fig:sampling_meteor}
\end{subfigure}
\caption{Histograms of (a) SBERT distances (lower is better), and (b) meteor scores (higher is better) for 2 sampling settings.}
  \label{fig:sampling_hist}
\vspace{-3mm}
\end{figure}

We show in~\autoref{fig:tradeoff_sampling_selective} different operating points based on varying $n$ from 1 to 40 samples. For each point, we show the relationship between bit accuracy and text utility (demonstrated by the averaged meteor score and SBERT distance). We found that the meteor score increases and the SBERT distance decreases with increasing the number of samples. Additionally, we show in~\autoref{fig:sampling_hist} a histogram of the SBERT distances and meteor scores for two sampling settings; only 1 sample (bit accuracy 97\%), and selecting the best from 30 samples (bit accuracy $\sim$85\%). In the latter case, the output is moving towards identical reconstruction. This analysis suggests that higher-quality output sentences can be acquired by sampling and that the language model metric also correlates with the meteor and SBERT ones.

\paragraph{Selective encoding} Alternatively, to provide further flexibility, we leave a percentage of sentences non-watermarked to reduce the overall change to the output text. 
The message decoder side does not need to know which sentences were watermarked as it can attempt to decode the message from all sentences in a document. The matching accuracy of non-watermarked sentences approximates the random chance while watermarked sentences will have a strong matching (we use the 1-sample output in~\autoref{tab:ablation}). We can then base the decision on the matching of the whole decoded sequence of messages (i.e., using null-hypothesis testing as we show in Section~\ref{agg}). We decide which sentences to leave based on setting a threshold on the increase of the language model loss compared to the original sentence. We examine different thresholds that encode different quantiles of the test set sentences (from 75\% to 100\%). We perform this experiment by sampling only 1 sample from the model. We show in~\autoref{fig:tradeoff_sampling_selective} the mean meteor and SBERT distance versus bit accuracy at each quantile. Besides the flexibility and utility advantage, selective encoding hinders the adversary effort to localize the watermark as not all sentences are watermarked.

\subsubsection{Watermark verification by sentence aggregation} \label{agg}
The previous strategies help to improve the output's quality. However, they reduce the bit accuracy. Therefore, in this section, 
we discuss the relationship between the verification confidence and bit accuracy at different input lengths. 

To allow a large number of watermarks and support an article-level watermarking, a longer watermark can be composed of multipliers of 4 bits messages; each 4 bits are embedded into one text segment. If the total text length is longer than the watermark, the long watermark sequence can be repeated partially or fully. The length of the unique long watermark can be determined based on the expected minimum text length. The decoded messages can be then verified against the sequence. Thus, we accumulate observations from all messages in the document to perform a null hypothesis test based on the number of matching bits~\cite{venugopal2011watermarking}. We assume that the null hypothesis ($H_0$) is getting this number of matching bits by chance. Under the null hypothesis, the probability of matching bits (random variable $X$) follows a binomial distribution; the number of trials is the number of bits in the sequence ($n$), $k$ is the number of successes (matching bits), and each bit has a 0.5 probability of success. We then compute the $p$-value of the hypothesis test by computing the probability of getting $k$ or higher matching bits under the null hypothesis: $$ Pr(X>k|H_0) = \sum_{i=k}^{n} \binom{n}{i} 0.5^n $$

The watermark is verified if the $p$-value is smaller than a threshold $\mathcal T$; meaning that it is not very likely to get this sequence by chance. This allows a soft matching of the decoded watermark instead of an exact one. We evaluate the thresholds of 0.05 and 0.01~\cite{venugopal2011watermarking}.
\begin{figure}[!b]
\centering
\includegraphics[width=0.95\linewidth]{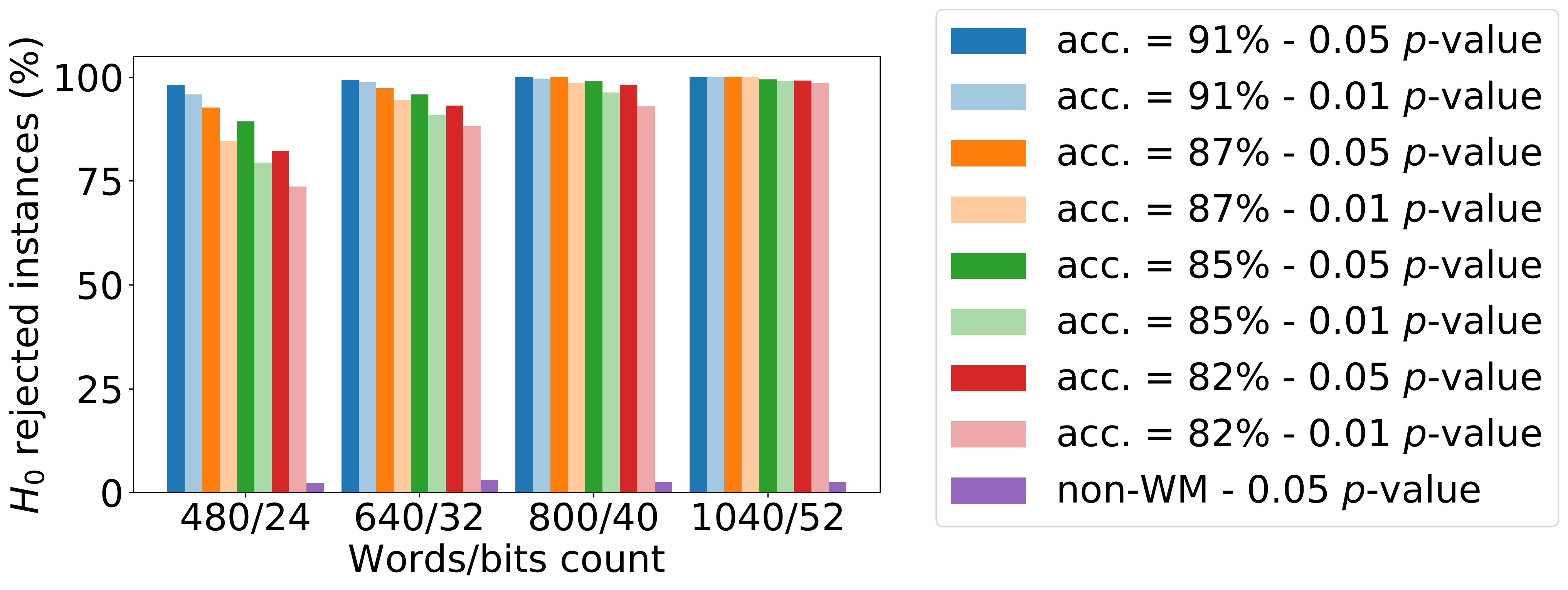}
\vspace{-3pt}
\caption{Percentage of instances where the null hypothesis (no watermarking) is rejected (for 0.05 and 0.01 $p$-value thresholds) versus text and bit lengths (words/bits), done for different operating points (i.e., bit accuracy)\new{, and real text.}} 
\label{fig:long_seq}
\vspace{-3mm}
\end{figure}

We empirically find the percentage of instances where the null hypothesis can be rejected (i.e., the watermark is correctly verified), and its relationship with the text length (i.e., the number of bits in the sequence). We perform this at different operating points that vary in their bit accuracy. We demonstrate this experiment in~\autoref{fig:long_seq}; when increasing the text length, we observe more correct observations, and thus, can reject the null hypothesis. Therefore, the use of operating points can be flexibly determined by the expected text length; at longer lengths, it is affordable to use an operating point with lower bit accuracy (i.e., higher utility). We validate that the bit accuracy is close to \textit{chance level (49.9\%) when the input is non-watermarked (real) text}, which results, naturally, in high $p$-values (and low false-positive rates).
\begin{figure}[!t]
\centering
\includegraphics[width=0.6\linewidth]{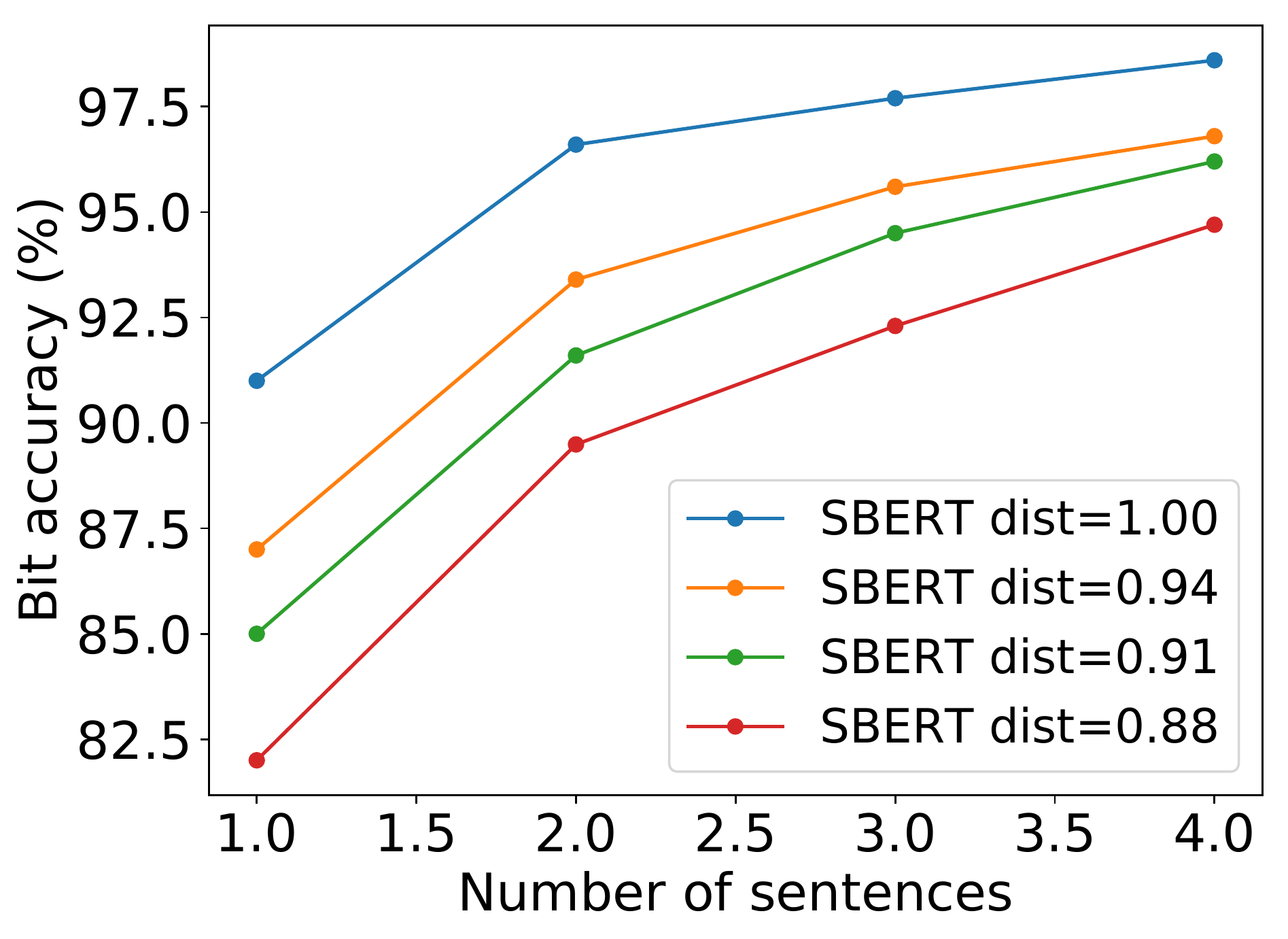}
\caption{Bit accuracy for 4 sampling operating points when averaging the posterior probabilities of multiple sentences encoded with the same message.} 
\label{fig:msg_avg}
\vspace{-3mm}
\end{figure}
\subsubsection{Decoding by averaging} \label{sec:avg}
We here aim to improve the bit accuracy of the best-of-many samples encoding strategy, this can be needed in applications where one is interested in decoding the message itself, rather than watermarking by concatenating segments from the whole document. 
We encode multiple text segments/sentences with the same binary message, decode each sentence independently, and then average their posterior probabilities. We demonstrate in~\autoref{fig:msg_avg} the performance gain when averaging up to 4 sentences, compared to using only 1 sentence. We perform this analysis for 4 different operating points that vary in the number of samples. As can be observed, using only 2 sentences can increase the bit accuracy for all operating points. Increasing the number of sentences can still further improve the accuracy. This strategy can be used by repeating the messages in the document with an agreed-upon sequence.  
\subsubsection{Qualitative analysis} \label{qual}
We qualitatively analyse the model's output. We first compare different variants, we then discuss the implications of the used metrics. 
Lastly, we visualize and analyse the changes performed by the model.
\newcolumntype{L}{>{\arraybackslash}m{5cm}}
\renewcommand{\arraystretch}{1.3}
\begin{table} [!b]
\centering
\resizebox{0.95\linewidth}{!}{%
\begin{tabular}{L|L}
\toprule
\textbf{Input} & \textbf{$-$ discriminator output} \\\midrule
He was appointed \underline{\textit{the}} commanding officer. & He was appointed \underline{\textit{\hlc[light_red]{Bunbury}}} commanding officer.\\
one of \underline{\textit{the}} most fascinating characters in \underline{\textit{the}} series & one of \underline{\textit{\hlc[light_red]{Milton}}} most fascinating characters in \underline{\textit{\hlc[light_red]{Milton}}} series \\\bottomrule
\end{tabular}}
\caption{Examples of input and output pairs of the model trained without adversarial training showing systematic fixed changes that insert less likely tokens.} \label{tab:ex_noadv_model}
\vspace{-3mm}
\end{table}

\paragraph{Model's variants} To examine the effect of the adversarial training, we show in~\autoref{tab:ex_noadv_model} examples of input and output pairs of the model trained with text reconstruction only (the third row in~\autoref{tab:ablation}). We observed that there are two main problems with this model: first, it performs systematic and fixed modifications that alter the text statistics, e.g., the word ``the'' is often changed. Second, it encodes the message with tokens that have low occurrences count in the natural text (possibly, since there are no other constraints on the naturalness, the model exploits this shortcut as a trivial solution as these rare tokens would be clearly distinctive of the message). These two problems could make the watermark easily detectable by adversaries (and thus removable). It also makes the output less natural and reduces the semantic correctness (which is indicated by the higher SBERT distance in~\autoref{tab:ablation}, supporting the use of an additional metric besides the meteor). 
\begin{figure}[!t]
\centering
\includegraphics[width=0.85\linewidth]{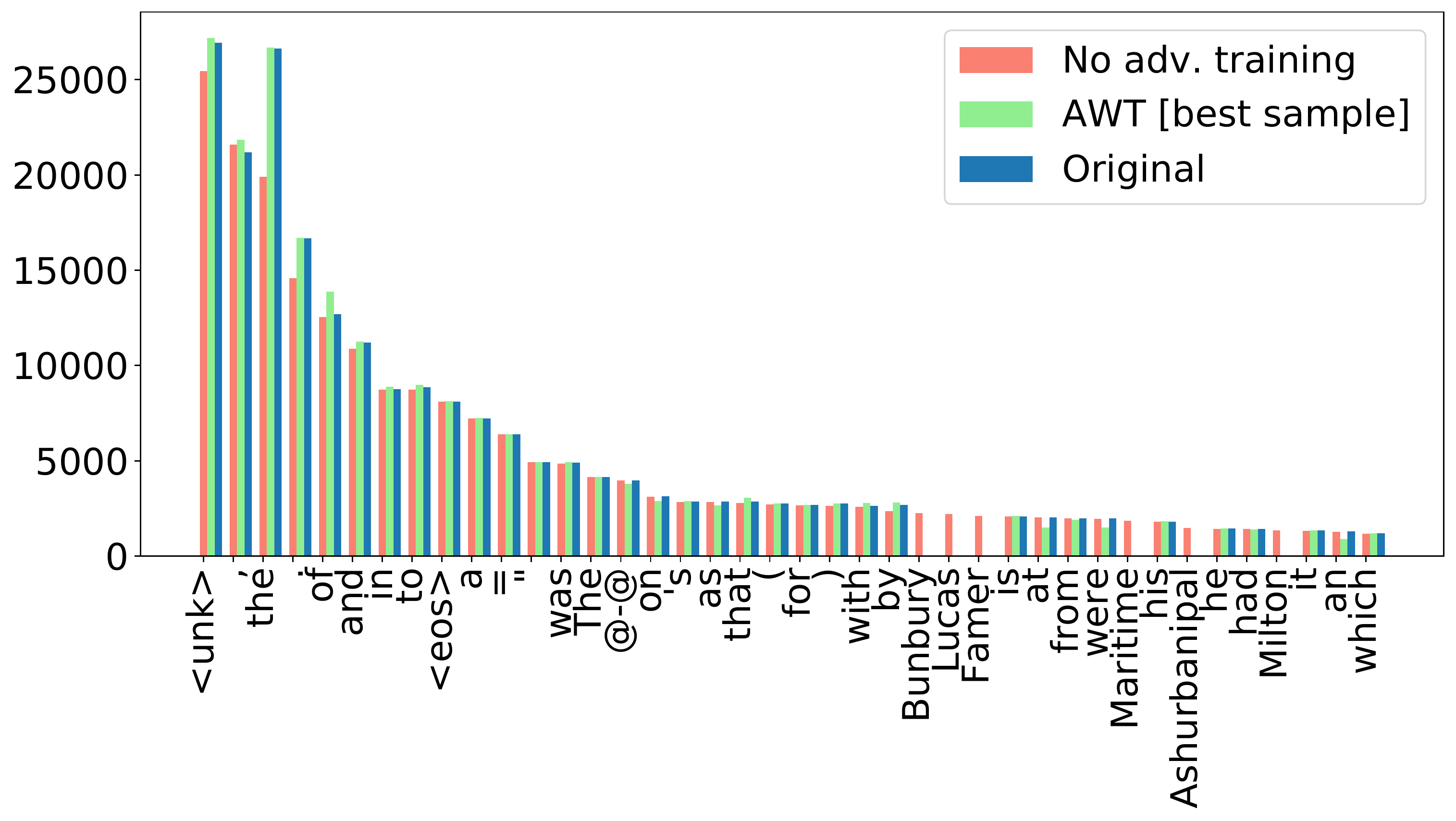}
\caption{Top words' count in the model trained without adversarial training compared to their counts in \model{} output and the original dataset.} 
\label{fig:hist_words}
\vspace{-4mm}
\end{figure}
To validate this observation, we show in~\autoref{fig:hist_words} the occurrences of the top words in this model compared to their occurrences in the \model{} model and the original text. Unlike \model{}, this model's variant pushes unlikely words to the top and decreases the count of more likely words (e.g., ``the''), introducing clear artifacts. In contrast, \model{} keeps the distribution of top words similar and encodes the message with also likely words, providing better concealing. The model without fine-tuning also keeps the top words' counts similar (not shown in the figure), but it still shows syntactic inconsistencies, e.g., using the end-of-sentence token in the middle of the sentence. We observed that fine-tuning the model helps to reduce these inconsistencies, examples are shown in~\autoref{tab:without_ft}.

\newcolumntype{L}{>{\arraybackslash}m{3.2cm}}
\renewcommand{\arraystretch}{1.5}
\begin{table} [!b]
\centering
\resizebox{0.95\linewidth}{!}{%
\begin{tabular}{L|L|L}
\toprule
\textbf{Input} & \textbf{$-$ fine-tuning output} & \textbf{\model{} output} \\\midrule
the Business Corporation, \underline{\textit{which}} was formed by a group of leaders \underline{\textit{from}} the area. & the Business Corporation, \underline{\textit{\hlc[light_red]{$<$eos$>$}}} was formed by a group of leaders from the area. & the Business Corporation, which was formed by a group of leaders \underline{\textit{\hlc[light_green]{at}}} the area. \\

The railroads provided a means of transportation and \underline{\textit{an}} influx of industries & The railroads provided a means of transportation and \underline{\textit{\hlc[light_red]{$<$eos$>$}}} influx of industries & The railroads provided a means of transportation and \underline{\textit{\hlc[light_green]{that}}} influx of industries \\

the measurements indicated that a segment of M @-@ 82 west of $<$unk$>$ \underline{\textit{had}} the peak volume for the highway & the measurements indicated that a segment of M @-@ 82 west of $<$unk$>$\underline{\textit{\hlc[light_red]{'s}}} the peak volume for the highway & the measurements indicated that a segment of M @-@ 82 west of $<$unk$>$ \underline{\textit{\hlc[light_green]{were}}} the peak volume for the highway 
\\\bottomrule
\end{tabular}}
\caption{Comparison between two variants of the model: before and after fine-tuning. The fine-tuned model shows better syntactic consistency.} \label{tab:without_ft}
\vspace{-4mm}
\end{table}

\newcolumntype{L}{>{\arraybackslash}m{5cm}}
\renewcommand{\arraystretch}{1.3}
\begin{table} [!t]
\centering
\resizebox{0.95\linewidth}{!}{%
\begin{tabular}{L|L}
\toprule
\textbf{Input} & \textbf{\model{} output} \\\midrule
In 1951 , a small airstrip was built \underline{\textit{at}} the ruins & In 1951 , a small airstrip was built \underline{\textit{\hlc[light_green]{on}}} the ruins \\ 
It is the opening track \underline{\textit{from}} their 1987 album & It is the opening track \underline{\textit{\hlc[light_green]{of}}} their 1987 album \\ 
the ancient city is built \underline{\textit{from}} limestone & the ancient city is built \underline{\textit{\hlc[light_green]{with}}} limestone \\ 
He \underline{\textit{also}} performed as an actor and a singer & He \underline{\textit{\hlc[light_green]{had}}} performed as an actor and a singer \\ 
While $<$unk$>$ \underline{\textit{had}} retained some control of the situation & While $<$unk$>$ \underline{\textit{\hlc[light_green]{also}}} retained some control of the situation \\ 
It is bordered \underline{\textit{on}} the east side by identical temples & It is bordered \underline{\textit{\hlc[light_green]{at}}} the east side by identical temples \\ 
a family that \underline{\textit{'s}} half black , half white , half American , half British & a family that \underline{\textit{\hlc[light_green]{was}}} half black , half white , half American , half British \\
they called out to the other passengers , who they thought \underline{\textit{were}} still alive . & they called out to the other passengers , who they thought \underline{\textit{\hlc[light_green]{,}}} still alive .\\ 
, \underline{\textit{but}} the complex is broken up by the heat of cooking & , \underline{\textit{\hlc[light_green]{and}}} the complex is broken up by the heat of cooking \\ \bottomrule
\end{tabular}}
\caption{Examples of input and output pairs using \model{} where the meaning and correctness are preserved.} \label{tab:good_examples}
\vspace{-4mm}
\end{table}

We also show in~\autoref{tab:good_examples} examples of input and output pairs obtained using \model{} and the best-of-many sampling strategy ($n$ = 20 samples). The hidden information in these examples was encoded using common tokens (e.g., preposition, articles, or auxiliary verbs), correct structure, and with a very comparable meaning to the input sentence.   

Even though fine-tuning and sampling improve the quality of the output to a large extent, we still observed some failure cases of incorrect replacements that cause grammatical and syntactic mistakes. Examples of such cases are shown in~\autoref{tab:failures}. One common failure mode happens when the type of the word changes. However, this cannot be entirely generalized as a failure case, e.g., some examples in~\autoref{tab:good_examples} removed a verb (``had'') with an adverb (``also'') while still being grammatically correct and also semantically consistent.

\newcolumntype{L}{>{\arraybackslash}m{5cm}}
\renewcommand{\arraystretch}{1.3}
\begin{table} [!b]
\centering
\resizebox{\linewidth}{!}{%
\begin{tabular}{L|L}
\toprule
\textbf{Input} & \textbf{\model{} output} \\\midrule
He is \underline{\textit{also}} present in the third original video animation & He is \underline{\textit{\hlc[light_red]{could}}} present in the third original video animation \\ 
resulting in a population decline \underline{\textit{as}} workers left for other areas & resulting in a population decline \underline{\textit{\hlc[light_red]{an}}} workers left for other areas \\
government officials had \underline{\textit{been}} suspected & government officials \underline{\textit{\hlc[light_red]{at}}} been suspected \\
who has \underline{\textit{been}} in office since 2009 & who has \underline{\textit{\hlc[light_red]{were}}} in office since 2009 \\
The M @-@ 82 designation was truncated \underline{\textit{at}} this time & The M @-@ 82 designation was truncated \underline{\textit{\hlc[light_red]{were}}} this time \\
\bottomrule
\end{tabular}}
\caption{Examples of failure modes showing input and output pairs with grammatical errors.} \label{tab:failures}
\vspace{-3mm}
\end{table}

\paragraph{Metrics Analysis}
We use the SBERT distance as an evaluation metric in addition to using the language model likelihood as a sorting metric. Therefore, we validate them by evaluating their recall of the best sample. On a subset of 100 input sentences, we use \model{} to generate 10 samples for each input sentence. We examine the possible sentences to find the best sample (in terms of both semantic similarity and grammatical correctness).  
For 92 out of 100 sentences, we found that the best sample is retrieved by either one or both metrics. This suggests that these two evaluation methods correlate with human annotation.

Since we use the language model to sort samples, we compare the best sample by the SBERT versus the best sample by the language model. On a subset of 200 sentences: the two metrics yielded the same sample in 44\% of the cases, while they yielded comparable samples in 25\%. The SBERT metric had a better sample in 9\%, while the language model had a better sample in 22\%. This shows that they have comparable performance, however, the language model was slightly better and more sensitive to grammar correctness, see Appendix~\ref{appendix_metrics} for such cases and for more qualitative analysis of the SBERT distance metric.

\begin{figure}[!b]
\centering
\vspace{-3mm}
\includegraphics[width=\linewidth]{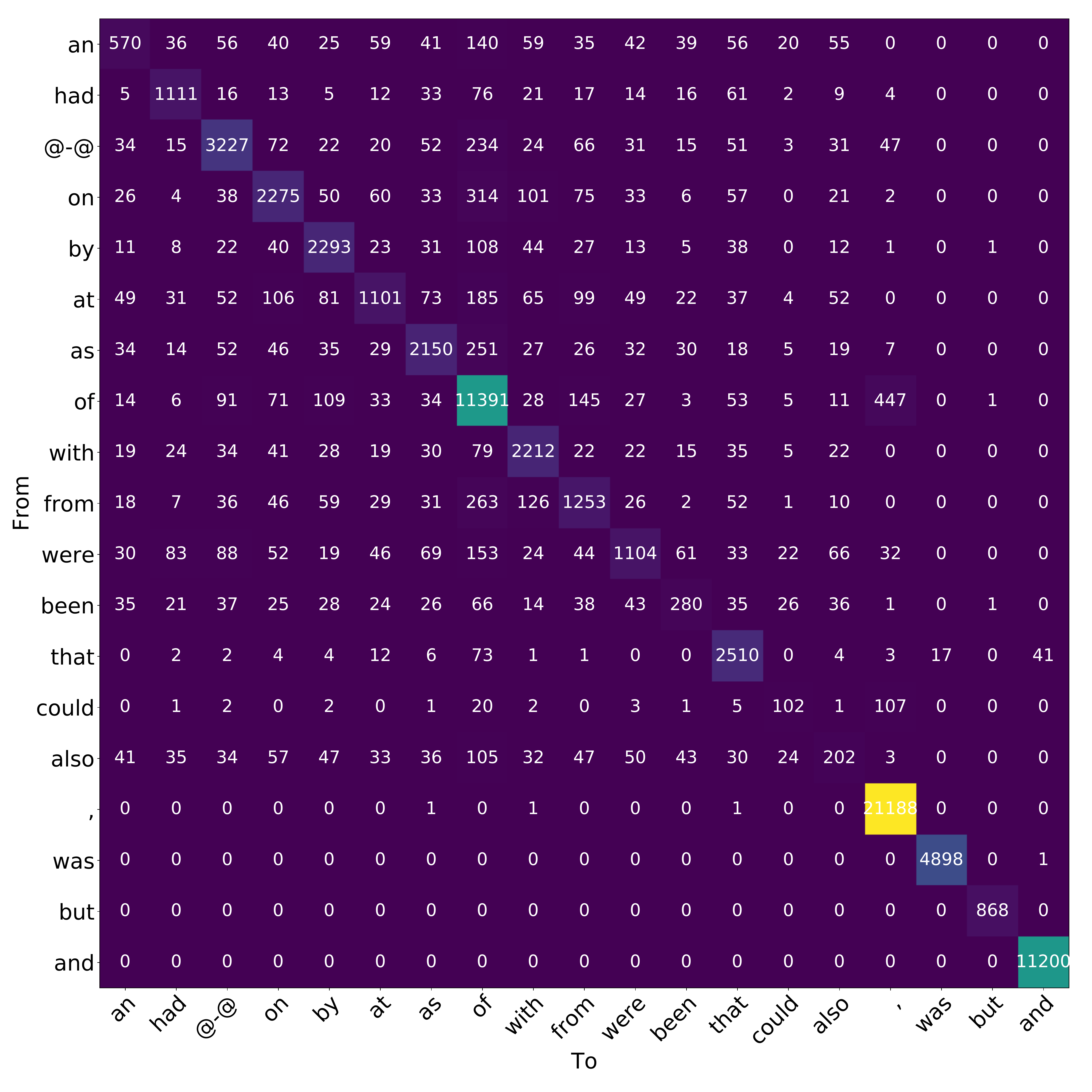}
\caption{A matrix of word changes' count from the original text to modified text using \model{}. \new{We show the no-diagonal transitions only in Appendix~\ref{vis_analysis}}.} 
\label{fig:matrix_words}
\vspace{-4mm}
\end{figure}

\paragraph{Visualizations and analysis} To further visualize the types of changes performed by the model at scale, we analyzed the count of transitions between words in the input to output text, as shown in~\autoref{fig:matrix_words}. We performed this analysis on the most commonly changed words (or changed to), shown in Appendix~\ref{vis_analysis}. Based on this analysis, we highlight the following observations: 1) Words are not consistently replaced since the diagonal line has a high count, meaning that in most occurrences, the model keeps these most commonly changed words unchanged. 2) There are no clear sparse transitions between words; meaning that a word is not always replaced by a specific word. 3) These message-holding words are not exclusive to the watermark occurrence. 4) These words are all from the most occurring words in the dataset (see~\autoref{fig:hist_words}).

These observations suggest that the model does not produce obvious artifacts or telltale signs in terms of changing the statistics of top words. In addition, there are no fixed rules that could describe or substitute the changes since it does not perform systematic changes between pairs of words. Thus, these factors contribute to the hidden encoding of information.

\subsection{Secrecy Evaluation} \label{secrecy_sec}
Our analysis shows that the model hides the information in a natural way by optimizing the translations to be as close and correct as possible to the original text. Nevertheless, we systematically study the secrecy by training classifiers to discriminate between encoded and non-encoded text. Although secret communication is not the primary task, the adversary might train a classifier in order to help identify and possibly remove the watermark (e.g., by adversarial training). Besides, the secrecy evaluation is another proxy for the naturalness of the encoding.
\begin{table} [!b]
\begin{center}
\resizebox{0.8\linewidth}{!}{%
\begin{tabular}{lll|l}
\toprule
\textbf{Model's variant} & \textbf{Sampling} & \textbf{Classifier} & \textbf{F1 score} \\ \midrule
\multirow{3}{*}{$-$ discriminator} & \multirow{3}{*}{1 sample} & Transformer & 0.89 \\  
& & \small{LSTM} &  0.80 \\ 
& & \small{BoW+SVM} &  0.98 \\ \hline
\multirow{4}{*}{$-$ fine-tuning} & \multirow{3}{*}{1 sample} & Transformer & 0.65 \\ 
& &\small{LSTM} & 0.56 \\ 
& &\small{BoW+SVM} & 0.63 \\ \cline{2-3}
& Best of 20 samples & Transformer & 0.59 \\  \hline 
\model{} & Best of 20 samples & Transformer & \textbf{0.53} \\
\bottomrule
\end{tabular}} 
\caption{Secrecy evaluation of different model's variants indicated by the F1 score of the adversary.} \label{tab:secrecy}
\end{center}
\vspace{-4mm}
\end{table}

We train the classifiers on non-paired (i.e., disjoint) watermarked and non-watermarked text, that is a reasonable assumption since the adversary does not have access to text before watermarking. We split the training data into two equal parts and use \model{} to watermark one of them by different messages sampled randomly. Similarly, we watermark the test and validation sets, but we use the entire sets for each class. We measure the classifiers' success by the F1 score; \textit{ideally, it should be at a chance level performance (0.5)}, denoting the adversary's failure.   

We compare the classifiers' success on different model's variants; the no-discriminator model, the no fine-tuning model, and the full \model{}. Since we use a transformer-based discriminator in \model{}, we also study the generalization with respect to different classifiers (transformer, LSTM, and Bag-of-Words with Support Vector Machines~\cite{suykens1999least}). The LSTM and transformer classifiers are trained on single length-varying sentences (similar to \model{} training). The BoW+SVM classifier is trained on word occurrences' feature vectors in paragraphs (up to 800 words). For each classifier, we optimize the hyperparameters and show the results of the most adversarial ones. Finally, we examine the effect of improving the quality (by sampling) on the classifiers' performance. 

As depicted in~\autoref{tab:secrecy}, we summarize our main conclusions as follows: 1) The no-discriminator model shows very high F1 scores (especially on longer sequences) indicating poor hiding capability, which supports our previous qualitative observations. 2) The adversarially trained models show considerably lower F1 scores across different classifiers. 3) Improving the quality (by sampling) helps to further reduce the classifiers' success. 4) The full \model{} with sampling model has a very close to the ideal random chance performance. 
\subsection{Robustness Evaluation} \label{sec:robustness}
Regardless of the secrecy, the adversary might apply different attacks on the watermarked text. 
The attacker's goal is to \textit{tamper with (remove) the watermark} with minimal changes to the text, in order to still have a high utility. Therefore, for each attack, we study the relationship between the drop in bit accuracy and text similarity. We focus on automated attacks without human inspection or modification~\cite{zellers2019defending,topkara2006hiding}. In Sections~\ref{sec:robust_random} and~\ref{sec:robust_denoising}, we study two attacks that assume that the adversary has general knowledge about using a translation-based watermarking scheme. However, \model{} is not explicitly trained or tailored to defend against these attacks. \new{We study adaptive attacks in Section~\ref{sec:robust_piracy} and~\ref{sec:dewatermarking} where the attacker can train their own watermarking model (\model{}$_{\text{adv}}$).}
\begin{figure}[!b]
\centering
\begin{subfigure}{0.45\columnwidth}
  \centering
  \includegraphics[width=\linewidth]{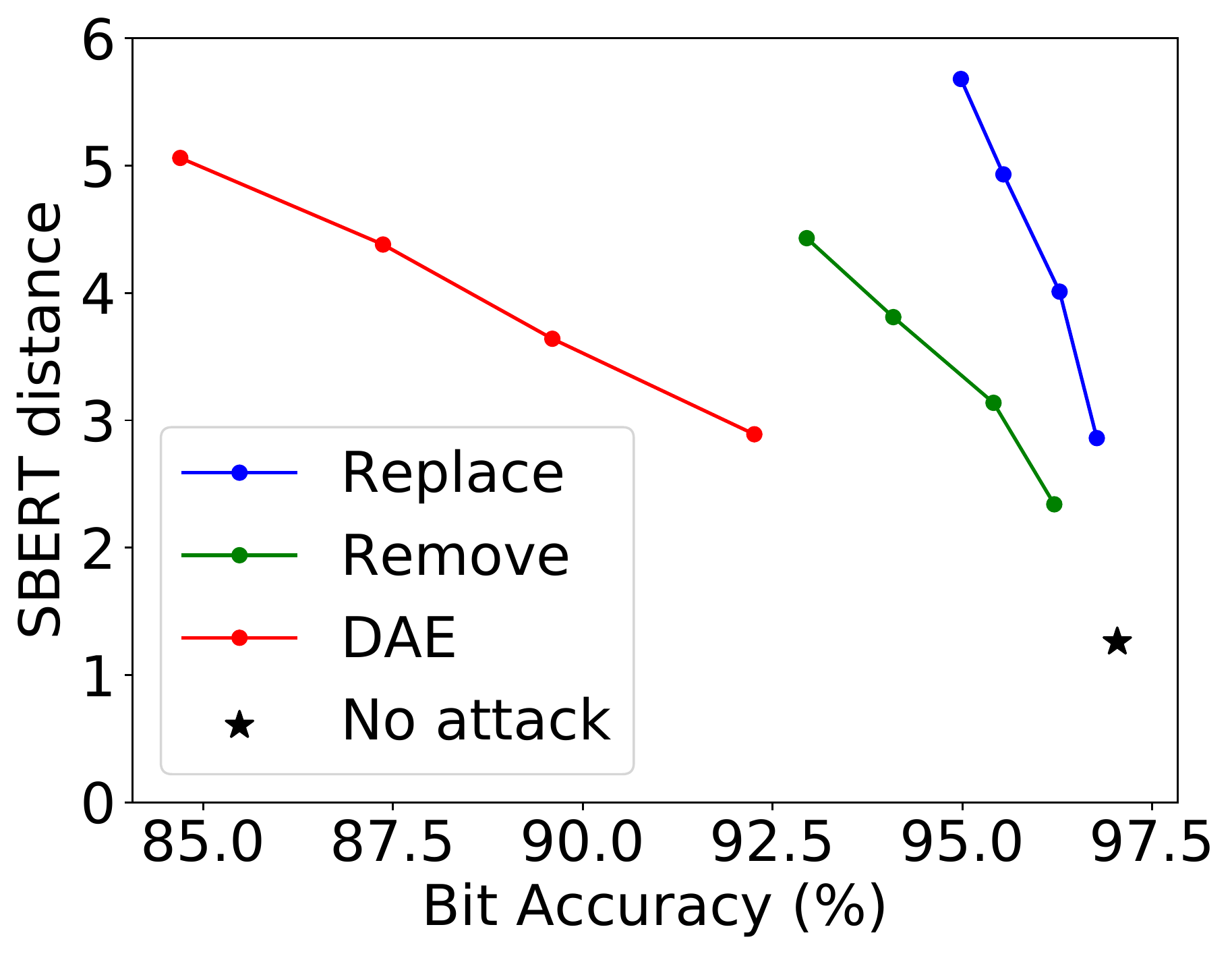}  
  \caption{1 sample}
  \label{fig:attacks-sample1}
\end{subfigure}
\begin{subfigure}{0.45\columnwidth}
  \centering
  \includegraphics[width=\linewidth]{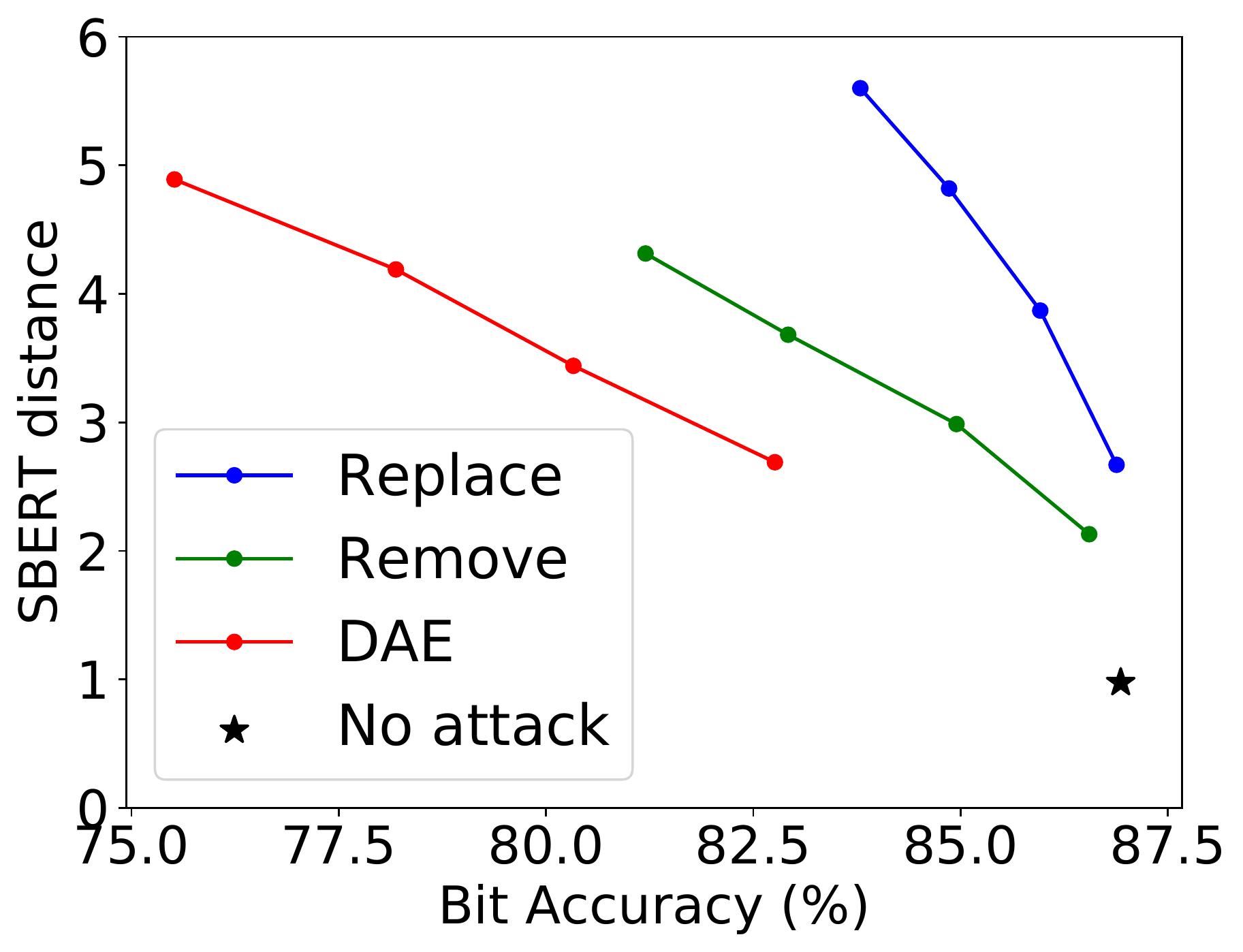}  
  \caption{Best of 20 samples}
  \label{fig:attacks-sample20}
\end{subfigure}
\caption{Random attacks (replacing and removing words) and denoising attack (applied to noisy text).} \label{fig:attacks}
\vspace{-3mm}
\end{figure}
\subsubsection{Random changes} \label{sec:robust_random} We consider two types of random changes to the watermarked text: removing words and replacing words with synonyms. 
For each attack, we change each word with a probability $p$ that we vary from 0.05 to 0.2. For each case, we compute the bit accuracy and SBERT distance. For synonym substitution, we use WordNet as a commonly used lexical database to find synonyms for words to be replaced. Instead of the naive random replacement, we assume that the attacker has access to a model like SBERT and uses it to select the synonym that gives the lowest distance from the set of possible synonyms.

We demonstrate the results of these two attacks in~\autoref{fig:attacks}. We perform these attacks on the output of \model{} using 1 sample in~\autoref{fig:attacks-sample1}, and 20 samples in~\autoref{fig:attacks-sample20}. The `remove' attack was found to be slightly more successful than the `replace' attack since not all words used to encode the message have synonyms in WordNet. However, For both the two attacks and the two operating points, the bit accuracy decreased by 0.05\% up to 6.5\%, while on the other hand, the SBERT increased by 86\% up to 577\%. This shows that the bit accuracy is robust to local changes and that the adversary needs to substantially change the text by random changes in order to make the watermark not usable. 

\subsubsection{Denoising} \label{sec:robust_denoising} 

Instead of random changes, a more knowledgeable adversary might train counter-models to \textit{reconstruct the text}. We train a transformer-based denoising autoencoder (DAE)~\cite{wang2019denoising} (sequence-to-sequence model) that is tasked to denoise an input sequence. We apply two types of noise to the input sequence ($S$): embedding dropout, and random word replacement, to form a corrupted sequence ($\hat{S}$). The noise is applied with a 5\% probability. $\hat{S}$ is then fed to the encoder. The decoder is tasked to reconstruct the original sequence $S$, and is fed the shifted $S$. The denoising maximizes $p(S|\hat{S})$, which can be described as~\cite{wang2019denoising}: $$ p(S|\hat{S}) = \prod_{i = 1}^{n} p(W_i|\hat{S},W_{<i}) $$ 
That is: predicting $W_i$ is conditioned on the full corrupted sequence $\hat{S}$ and the left side non-noisy sequence $W_{<i}$.

We perform the DAE training on non-watermarked text, and use the trained DAE to \textit{denoise the watermarked text} at test time. 
If the DAE was trained on watermarked text, it would be tasked to reconstruct it and therefore would not change the watermark. In contrast, with the current setup, the watermark could approximate the noise applied during the DAE training. The word replacement noise is in line with our watermarking scheme that is also based on word replacement, imitating an adversary with prior knowledge about our approach.

We hypothesize that a less natural encoding of the information would be more vulnerable to denoising than a more natural one. To validate this, we apply the DAE on the output of the three model's variants that we previously discussed, \textit{without} applying additional noise. We demonstrate this experiment in~\autoref{tab:denoising} in which we show the bit accuracy drop and the SBERT relative change. We summarize our interpretation as follows: 1) Improving the quality makes the denoising attack less effective; the `no-discriminator' model had a huge drop in bit accuracy and it reached a chance level, while it decreased slightly for the other variants, in particular, the better-quality \model{} model. 2) The DAE does not perfectly reconstruct the sentences and still introduces other changes besides the watermark's changes, this increased the SBERT distance for the two adversarially trained models. 3) On the other hand, the changes introduced to the `no-discriminator' model reduced the SBERT, indicating more successful denoising. We show examples of these different cases and more details about the DAE in Appendix~\ref{denoise}.
\begin{table} [!b]
\centering
\resizebox{0.7\linewidth}{!}{%
\begin{tabular}{c|cc}
\toprule
\textbf{Model} & \textbf{Bit accuracy drop} & \textbf{SBERT change} \\ \midrule
\model{} & 1.93\%\rpm0.19 & 30.77\%\rpm1.03\textcolor{red}{$\uparrow$}\\ 
$-$ fine-tuning & 5.21\%\rpm0.12 & 14.20\%\rpm1.11\textcolor{red}{$\uparrow$} \\
$-$ discriminator & 47.92\%\rpm0.44 & 15.93\%\rpm0.94\textcolor{dark_green}{$\downarrow$}\\ \bottomrule
\end{tabular}}
\caption{The relative performance of denoising attack applied to the 1-sample output. The no-attack performance is in~\autoref{tab:ablation}. 
} \label{tab:denoising}
\vspace{-4mm}
\end{table}

We then study a different attack variant where we introduce additional noise to the watermarked text before applying the DAE. This is, instead of applying random word replacement solely as an attack, we apply these random changes that might remove the watermark, and then use the DAE to generate a more realistic/smoothed sentence than the corrupted one. Similarly, we vary the probability of the noise and study the relationship between bit accuracy and SBERT distance. We show in~\autoref{fig:attacks} the performance of this attack in comparison with random changes alone. We found that this variant is more effective than using random changes; at the same level of SBERT, the drop in bit accuracy is higher. However, it still causes a significant increase in the SBERT distance (e.g., at a 10\% drop in bit accuracy, the SBERT increased by 319\%).
\subsubsection{Re-watermarking} \label{sec:robust_piracy} 
Watermark piracy~\cite{li2019piracy,fan2019rethinking} is an attack in model watermarking where the adversary's goal is to dispute or claim ownership of a stolen watermarked model by inserting their own watermark (to corrupt, exist alongside, or replace the original~\cite{li2019piracy}). We adapt re-watermarking as an attack on our method. Our threat model targets misuse instead of model stealing. Thus, we assume that the adversary's goal is to use the service/APIs without getting detected, instead of claiming ownership, i.e., to \textit{corrupt or tamper} with the owner's watermark and reduce its decoding accuracy.

We assume a strong adversary who has full knowledge about \model{} architecture, training details, access to the same training data, and the granularity of input sentences. \new{In our threat model, we consider a black-box scenario in which the adversary can train their own model and use it to insert a random watermark into the watermarked text, in hope of corrupting the original watermark and confusing the decoder. For completeness, we also show the less realistic white-box case when the re-watermarking is done using the same model.}

To run the \new{black-box attack, we train another model \model{}$_{\text{adv}}$ that is only different in initialization and reaches a comparable performance to \model{}.} We first watermark the text with \model{}, then we re-watermark it with a random message using \model{}$_{\text{adv}}$(\new{using the same or a different message was comparable}). We use the message decoder of \model{} (i.e., the first model) to decode the re-watermarked text and compute the matching with the original watermarks. As shown in~\autoref{tab:adaptive_attacks}, re-watermarking is stronger than denoising (\autoref{tab:denoising}) in decreasing the accuracy, but it also affects utility and perturbs the text due to double watermarking. This is in contrast with model watermarking where piracy can mostly retain the task performance~\cite{li2019piracy}. Also, the new watermarks did not completely corrupt the original ones \new{(i.e., the matching accuracy dropped to $\sim$85\%, while the accuracy of non-watermarked text is $\sim$50\%).} \new{A possible interpretation is that \model{}$_{\text{adv}}$ (i.e., another instance) does not necessarily use the same patterns (e.g., words to be replaced, added words, and locations) to encode the information and so it does not completely replace the original changes or confuse the first model's decoder.} We validated this by decoding one model's translation by the other model's decoder (\model{} and \model{}$_{\text{adv}}$ with no re-watermarking) and the matching accuracy was close to random chance (51.8\% and 53.2\%). \new{Our observation that different models produce different patterns is also consistent with previous data hiding studies in images (e.g.,~\cite{zhu2018hidden}).}
\begin{table} [!b]
\centering
\resizebox{0.85\linewidth}{!}{%
\begin{tabular}{cc|cc}
\toprule
\textbf{Attack} & & \textbf{Bit accuracy drop} & \textbf{SBERT change} \\ \midrule
\multirow{2}{*}{Re-watermarking} & white-box & 46.8\%\rpm0.46 & 23.4\%\rpm0.45\textcolor{red}{$\uparrow$}\\ 
& \textbf{black-box} & 12.6\%\rpm0.38 & 66.1\%\rpm1.89\textcolor{red}{$\uparrow$} \\ \midrule

\multirow{2}{*}{De-watermarking} & white-box & 41.6\%\rpm0.34 & 55.2\%\rpm0.39\textcolor{dark_green}{$\downarrow$}\\
 & \textbf{black-box} & 11.5\%\rpm0.32 & 11.3\%\rpm0.53\textcolor{red}{$\uparrow$} \\ \bottomrule
\end{tabular}}
\caption{\new{The relative performance of adaptive attacks that are applied to the 1-sample output in the white-box and black-box (which we mainly consider) settings.}} \label{tab:adaptive_attacks}
\vspace{-3mm}
\end{table}

Although the new watermarks in the re-watermarked text have high matching accuracy by the decoder of \model{}$_{\text{adv}}$ ($\sim$96\%), the adversary has no strong incentive or evidence to dispute provenance since 1) human-written text/news is mostly non-watermarked. 2) the presence of the original watermark by the decoder of \model{} indicates that the text was re-watermarked because otherwise, it should have a random chance matching.

\new{Finally, in the less realistic white-box case, re-watermarking with a different message overrides the original watermarks. We found that this is mainly because the model very often undoes the same changes done by the first watermarking step. A more detailed discussion on re-watermarking is in Appendix~\ref{sec:append_adaptive}.}

\subsubsection{De-watermarking} \label{sec:dewatermarking}
\new{Our last attack assumes that the adversary could use their knowledge about \model{} to \textit{de-watermark} the text, instead of adding a new watermark. Ideally, training an inverse de-watermarking model requires paired training data of the same text before and after watermarking, which is not feasible in our black-box scenario. To circumvent this, the adversary might try to train a denoising autoencoder (DAE$_\text{paired}$) on the paired data of \model{}$_{\text{adv}}$. The DAE$_\text{paired}$ takes the watermarked sentence as an input, with no additional noise, and should reconstruct the original non-watermarked sentence.}

\new{In~\autoref{tab:adaptive_attacks}, as a sanity check, we first evaluate the white-box case when the DAE$_\text{paired}$ is applied to \model{}$_{\text{adv}}$. This significantly reduced the bit accuracy (dropped to $\sim$55\%) and also the SBERT distance indicating a successful reconstruction. This is mainly because the DAE$_\text{paired}$ was exposed to the patterns the model \model{}$_{\text{adv}}$ frequently uses. In contrast, The black-box attack is significantly less successful (bit accuracy dropped to $\sim$86\%). However, in terms of the trade-off (i.e., decreasing bit accuracy with minimal changes), it may be the most effective one among the attacks we considered since it increased SBERT by $\sim$11\%, while re-watermarking increased it by $\sim$66\% with a comparable drop in accuracy.}

\new{The cases where the attack succeeded in the black-box setting were mainly either: 1) sentences with lower syntactic correctness or 2) similar changes to \model{}$_{\text{adv}}$. Otherwise, the attack was not successful due to the differences between the two models and the subtle encoding. Further improving the quality and diversity of watermarks both within and across models could help to defend against adaptive attacks, we leave that to future work. A detailed discussion is in Appendix~\ref{sec:append_adaptive}.}

\begin{table} [!b]
\begin{center}
\centering
\resizebox{0.6\linewidth}{!}{%
\begin{tabular}{l|ll|l}
\toprule
\textbf{Model}  & \textbf{Acc.} & \textbf{SBERT} & \textbf{F1} \\ \midrule
Synonym & 83.28\%\rpm0.62 & 3.62\rpm0.004 & 0.98  \\ 
\model{} & \textbf{86.3\%\rpm 0.99} & \textbf{0.944\rpm0.02} & \textbf{0.53}  \\
\bottomrule
\end{tabular}}
\caption{Comparing \model{} and synonym substitution in terms of bit accuracy, SBERT distance (showing the average and standard deviation of different runs), and F1 score.} \label{tab:baseline}
\end{center}
\vspace{-2mm}
\end{table}
\subsection{Baselines}
In this section, we compare \model{} against baselines. First, we implement a rule-based synonym substitution method that adopts the method in~\cite{topkara2006hiding}. Second, as an alternative to translation-based data hiding, we train an autoregressive language model, while simultaneously optimizing the message encoding and decoding. 

\subsubsection{Synonym substitution} \label{baseline}
The method in~\cite{topkara2006hiding} uses synonyms from WordNet to encode binary bits. The authors relied on ambiguity to make it hard for the adversary to perform automatic disambiguation. The ambiguity comes from encoding the message by synonyms that are ``homographs'' (having multiple meanings). 

We first form a list of words (in the dataset vocabulary) to be replaced by finding the words that have homographs (at least 2) in their synonym sets. We randomly divide each homograph set into two disjoint sets to encode `1' and `0' bits (bit-holding words). To have a unique encoding and decoding, we make sure no single word is assigned multiple values by being found in different words' synonym sets. Therefore, we skip a word if it was already assigned a value.

To encode the message, we find the occurrences of this list of words in the sentence. We replace each word with a `1' or `0' synonym according to the current bit in the message. We repeat until all bits are encoded. 
The decoding is then done by simple dictionary lookups. We use a message length of 4 bits similar to our setup. To have unique decoding, we replace any accidental occurrences of the `bit-holding' words in the original text with their corresponding synonym in the `replace' list. This prevents unintentional encoding. We highlight this important advantage of our model; \textit{\model{} does not impose such restrictions on the used words} since there are no words that are exclusive to the message encoding (as per~\autoref{fig:matrix_words}).  

\begin{figure} [!b]
\vspace{-3mm}
    \centering
    \includegraphics[width=0.8\linewidth]{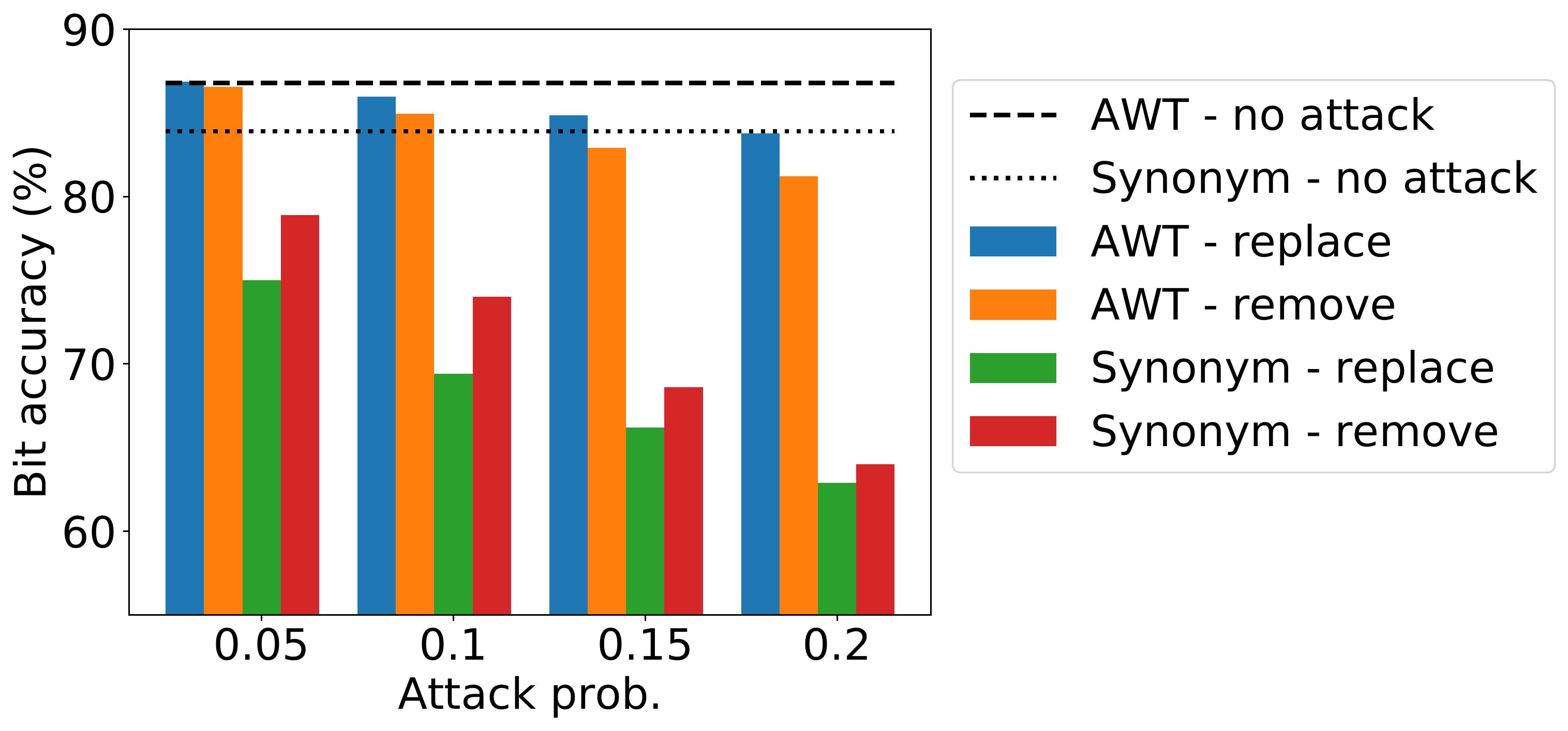}
    \caption{Comparing \model{} and the synonym substitution baseline bit accuracy under `remove' and `replace' attacks.}
    \label{fig:baseline_attacks}
\end{figure}

We again evaluate this baseline across the different evaluation axes: \textbf{effectiveness} (utility and bit accuracy), \textbf{secrecy}, and \textbf{robustness}. For effectiveness, we compute the bit accuracy and SBERT distance. For secrecy, we train a transformer-based classifier with the same setup as in Section~\ref{secrecy_sec}. We show a summary of these two evaluation factors in~\autoref{tab:baseline}. We compare the baseline against \model{} at a comparable bit accuracy level (resulted from sampling from the model) for a fair comparison. We summarize our findings as follows: 1) The message encoding was not successful in all sentences since not all sentences have words from the fixed `replace' list. 2) At an even higher bit accuracy level, \model{} has a considerably lower SBERT distance. 3) The baseline has a very high F1 score compared to the F1 score of \model{}.

For robustness, we apply the words removing and replacing attacks as in Section~\ref{sec:robustness}. We do not apply the DAE attack since some words used in the baseline method might be Out-of-Vocabulary words with respect to the DAE. As shown in~\autoref{fig:baseline_attacks}, the baseline is more sensitive to attacks since the encoding changes a larger amount of words compared to \model{}. The `replace' attack is even stronger than the `remove' attack; not only can it remove the original `bit holding' words, but it can also introduce accidental wrong encoding by adding other `bit holding' words instead of regular words. This analysis shows that \model{} achieves a significantly better trade-off between the three different evaluation axes.

\subsubsection{Generation-based hiding} \label{lstm}
An alternative strategy to the translation-based data hiding of the generated text (as a post-processing step) is to generate text that is already encoded with the input message~\cite{fang2017generating}. Unlike previous generation-based steganography work that relied on masking~\cite{fang2017generating}, we jointly train a language model (in contrast to \model{}, an autoencoder and thus bidirectional) with a message decoder. We used the same AWD-LSTM language model in~\cite{merity2017regularizing}. In our case, it takes the input word added to the input message at each time step and is trained to predict the next word given previous words. The message decoder takes the generated sequence and is trained to reconstruct the input message. The model is trained jointly with both losses. More details are in Appendix~\ref{apendix_gen}.
\begin{figure}[!t]
    \centering
    \includegraphics[width=0.85\linewidth]{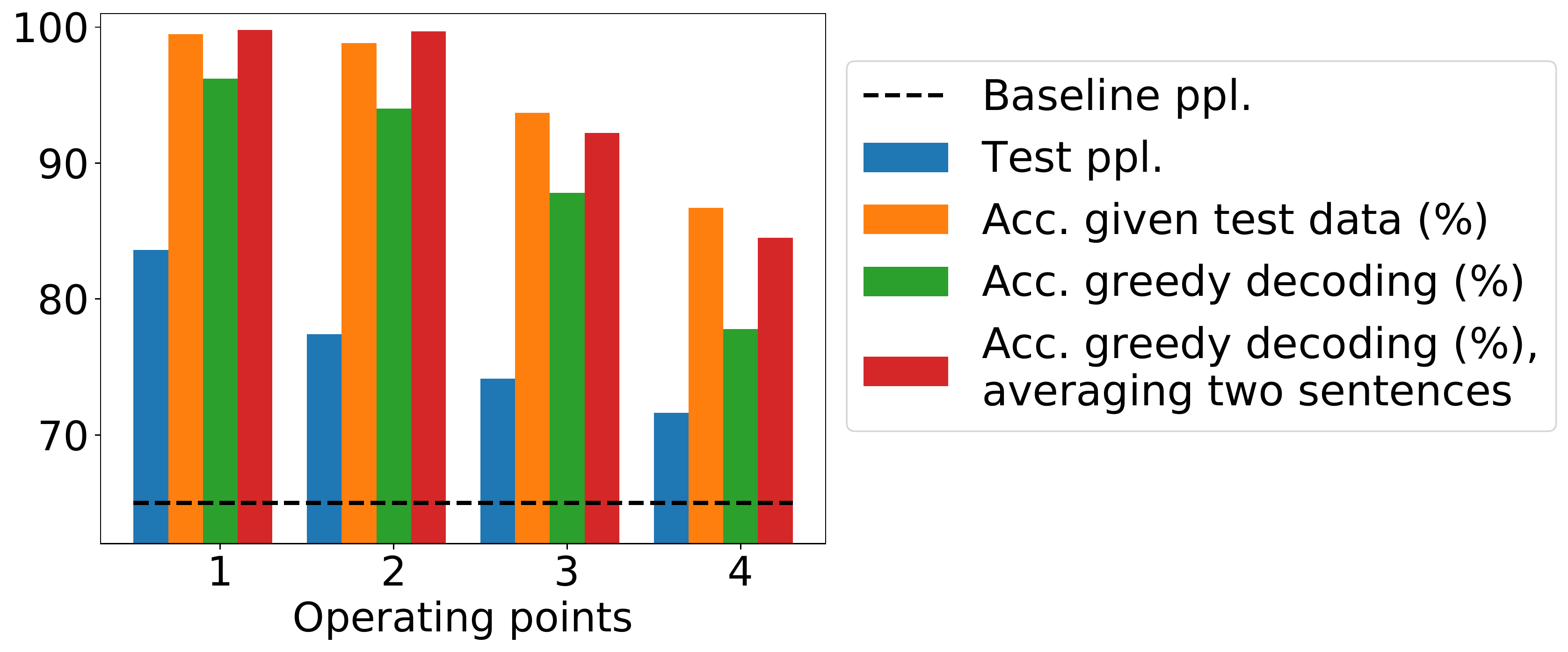}
    \caption{AWD-LSTM with data hiding showing different operating points that vary in perplexity and bit accuracy. The baseline perplexity is the AWD-LSTM without data hiding.} 
    \label{fig:lstm_gen_results}
\vspace{-3mm}
\end{figure}
We evaluate the model using the perplexity (i.e., exponential of the model loss, lower is better) and the bit accuracy. The ideal perplexity would be the perplexity of the AWD-LSTM without data hiding. As shown in~\autoref{fig:lstm_gen_results}, a very high bit accuracy can be achieved with around 12 points increase in perplexity (second operating point). The perplexity could be further reduced by tuning the weights between the two losses, which also decreases the bit accuracy. 

However, the main limitation is that message accuracy further drops during inference using recursive greedy decoding. Although it improves with averaging 2 sentences, it indicates that it would be even harder to retain high accuracy using other decoding strategies that introduce more variation in generations, such as top-$k$ or top-$p$ sampling~\cite{zellers2019defending,radford2019language,ippolito2020automatic,holtzman2019curious}. These strategies are typically used in open-ended generations due to having higher quality output~\cite{holtzman2019curious}. In contrast, \model{} does not suffer from these discrepancies since it can be applied agnostically on the generated sequence regardless of the decoding strategies and the language model.

\subsection{Human Evaluation}
It is common for machine translation and generation tasks to use human evaluation as an auxiliary evaluation besides the other metrics~\cite{shetty2018a4nt,zellers2019defending}. Therefore, we conducted a user study in order to evaluate the naturalness and correctness of our model, as a proxy to measure the stealthiness of the watermark. 

The study is conducted on the best variant of the model (with fine-tuning) with the best-of-20 samples strategy (bit accuracy: $\sim$86\%) and on the synonym baseline in Section~\ref{baseline} (bit accuracy: $\sim$83\%). It was performed by 6 judges who were asked to rate sentences with a Likert scale from 0 (lowest) to 5 (highest). The ratings are described with instructions that range from: \textit{`This sentence is completely understandable, natural, and grammatically correct'}, to: \textit{`This sentence is completely not understandable, unnatural, and you cannot get its main idea'}. We included different random sentences from \model{}, the synonym-based baseline, and the original non-watermarked text, displayed in a randomized order. The non-watermarked text works as a reference to the two approaches as the rating of the original text might not always be `5', since the dataset has processing tokens that might make it ambiguous. 
\begin{table} [!t]
\centering
\resizebox{0.65\linewidth}{!}{%
\begin{tabular}{lll}
\toprule
\textbf{\model{}} & \textbf{Synonym-baseline} & \textbf{Non-wm Dataset} \\ \midrule
\textcolor{black}{4.5\rpm0.76} & \textcolor{black}{3.42\rpm1.16} & \textcolor{black}{4.65\rpm0.62}\\ \bottomrule
\end{tabular}}
\caption{The results of a user study to rate (0 to 5) sentences from \model{}, the baseline, and non-watermarked text.} \label{tab:study_results} 
\vspace{-3mm}
\end{table}
We show the average rating for each case in~\autoref{tab:study_results}. \model{} had both higher ratings and less variance than the baseline. The high variance in the case of the baseline can be attributed to the observation that not all sentences were successfully encoded with the full 4 bits, and therefore, some of the sentences did not have a lot of changes. In the case of successful encoding, the sentence generally undergoes a lot of changes compared to \model{}, where usually not all of them are consistent. More details about the study are in Appendix~\ref{app_study}.

\section{Discussion} \label{sec:discussion}
We here discuss other several aspects of our work, other assumptions, scope, and limitations. \paragraph{Granularity}
We focus on the threat scenario of news articles that have a large number of tokens~\cite{zellers2019defending}. While other threats such as misinformation on Twitter are important~\cite{castillo2011information}, they are less relevant for machine-generated text that requires longer context for generation or detection (e.g., up to 1024 tokens in~\cite{zellers2019defending} or at least 192 tokens in~\cite{ippolito2020automatic}). Although it is possible to encode 4 bits in a short text using our approach, this short message is not enough for confidence calculation. Verification on short text would require a longer watermark and thus, severely affect the text, as the task of data hiding in text is inherently more difficult than its counterpart in images.

\paragraph{False positives} When concatenating several 4 bits messages, the false positives can be directly controlled by the $p$-value threshold~\cite{venugopal2011watermarking}, \new{since the accuracy of non-watermarked text is at the chance level}. We evaluated the thresholds of 0.05 and 0.01 (\autoref{fig:long_seq}). One possible way to improve false positives is to use multiple confidence thresholds with an increasing alarm for false positives. \new{Then, if the watermark verification is in the low-confidence range, our solution could potentially be combined with other previously introduced fake news defenses (e.g., discriminators~\cite{zellers2019defending,tan2020detecting,ippolito2020automatic}, automated fact-checking and stance detection~\cite{thorne2017fake}). On the other hand, human fact-checking is still a standard solution for news verification~\cite{hassan2017toward}, while automated solutions aim to reduce these human efforts, humans can still be kept in-the-loop for verifying low-confidence instances, reducing the otherwise full effort to verify all articles.}

\paragraph{Human editing} The black-box APIs might be used legitimately for partial text completion or suggestions to some of the sentences with further interactive human editing. However, the main threat we consider is misusing these models in an unintended way to generate entire articles at scale, possibly conditioned on a headline or a context. Although the threat of combining the generation with human editing is conceivable, it is a limited use-case for the adversary since it reduces the scalability and adds manual time-consuming efforts, largely reducing the advantages of using machine-generated text.

\paragraph{Possible release of models} We assume black-box access to the language model, however, it is still an important step towards defending against misuse. While GPT-2 was released after a staged release, this might not be the case for future models. By the time of writing this paper, OpenAI is not open-sourcing GPT-3, and it is only available through commercial APIs~\cite{openaigpt3}, where one of the announced reasons is to prevent or limit misuse. Additionally, our solution is also helpful for scenarios where a general language model like GPT-2 is fine-tuned by a service for specific tasks or domains.

\paragraph{Training in-house language models} Another option for the adversary to circumvent defenses is to train their own language model. However, training modern state-of-the-art language models, including massive datasets collection, is a very expensive and time-consuming process that requires significant technical expertise, and the cost is progressively increasing. Training Grover~\cite{zellers2019defending} requires around \$35k using AWS. Training a 1.5 billion parameter model is estimated at \$1.6m~\cite{sharir2020cost}. 
The 175B GPT-3 training cost is estimated at \$4.6m~\cite{lambda}. Final actual costs could be even higher due to multiple runs of hyperparameters tuning.

\paragraph{Watermarks regulation} Since we use a multi-bit watermarking scheme, our scenario can be extended to watermarking multiple models offered by different owners. However, this would require further cooperation of models' owners or a potential regulation by a trusted regulatory third party that handles the distribution of watermarks, and sharing the watermarks' encoder and decoder. We hope that our work opens follow-up future research and discussions on the regulation and proactive protective release strategies of such technologies.

\section{Conclusion}
In this paper, we present \model{}, a new framework for language watermarking as a potential solution towards marking and tracing the provenance of machine-generated text. \model{} is the first end-to-end data hiding solution for natural text and is optimized to unobstructively encode the cover text by adversarial training and other smoothing auxiliary losses. 

\model{} achieves more flexibility and a significantly better trade-off between the different evaluation axes (effectiveness, secrecy, and robustness), in terms of quantitative, qualitative, and human evaluations, compared to a rule-based synonym substitution baseline. Our work offers a new research area towards improving and robustifying automatic data hiding in natural language, similar to its precedent in images. 

\bibliographystyle{IEEEtran}
\bibliography{ref.bib}

\begin{thebibliography}{10}
\providecommand{\url}[1]{#1}
\csname url@samestyle\endcsname
\providecommand{\newblock}{\relax}
\providecommand{\bibinfo}[2]{#2}
\providecommand{\BIBentrySTDinterwordspacing}{\spaceskip=0pt\relax}
\providecommand{\BIBentryALTinterwordstretchfactor}{4}
\providecommand{\BIBentryALTinterwordspacing}{\spaceskip=\fontdimen2\font plus
\BIBentryALTinterwordstretchfactor\fontdimen3\font minus
  \fontdimen4\font\relax}
\providecommand{\BIBforeignlanguage}[2]{{%
\expandafter\ifx\csname l@#1\endcsname\relax
\typeout{** WARNING: IEEEtran.bst: No hyphenation pattern has been}%
\typeout{** loaded for the language `#1'. Using the pattern for}%
\typeout{** the default language instead.}%
\else
\language=\csname l@#1\endcsname
\fi
#2}}
\providecommand{\BIBdecl}{\relax}
\BIBdecl

\bibitem{vaswani2017attention}
A.~Vaswani, N.~Shazeer, N.~Parmar, J.~Uszkoreit, L.~Jones, A.~N. Gomez,
  {\L}.~Kaiser, and I.~Polosukhin, ``Attention is all you need,'' in
  \emph{Advances in Neural Information Processing Systems}, 2017.

\bibitem{conneau2019cross}
A.~Conneau and G.~Lample, ``Cross-lingual language model pretraining,'' in
  \emph{Advances in Neural Information Processing Systems}, 2019.

\bibitem{yang2019xlnet}
Z.~Yang, Z.~Dai, Y.~Yang, J.~Carbonell, R.~R. Salakhutdinov, and Q.~V. Le,
  ``Xlnet: Generalized autoregressive pretraining for language understanding,''
  in \emph{Advances in Neural Information Processing Systems}, 2019.

\bibitem{devlin2018bert}
J.~Devlin, M.-W. Chang, K.~Lee, and K.~Toutanova, ``Bert: Pre-training of deep
  bidirectional transformers for language understanding,'' in \emph{North
  American Chapter of the Association for Computational Linguistics (NAACL)},
  2019.

\bibitem{zellers2019defending}
R.~Zellers, A.~Holtzman, H.~Rashkin, Y.~Bisk, A.~Farhadi, F.~Roesner, and
  Y.~Choi, ``Defending against neural fake news,'' in \emph{Advances in Neural
  Information Processing Systems}, 2019.

\bibitem{peters2018deep}
M.~E. Peters, M.~Neumann, M.~Iyyer, M.~Gardner, C.~Clark, K.~Lee, and
  L.~Zettlemoyer, ``Deep contextualized word representations,'' in \emph{North
  American Chapter of the Association for Computational Linguistics: Human
  Language Technologies (NAACL-HLT)}, 2018.

\bibitem{howard2018universal}
J.~Howard and S.~Ruder, ``Universal language model fine-tuning for text
  classification,'' in \emph{the 56th Annual Meeting of the Association for
  Computational Linguistics (ACL)}, 2018.

\bibitem{radford2019language}
A.~Radford, J.~Wu, R.~Child, D.~Luan, D.~Amodei, and I.~Sutskever, ``Language
  models are unsupervised multitask learners,'' OpenAI, Tech. Rep., 2019.

\bibitem{radfordimproving}
A.~Radford, K.~Narasimhan, T.~Salimans, and I.~Sutskever, ``Improving language
  understanding by generative pre-training,'' OpenAI, Tech. Rep., 2018.

\bibitem{wang2019denoising}
L.~Wang, W.~Zhao, R.~Jia, S.~Li, and J.~Liu, ``Denoising based
  sequence-to-sequence pre-training for text generation,'' in \emph{Empirical
  Methods in Natural Language Processing and the 9th International Joint
  Conference on Natural Language Processing (EMNLP-IJCNLP)}, 2019.

\bibitem{brown2020language}
T.~B. Brown, B.~Mann, N.~Ryder, M.~Subbiah, J.~Kaplan, P.~Dhariwal,
  A.~Neelakantan, P.~Shyam, G.~Sastry, A.~Askell \emph{et~al.}, ``Language
  models are few-shot learners,'' \emph{arXiv preprint arXiv:2005.14165}, 2020.

\bibitem{solaiman2019release}
I.~Solaiman, M.~Brundage, J.~Clark, A.~Askell, A.~Herbert-Voss, J.~Wu,
  A.~Radford, and J.~Wang, ``Release strategies and the social impacts of
  language models,'' \emph{arXiv preprint arXiv:1908.09203}, 2019.

\bibitem{ippolito2020automatic}
D.~Ippolito, D.~Duckworth, C.~Callison-Burch, and D.~Eck, ``Automatic detection
  of generated text is easiest when humans are fooled,'' in \emph{the 58th
  Annual Meeting of the Association for Computational Linguistics (ACL)}, 2020.

\bibitem{adelani2020generating}
D.~I. Adelani, H.~Mai, F.~Fang, H.~H. Nguyen, J.~Yamagishi, and I.~Echizen,
  ``Generating sentiment-preserving fake online reviews using neural language
  models and their human-and machine-based detection,'' in \emph{International
  Conference on Advanced Information Networking and Applications}.\hskip 1em
  plus 0.5em minus 0.4em\relax Springer, 2020.

\bibitem{openaigpt3}
\BIBentryALTinterwordspacing
OpenAI, ``Openai api.'' [Online]. Available:
  \url{https://openai.com/blog/openai-api/}
\BIBentrySTDinterwordspacing

\bibitem{krishna2019thieves}
K.~Krishna, G.~S. Tomar, A.~P. Parikh, N.~Papernot, and M.~Iyyer, ``Thieves on
  sesame street! model extraction of bert-based apis,'' in \emph{International
  Conference on Learning Representations (ICLR)}, 2020.

\bibitem{orekondy2019knockoff}
T.~Orekondy, B.~Schiele, and M.~Fritz, ``Knockoff nets: Stealing functionality
  of black-box models,'' in \emph{the IEEE Conference on Computer Vision and
  Pattern Recognition (CVPR)}, 2019.

\bibitem{tramer2016stealing}
F.~Tram{\`e}r, F.~Zhang, A.~Juels, M.~K. Reiter, and T.~Ristenpart, ``Stealing
  machine learning models via prediction apis,'' in \emph{25th USENIX Security
  Symposium (USENIX Security 16)}, 2016.

\bibitem{papernot2017practical}
N.~Papernot, P.~McDaniel, I.~Goodfellow, S.~Jha, Z.~B. Celik, and A.~Swami,
  ``Practical black-box attacks against machine learning,'' in \emph{the ACM
  Asia Conference on Computer and Communications Security (AsiaCCS)}, 2017.

\bibitem{topkara2006hiding}
U.~Topkara, M.~Topkara, and M.~J. Atallah, ``The hiding virtues of ambiguity:
  quantifiably resilient watermarking of natural language text through synonym
  substitutions,'' in \emph{the 8th Workshop on Multimedia and Security}, 2006.

\bibitem{chang2010practical}
C.~Y. Chang and S.~Clark, ``Practical linguistic steganography using contextual
  synonym substitution and vertex colour coding,'' in \emph{Empirical Methods
  in Natural Language Processing (EMNLP)}, 2010.

\bibitem{topkara2006words}
M.~Topkara, U.~Topkara, and M.~J. Atallah, ``Words are not enough: sentence
  level natural language watermarking,'' in \emph{the 4th ACM International
  Workshop on Contents Protection and Security}, 2006.

\bibitem{meral2007syntactic}
H.~M. Meral, E.~Sevinc, E.~{\"U}nkar, B.~Sankur, A.~S. {\"O}zsoy, and
  T.~G{\"u}ng{\"o}r, ``Syntactic tools for text watermarking,'' in
  \emph{Security, Steganography, and Watermarking of Multimedia Contents
  IX}.\hskip 1em plus 0.5em minus 0.4em\relax International Society for Optics
  and Photonics, 2007.

\bibitem{chiang2003natural}
Y.-L. Chiang, L.-P. Chang, W.-T. Hsieh, and W.-C. Chen, ``Natural language
  watermarking using semantic substitution for chinese text,'' in
  \emph{International Workshop on Digital Watermarking}.\hskip 1em plus 0.5em
  minus 0.4em\relax Springer, 2003.

\bibitem{halvani2013natural}
O.~Halvani, M.~Steinebach, P.~Wolf, and R.~Zimmermann, ``Natural language
  watermarking for german texts,'' in \emph{the first ACM Workshop on
  Information Hiding and Multimedia Security}, 2013.

\bibitem{Ingemar2007digital}
I.~Cox, M.~Miller, J.~Bloom, J.~Fridrich, and T.~Kalker, \emph{Digital
  Watermarking and Steganography}, 2nd~ed.\hskip 1em plus 0.5em minus
  0.4em\relax Morgan Kaufmann Publishers Inc., 2007.

\bibitem{zhu2018hidden}
J.~Zhu, R.~Kaplan, J.~Johnson, and L.~Fei-Fei, ``Hidden: Hiding data with deep
  networks,'' in \emph{European Conference on Computer Vision (ECCV)}, 2018.

\bibitem{baluja2017hiding}
S.~Baluja, ``Hiding images in plain sight: Deep steganography,'' in
  \emph{Advances in Neural Information Processing Systems}, 2017.

\bibitem{hayes2017generating}
J.~Hayes and G.~Danezis, ``Generating steganographic images via adversarial
  training,'' in \emph{Advances in Neural Information Processing Systems},
  2017.

\bibitem{vukotic2018deep}
V.~Vukoti{\'c}, V.~Chappelier, and T.~Furon, ``Are deep neural networks good
  for blind image watermarking?'' in \emph{the IEEE International Workshop on
  Information Forensics and Security (WIFS)}, 2018.

\bibitem{zhang2019invisible}
R.~Zhang, S.~Dong, and J.~Liu, ``Invisible steganography via generative
  adversarial networks,'' \emph{Multimedia Tools and Applications}, vol.~78,
  no.~7, pp. 8559--8575, 2019.

\bibitem{sutskever2014sequence}
I.~Sutskever, O.~Vinyals, and Q.~V. Le, ``Sequence to sequence learning with
  neural networks,'' in \emph{Advances in Neural Information Processing
  Systems}, 2014.

\bibitem{goodfellow2014generative}
I.~Goodfellow, J.~Pouget-Abadie, M.~Mirza, B.~Xu, D.~Warde-Farley, S.~Ozair,
  A.~Courville, and Y.~Bengio, ``Generative adversarial nets,'' in
  \emph{Advances in Neural Information Processing Systems}, 2014.

\bibitem{kamaruddin2018review}
N.~S. Kamaruddin, A.~Kamsin, L.~Y. Por, and H.~Rahman, ``A review of text
  watermarking: theory, methods, and applications,'' \emph{IEEE Access},
  vol.~6, pp. 8011--8028, 2018.

\bibitem{podilchuk2001digital}
C.~I. Podilchuk and E.~J. Delp, ``Digital watermarking: algorithms and
  applications,'' \emph{IEEE Signal Processing Magazine}, vol.~18, no.~4, pp.
  33--46, 2001.

\bibitem{singh2013survey}
N.~Singh, M.~Jain, and S.~Sharma, ``A survey of digital watermarking
  techniques,'' \emph{International Journal of Modern Communication
  Technologies and Research}, vol.~1, no.~6, p. 265852, 2013.

\bibitem{brassil1995electronic}
J.~T. Brassil, S.~Low, N.~F. Maxemchuk, and L.~O'Gorman, ``Electronic marking
  and identification techniques to discourage document copying,'' \emph{IEEE
  Journal on Selected Areas in Communications}, vol.~13, no.~8, pp. 1495--1504,
  1995.

\bibitem{topkara2005natural}
M.~Topkara, C.~M. Taskiran, and E.~J. Delp~III, ``Natural language
  watermarking,'' in \emph{Security, Steganography, and Watermarking of
  Multimedia Contents VII}.\hskip 1em plus 0.5em minus 0.4em\relax
  International Society for Optics and Photonics, 2005.

\bibitem{topkara2006natural}
M.~Topkara, G.~Riccardi, D.~Hakkani-T{\"u}r, and M.~J. Atallah, ``Natural
  language watermarking: Challenges in building a practical system,'' in
  \emph{Security, Steganography, and Watermarking of Multimedia Contents
  VIII}.\hskip 1em plus 0.5em minus 0.4em\relax International Society for
  Optics and Photonics, 2006.

\bibitem{meral2009natural}
H.~M. Meral, B.~Sankur, A.~S. {\"O}zsoy, T.~G{\"u}ng{\"o}r, and
  E.~Sevin{\c{c}}, ``Natural language watermarking via morphosyntactic
  alterations,'' \emph{Computer Speech \& Language}, vol.~23, no.~1, pp.
  107--125, 2009.

\bibitem{wilson2014linguistic}
A.~Wilson, P.~Blunsom, and A.~D. Ker, ``Linguistic steganography on twitter:
  hierarchical language modeling with manual interaction,'' in \emph{Media
  Watermarking, Security, and Forensics}.\hskip 1em plus 0.5em minus
  0.4em\relax International Society for Optics and Photonics, 2014.

\bibitem{wilson2015detection}
A.~Wilson, P.~Blunsom, and A.~Ker, ``Detection of steganographic techniques on
  twitter,'' in \emph{Empirical Methods in Natural Language Processing
  (EMNLP)}, 2015.

\bibitem{wilson2016avoiding}
A.~Wilson and A.~D. Ker, ``Avoiding detection on twitter: embedding strategies
  for linguistic steganography,'' \emph{Electronic Imaging}, vol. 2016, no.~8,
  pp. 1--9, 2016.

\bibitem{shirali2008new}
M.~H. Shirali-Shahreza and M.~Shirali-Shahreza, ``A new synonym text
  steganography,'' in \emph{the IEEE International Conference on Intelligent
  Information Hiding and Multimedia Signal Processing}, 2008.

\bibitem{fang2017generating}
T.~Fang, M.~Jaggi, and K.~Argyraki, ``Generating steganographic text with
  lstms,'' in \emph{the 55th Annual Meeting of the Association for
  Computational Linguistics-Student Research Workshop}, 2017.

\bibitem{li2019prove}
Z.~Li, C.~Hu, Y.~Zhang, and S.~Guo, ``How to prove your model belongs to you: a
  blind-watermark based framework to protect intellectual property of dnn,'' in
  \emph{the 35th Annual Computer Security Applications Conference (ACSAC)},
  2019.

\bibitem{lukas2019deep}
N.~Lukas, Y.~Zhang, and F.~Kerschbaum, ``Deep neural network fingerprinting by
  conferrable adversarial examples,'' \emph{arXiv preprint arXiv:1912.00888},
  2019.

\bibitem{adi2018turning}
Y.~Adi, C.~Baum, M.~Cisse, B.~Pinkas, and J.~Keshet, ``Turning your weakness
  into a strength: Watermarking deep neural networks by backdooring,'' in
  \emph{27th USENIX Security Symposium (USENIX Security 18)}, 2018.

\bibitem{le2020adversarial}
E.~Le~Merrer, P.~Perez, and G.~Tr{\'e}dan, ``Adversarial frontier stitching for
  remote neural network watermarking,'' \emph{Neural Computing and
  Applications}, vol.~32, no.~13, pp. 9233--9244, 2020.

\bibitem{uchida2017embedding}
Y.~Uchida, Y.~Nagai, S.~Sakazawa, and S.~Satoh, ``Embedding watermarks into
  deep neural networks,'' in \emph{International Conference on Multimedia
  Retrieval (ICMR)}, 2017.

\bibitem{chen2018deepmarks}
H.~Chen, B.~D. Rouhani, C.~Fu, J.~Zhao, and F.~Koushanfar, ``Deepmarks: A
  secure fingerprinting framework for digital rights management of deep
  learning models,'' in \emph{International Conference on Multimedia Retrieval
  (ICMR)}, 2019.

\bibitem{darvish2019deepsigns}
B.~Darvish~Rouhani, H.~Chen, and F.~Koushanfar, ``Deepsigns: an end-to-end
  watermarking framework for ownership protection of deep neural networks,'' in
  \emph{the 24th International Conference on Architectural Support for
  Programming Languages and Operating Systems}, 2019.

\bibitem{gu2017badnets}
T.~Gu, B.~Dolan-Gavitt, and S.~Garg, ``Badnets: Identifying vulnerabilities in
  the machine learning model supply chain,'' \emph{arXiv preprint
  arXiv:1708.06733}, 2017.

\bibitem{zhang2018protecting}
J.~Zhang, Z.~Gu, J.~Jang, H.~Wu, M.~P. Stoecklin, H.~Huang, and I.~Molloy,
  ``Protecting intellectual property of deep neural networks with
  watermarking,'' in \emph{the ACM Asia Conference on Computer and
  Communications Security (AsiaCCS)}, 2018.

\bibitem{jia2020entangled}
H.~Jia, C.~A. Choquette-Choo, and N.~Papernot, ``Entangled watermarks as a
  defense against model extraction,'' \emph{arXiv preprint arXiv:2002.12200},
  2020.

\bibitem{li2019piracy}
H.~Li, E.~Wenger, B.~Y. Zhao, and H.~Zheng, ``Piracy resistant watermarks for
  deep neural networks,'' \emph{arXiv preprint arXiv:1910.01226}, 2019.

\bibitem{yu2019attributing}
N.~Yu, L.~S. Davis, and M.~Fritz, ``Attributing fake images to gans: Learning
  and analyzing gan fingerprints,'' in \emph{the IEEE International Conference
  on Computer Vision (ICCV)}, 2019.

\bibitem{wang2020cnn}
S.-Y. Wang, O.~Wang, R.~Zhang, A.~Owens, and A.~A. Efros, ``Cnn-generated
  images are surprisingly easy to spot... for now,'' in \emph{the IEEE
  Conference on Computer Vision and Pattern Recognition (CVPR)}, 2020.

\bibitem{carlini2020evading}
N.~Carlini and H.~Farid, ``Evading deepfake-image detectors with white-and
  black-box attacks,'' in \emph{the IEEE Conference on Computer Vision and
  Pattern Recognition (CVPR) Workshops}, 2020.

\bibitem{zhang2020not}
B.~Zhang, J.~P. Zhou, I.~Shumailov, and N.~Papernot, ``Not my deepfake: Towards
  plausible deniability for machine-generated media,'' \emph{arXiv preprint
  arXiv:2008.09194}, 2020.

\bibitem{stern2019insertion}
M.~Stern, W.~Chan, J.~Kiros, and J.~Uszkoreit, ``Insertion transformer:
  Flexible sequence generation via insertion operations,'' in
  \emph{International Conference on Machine Learning (ICML)}, 2019.

\bibitem{caccia2018language}
M.~Caccia, L.~Caccia, W.~Fedus, H.~Larochelle, J.~Pineau, and L.~Charlin,
  ``Language gans falling short,'' in \emph{International Conference on
  Learning Representations (ICLR)}, 2020.

\bibitem{hosseini2017attacking}
H.~Hosseini, B.~Xiao, A.~Clark, and R.~Poovendran, ``Attacking automatic video
  analysis algorithms: A case study of google cloud video intelligence api,''
  in \emph{Multimedia Privacy and Security}, 2017.

\bibitem{mariconti2019you}
E.~Mariconti, G.~Suarez-Tangil, J.~Blackburn, E.~De~Cristofaro, N.~Kourtellis,
  I.~Leontiadis, J.~L. Serrano, and G.~Stringhini, ``"you know what to do"
  proactive detection of youtube videos targeted by coordinated hate attacks,''
  \emph{Human-Computer Interaction}, vol.~3, no. CSCW, pp. 1--21, 2019.

\bibitem{bahdanau2014neural}
D.~Bahdanau, K.~Cho, and Y.~Bengio, ``Neural machine translation by jointly
  learning to align and translate,'' in \emph{International Conference on
  Learning Representations (ICLR)}, 2015.

\bibitem{shetty2017speaking}
R.~Shetty, M.~Rohrbach, L.~Anne~Hendricks, M.~Fritz, and B.~Schiele, ``Speaking
  the same language: Matching machine to human captions by adversarial
  training,'' in \emph{the IEEE International Conference on Computer Vision
  (ICCV)}, 2017.

\bibitem{choi2019encoding}
K.~Choi, C.~Hawthorne, I.~Simon, M.~Dinculescu, and J.~Engel, ``Encoding
  musical style with transformer autoencoders,'' \emph{arXiv preprint
  arXiv:1912.05537}, 2019.

\bibitem{jang2016categorical}
E.~Jang, S.~Gu, and B.~Poole, ``Categorical reparameterization with
  gumbel-softmax,'' in \emph{International Conference on Learning
  Representations (ICLR)}, 2017.

\bibitem{kusner2016gans}
M.~J. Kusner and J.~M. Hern{\'a}ndez-Lobato, ``Gans for sequences of discrete
  elements with the gumbel-softmax distribution,'' \emph{arXiv preprint
  arXiv:1611.04051}, 2016.

\bibitem{merity2017regularizing}
S.~Merity, N.~S. Keskar, and R.~Socher, ``Regularizing and optimizing lstm
  language models,'' in \emph{International Conference on Learning
  Representations (ICLR)}, 2018.

\bibitem{inan2016tying}
H.~Inan, K.~Khosravi, and R.~Socher, ``Tying word vectors and word classifiers:
  A loss framework for language modeling,'' in \emph{International Conference
  on Learning Representations (ICLR)}, 2017.

\bibitem{shetty2018a4nt}
R.~Shetty, B.~Schiele, and M.~Fritz, ``A4nt: author attribute anonymity by
  adversarial training of neural machine translation,'' in \emph{27th USENIX
  Security Symposium (USENIX Security 18)}, 2018.

\bibitem{conneau2017supervised}
A.~Conneau, D.~Kiela, H.~Schwenk, L.~Barrault, and A.~Bordes, ``Supervised
  learning of universal sentence representations from natural language
  inference data,'' in \emph{Empirical Methods in Natural Language Processing
  (EMNLP)}, 2017.

\bibitem{bowman2015large}
S.~Bowman, G.~Angeli, C.~Potts, and C.~D. Manning, ``A large annotated corpus
  for learning natural language inference,'' in \emph{Empirical Methods in
  Natural Language Processing (EMNLP)}, 2015.

\bibitem{dai2019transformer}
Z.~Dai, Z.~Yang, Y.~Yang, J.~G. Carbonell, Q.~Le, and R.~Salakhutdinov,
  ``Transformer-xl: Attentive language models beyond a fixed-length context,''
  in \emph{the 57th Annual Meeting of the Association for Computational
  Linguistics (ACL)}, 2019.

\bibitem{carlini2019secret}
N.~Carlini, C.~Liu, {\'U}.~Erlingsson, J.~Kos, and D.~Song, ``The secret
  sharer: Evaluating and testing unintended memorization in neural networks,''
  in \emph{28th USENIX Security Symposium (USENIX Security 19)}, 2019.

\bibitem{merity2016pointer}
S.~Merity, C.~Xiong, J.~Bradbury, and R.~Socher, ``Pointer sentinel mixture
  models,'' in \emph{International Conference on Learning Representations
  (ICLR)}, 2017.

\bibitem{merity2018analysis}
S.~Merity, N.~S. Keskar, and R.~Socher, ``An analysis of neural language
  modeling at multiple scales,'' \emph{arXiv preprint arXiv:1803.08240}, 2018.

\bibitem{marcus1994penn}
M.~Marcus, B.~Santorini, and M.~A. Marcinkiewicz, ``Building a large annotated
  corpus of english: The penn treebank,'' \emph{Computational Linguistics},
  vol.~19, no.~2, pp. 313--330, 1993.

\bibitem{kingma2014adam}
D.~P. Kingma and J.~Ba, ``Adam: A method for stochastic optimization,'' in
  \emph{International Conference on Learning Representations (ICLR)}, 2015.

\bibitem{denkowski2014meteor}
M.~Denkowski and A.~Lavie, ``Meteor universal: Language specific translation
  evaluation for any target language,'' in \emph{the 9th Workshop on
  Statistical Machine Translation}, 2014.

\bibitem{miller1998wordnet}
G.~A. Miller, \emph{WordNet: An electronic lexical database}.\hskip 1em plus
  0.5em minus 0.4em\relax MIT press, 1998.

\bibitem{reimers2019sentence}
N.~Reimers and I.~Gurevych, ``Sentence-bert: Sentence embeddings using siamese
  bert-networks,'' in \emph{Empirical Methods in Natural Language Processing
  and the 9th International Joint Conference on Natural Language Processing
  (EMNLP-IJCNLP)}, 2019.

\bibitem{venugopal2011watermarking}
A.~Venugopal, J.~Uszkoreit, D.~Talbot, F.~J. Och, and J.~Ganitkevitch,
  ``Watermarking the outputs of structured prediction with an application in
  statistical machine translation,'' in \emph{Empirical Methods in Natural
  Language Processing (EMNLP)}, 2011.

\bibitem{suykens1999least}
J.~A. Suykens and J.~Vandewalle, ``Least squares support vector machine
  classifiers,'' \emph{Neural Processing Letters}, vol.~9, no.~3, pp. 293--300,
  1999.

\bibitem{fan2019rethinking}
L.~Fan, K.~W. Ng, and C.~S. Chan, ``Rethinking deep neural network ownership
  verification: Embedding passports to defeat ambiguity attacks,'' in
  \emph{Advances in Neural Information Processing Systems}, 2019.

\bibitem{holtzman2019curious}
A.~Holtzman, J.~Buys, L.~Du, M.~Forbes, and Y.~Choi, ``The curious case of
  neural text degeneration,'' in \emph{International Conference on Learning
  Representations (ICLR)}, 2020.

\bibitem{castillo2011information}
C.~Castillo, M.~Mendoza, and B.~Poblete, ``Information credibility on
  twitter,'' in \emph{the 20th international conference on World Wide Web},
  2011.

\bibitem{tan2020detecting}
R.~Tan, B.~Plummer, and K.~Saenko, ``Detecting cross-modal inconsistency to
  defend against neural fake news,'' in \emph{Empirical Methods in Natural
  Language Processing (EMNLP)}, 2020.

\bibitem{thorne2017fake}
J.~Thorne, M.~Chen, G.~Myrianthous, J.~Pu, X.~Wang, and A.~Vlachos, ``Fake news
  stance detection using stacked ensemble of classifiers,'' in \emph{the EMNLP
  Workshop: Natural Language Processing meets Journalism}, 2017.

\bibitem{hassan2017toward}
N.~Hassan, F.~Arslan, C.~Li, and M.~Tremayne, ``Toward automated fact-checking:
  Detecting check-worthy factual claims by claimbuster,'' in \emph{the ACM
  SIGKDD International Conference on Knowledge Discovery and Data Mining
  (KDD)}, 2017.

\bibitem{sharir2020cost}
O.~Sharir, B.~Peleg, and Y.~Shoham, ``The cost of training nlp models: A
  concise overview,'' \emph{arXiv preprint arXiv:2004.08900}, 2020.

\bibitem{lambda}
\BIBentryALTinterwordspacing
Lambda, ``Openai's gpt-3 language model: A technical overview.'' [Online].
  Available: \url{https://lambdalabs.com/blog/demystifying-gpt-3/}
\BIBentrySTDinterwordspacing

\end{thebibliography}
\section{Appendix}
\subsection{Metrics Analysis} \label{appendix_metrics}
We show more examples to examine and validate the metrics we use to evaluate or sort the output of the model.
\subsubsection{Sampling}
In Section~\ref{qual}, we discussed that the language model loss gives slightly better sentences in terms of syntactic correctness than SBERT, therefore, we used it to sort and select the best sample. In~\autoref{tab:sbert_vs_lm}, we show examples of such cases. Nevertheless, we still measure the semantic similarity using SBERT as a metric due to the benefits discussed below.

\newcolumntype{L}{>{\arraybackslash}m{3.8cm}}
\renewcommand{\arraystretch}{1.3}
\begin{table} [!b]
\centering
\resizebox{0.95\linewidth}{!}{%
\begin{tabular}{L|L|L}
\toprule
\textbf{Input} & \textbf{SBERT sample} & \textbf{LM sample} \\\midrule
The new M @-@ 120 designation replaced M @-@ 20 south \underline{\textit{of}} $<$unk$>$ . M @-@ 82 now ran \underline{\textit{from}} $<$unk$>$ to $<$unk$>$ only. & The new M @-@ 120 designation replaced M @-@ 20 south of $<$unk$>$ . M @-@ 82 now ran \underline{\textit{\hlc[light_red]{were}}} $<$unk$>$ to $<$unk$>$ only. & The new M @-@ 120 designation replaced M @-@ 20 south \underline{\textit{\hlc[light_green]{that}}} $<$unk$>$ . M @-@ 82 now ran from $<$unk$>$ to $<$unk$>$ only. \\

The city continued to grow thanks to a commission government's \underline{\textit{efforts}} to bring in a booming automobile industry in the 1920s. &  The city continued to grow thanks to a commission government's \underline{\textit{\hlc[light_red]{could}}} to bring in a booming automobile industry in the 1920s. & The city continued to grow thanks to a commission government's efforts to bring in a booming \underline{\textit{\hlc[light_green]{of}}} industry in the 1920s.
 \\\bottomrule
\end{tabular}}
\caption{Examples of input sentences, the best SBERT sample, and the best language model sample (slightly better).} \label{tab:sbert_vs_lm}
\vspace{-3mm}
\end{table}

\subsubsection{SBERT and Meteor}
In our analysis, we use the SBERT distance between the input and output sentences' embeddings as an auxiliary metric besides using the meteor score. We here demonstrate examples of sentences with high SBERT distance and the advantages of using it over meteor only.   

One of the cases that yields a high SBERT distance is when the output text has a changed sentiment (e.g., by using a negation), such as the two examples in~\autoref{tab:sbert_sentiment}. These examples do not have an extremely low meteor score since not a lot of words were changed. The first example also is grammatically correct (using ``are 't''). Despite that, they undesirably change the semantics of the input sentence, which is detected by the SBERT since it was trained on the NLI task. Additionally, we show in~\autoref{tab:sbert_vs_meteor} two samples for the same input sentence and comparable meteor scores, however, the one with the lower SBERT distance has more coherency. 

Given these observations, and the qualitative analysis we performed in Section~\ref{qual} (e.g., on the `no-discriminator' model), we found that using SBERT is an effective metric to approximate semantic similarity and adds more information than using meteor alone.   

\newcolumntype{L}{>{\arraybackslash}m{4.0cm}}
\renewcommand{\arraystretch}{1.3}
\begin{table} [!t]
\centering
\resizebox{0.95\linewidth}{!}{%
\begin{tabular}{L|L|l|l}
\toprule
\textbf{Input} & \textbf{Output} & \textbf{SBERT} & \textbf{Meteor}\\ \midrule

there \underline{\textit{are}} also many species of $<$unk$>$. There are three main routes which ascend the mountain , all of which gain over 4 \underline{\textit{@,@}} 100 feet ( 1 \underline{\textit{@,@}} 200 m ) of elevation. & there \underline{\textit{\hlc[light_red]{are 't}}} many species of $<$unk$>$. There are three main routes which ascend the mountain , all of which gain over 4 \underline{\textit{by}} 100 feet ( 1 \underline{\textit{by}} 200 m ) of elevation. & 7.5 & 0.93 \\  

Her family \underline{\textit{had}} originally come from Poland and Russia . $<$unk$>$ 's parents had both acted as children . $<$eos$>$ In a 2012 interview , $<$unk$>$ stated : " There was never [ religious ] faith in the house & Her family \underline{\textit{as}} originally come with Poland and Russia . $<$unk$>$ 's parents had both acted by children . $<$eos$>$ In a 2012 interview , $<$unk$>$ stated : " There was \underline{\textit{\hlc[light_red]{with}}} [ religious ] faith in the house & 7.19 & 0.93 \\ \bottomrule
\end{tabular}}
\caption{Examples in which introducing negation resulted in a relatively high SBERT distance.} \label{tab:sbert_sentiment}
\end{table}

\renewcommand{\arraystretch}{1.3}
\begin{table} [!htbp]
\centering
\resizebox{0.95\linewidth}{!}{%
\begin{tabular}{L|L|l|l}
\toprule
\textbf{Input} & \textbf{Output} & \textbf{SBERT} & \textbf{Meteor}\\ \midrule

This allegation became more widely known when $<$unk$>$ Alexander was featured in the documentary \underline{\textit{The}} Search for $<$unk$>$ , which has \underline{\textit{been}} cited by several authors including Gerald $<$unk$>$ , \underline{\textit{an}} expert on $<$unk$>$ . Towards the end of the song , there is a line " Feeding off the screams of the $<$unk$>$ he 's creating " , which was taken from the film \underline{\textit{The}} Boys from Brazil in which Dr. $<$unk$>$ was the villain. & 

This allegation became more widely known when $<$unk$>$ Alexander was featured in the documentary \underline{\textit{of}} Search for $<$unk$>$ , which has \underline{\textit{\hlc[light_red]{was}}} cited by several authors including Gerald $<$unk$>$ , \underline{\textit{from}} expert on $<$unk$>$ . Towards the end of the song , there is a line " Feeding off the screams of the $<$unk$>$ he 's creating " , which was taken from the film \underline{\textit{from}} Boys from Brazil in which Dr. $<$unk$>$ was the villain .
& \textcolor{red}{1.55} & 0.941 \\ \hline

This allegation became more widely known when $<$unk$>$ Alexander was featured in the documentary \underline{\textit{The}} Search for $<$unk$>$ , which has been cited by several authors including Gerald $<$unk$>$ , an expert on $<$unk$>$ . $<$eos$>$ Towards the end of the song , there is a line " Feeding off the screams of the $<$unk$>$ he 's creating " , which was taken from the film The Boys from Brazil in which Dr. $<$unk$>$ was the villain . &
 
This allegation became more widely known when $<$unk$>$ Alexander was featured in the documentary \underline{\textit{\hlc[light_green]{of}}} Search for $<$unk$>$ , which has been cited by several authors including Gerald $<$unk$>$ , an expert on $<$unk$>$ . Towards the end of the song , there is a line " Feeding off the screams of the $<$unk$>$ he 's creating " , which was taken from the film \underline{\textit{\hlc[light_green]{of}}} Boys from Brazil $<$unk$>$ which Dr. $<$unk$>$ was the villain . & \textcolor{dark_green}{1.17} & 0.939 \\
\bottomrule
\end{tabular}}
\caption{Two samples for the same input text segment. Although they have comparable meteor scores, the sample with the lower SBERT distance shows better coherence.} \label{tab:sbert_vs_meteor}
\vspace{-4mm}
\end{table}
\begin{table} [!b]
\vspace{-3mm}
\centering
\resizebox{0.4\linewidth}{!}{%
\begin{tabular}{l|ll}
\toprule
\textbf{Text} & \textbf{Meteor} & \textbf{SBERT} \\ \midrule 
Corrupted & 0.947 & 2.7 \\
Denoised & 0.956 & 2.25 \\ \bottomrule
\end{tabular}}
\caption{The similarity to the original sequence in the case of the corrupted and denoised text.} \label{tab:denoising_nonwm}
\vspace{-4mm}
\end{table}
\subsection{Denoising} \label{denoise}
For the denoising autoencoder (DAE), we used 6 encoding and decoding transformer layers in the encoder and decoder, respectively. We also share the embeddings of the encoder, decoder, and the pre-softmax layer (dimension: 512). The decoder has a masked self-attention and it attends to the output of the encoder.

\paragraph{Denoising non-watermarked text} We evaluate the DAE, regardless of the watermark, by applying the noise to the non-watermarked test set. We compare the similarity to the original text before and after denoising using the meteor and SBERT scores as shown in~\autoref{tab:denoising_nonwm}. We observed that denoising partially reconstructs the original sentence, but, it can introduce additional changes. We illustrate by the examples in~\autoref{tab:denoising_ex} that we categorize into three parts. In the first one, we show examples where the denoised sequence matches the original sequence; this was mainly for sentences with syntactic inconsistencies that removed common/likely words. In the second part, the DAE removed the added noise with more likely sequences, yet, it did not restore the original one which might cause semantic differences. In the third part, the noise words were not changed in the denoised text. This analysis suggests that the DAE is more likely to change sequences with clear flaws, but it is also likely to cause other changes that were not corrupted. We validate this observation by examining the denoising output of the watermarked text. 

\renewcommand{\arraystretch}{1.3}
\begin{table} [!t]
\centering
\resizebox{\linewidth}{!}{%
\begin{tabular}{L|L|L}
\toprule
\textbf{Input} & \textbf{Corrupted} & \textbf{Denoised} \\ \midrule
pair \underline{\textit{of}} claws & pair \underline{\textit{1941}} claws & pair \underline{\textit{\hlc[light_green]{of}}} claws \\
when you \underline{\textit{don}} 't & when you \underline{\textit{tendencies}} 't & when you \underline{\textit{\hlc[light_green]{don}}} 't \\ 
his earliest surviving poem \underline{\textit{,}} & his earliest surviving poem \underline{\textit{bill}} & his earliest surviving poem \underline{\textit{\hlc[light_green]{,}}}\\ 
he \underline{\textit{was}} arrested & He \underline{\textit{demolition}} arrested & He \underline{\textit{\hlc[light_green]{was}}} arrested  \\ \midrule

attempted to \underline{\textit{join}} the court & attempted to \underline{\textit{Desiree}} the court & attempted to \underline{\textit{\hlc[light_yellow]{take}}} the court \\ 
\underline{\textit{He}} next \underline{\textit{spent}} around six weeks & \underline{\textit{Dreamers}} next \underline{\textit{Punch}} around six weeks&\underline{\textit{\hlc[light_yellow]{The}}} next \underline{\textit{\hlc[light_yellow]{day}}} around six weeks  \\ \midrule
\underline{\textit{He}} appeared to be a $<$unk$>$ son & \underline{\textit{police}} appeared to be a $<$unk$>$ son & \underline{\textit{\hlc[light_red]{police}}} appeared to be a $<$unk$>$ son\\
Like many other poems in \underline{\textit{the}} Tang & Like many other poems in \underline{\textit{roof}} Tang & Like many other poems in \underline{\textit{\hlc[light_red]{roof ,}}} \\ 
The \underline{\textit{tenor}} of his work changed & The \underline{\textit{luck}} of his work changed & The \underline{\textit{\hlc[light_red]{luck}}} of his work changed \\ 
 \bottomrule
\end{tabular}}
\caption{DAE output when applying word replacement noise to the non-watermarked test set.} \label{tab:denoising_ex}
\end{table}

\paragraph{Denoising watermarked text} In~\autoref{tab:denoising_wm_ex}, we show examples when applying the DAE to watermarked text without additional noise (the results in~\autoref{tab:denoising}). We categorize these examples into three parts; the first is the examples where the watermarking changes were not changed by the DAE. Second, we show examples where they were changed; these examples are from different variants of the model, and they generally cause clear flaws, this explains the large drop in the `no-discriminator' model since this variant generally had lower quality output. Third, we show examples where the DAE introduced additional changes to sequences that were not originally changed by the watermarking model, this increased the SBERT distance in the first two rows in~\autoref{tab:denoising}.

\renewcommand{\arraystretch}{1.4}
\begin{table} [!t]
\centering
\resizebox{\linewidth}{!}{%
\begin{tabular}{L|L|L}
\toprule
\textbf{Input} & \textbf{Watermarked} & \textbf{Denoised} \\ \midrule
The eggs hatch \underline{\textit{at}} night & The eggs hatch \underline{\textit{\hlc[light_green]{with}}} night & The eggs hatch \underline{\textit{with}} night \\ 
and a mass \underline{\textit{of}} 6 kilograms & and a mass \underline{\textit{\hlc[light_green]{as}}} 6 kilograms & and a mass \underline{\textit{as}} 6 kilograms \\ 
several years writing for the television \underline{\textit{sitcoms}} Grace Under Fire & several years writing for the television \underline{\textit{\hlc[light_green]{of}}} Grace Under Fire & several years writing for the television \underline{\textit{of}} Grace Under Fire \\ 
He \underline{\textit{also}} performed as an actor and a singer & He \underline{\textit{\hlc[light_green]{had}}} performed as an actor and a singer & He \underline{\textit{had}} performed as an actor and a singer \\ \midrule

he \underline{\textit{took}} the civil service exam & he \underline{\textit{\hlc[light_red]{an}}} the civil service exam & he \textit{\underline{was}} the civil service exam  \\ 
\underline{\textit{The}} first RAAF helicopters were committed to  & . \underline{\textit{\hlc[light_red]{with}}} first RAAF helicopters were committed to & . \underline{\textit{The}} first RAAF helicopters were committed to \\ 
consisting of \underline{\textit{an}} infantry battalion & consisting of \underline{\textit{\hlc[light_red]{been}}} infantry battalion & consisting of \underline{\textit{two}} infantry battalion \\ 
\underline{\textit{,}} but the species is also widely known as & \underline{\textit{\hlc[light_red]{Bunbury}}} but the species is also widely known as & \underline{\textit{,}} but the species is also widely known as \\ \midrule

This occurs because , in \underline{\textit{life}} , the \underline{\textit{red}} pigment & This occurs because , in \underline{\textit{life}} , the \underline{\textit{red}} pigment & This occurs because , in \underline{\textit{\hlc[light_yellow]{particular}}} , the \underline{\textit{\hlc[light_yellow]{small}}} pigment \\ 
and \underline{\textit{adopts}} a $<$unk$>$ lifestyle & and \underline{\textit{adopts}} a $<$unk$>$ lifestyle & and \underline{\textit{\hlc[light_yellow]{has}}} a $<$unk$>$ lifestyle \\ 
The last \underline{\textit{distinct}} population & The last \underline{\textit{distinct}} population & The last \underline{\textit{\hlc[light_yellow]{major}}} population \\
\bottomrule
\end{tabular}}
\caption{DAE output when applied to the watermarked text (from different model's variants).} \label{tab:denoising_wm_ex}
\vspace{-4mm}
\end{table}

We observed other cases where the watermarking changes were not altered by the DAE even when having other grammatical mistakes, these changes might be removed by training a stronger DAE (e.g., larger model or larger dataset), however, this requires an even more experienced attacker with more technical knowledge and powerful computational resources.      
\subsection{Visualizations} \label{vis_analysis}
\begin{figure}[!t]
\centering
\begin{subfigure}{0.45\columnwidth}
    \centering
    \includegraphics[width=\linewidth]{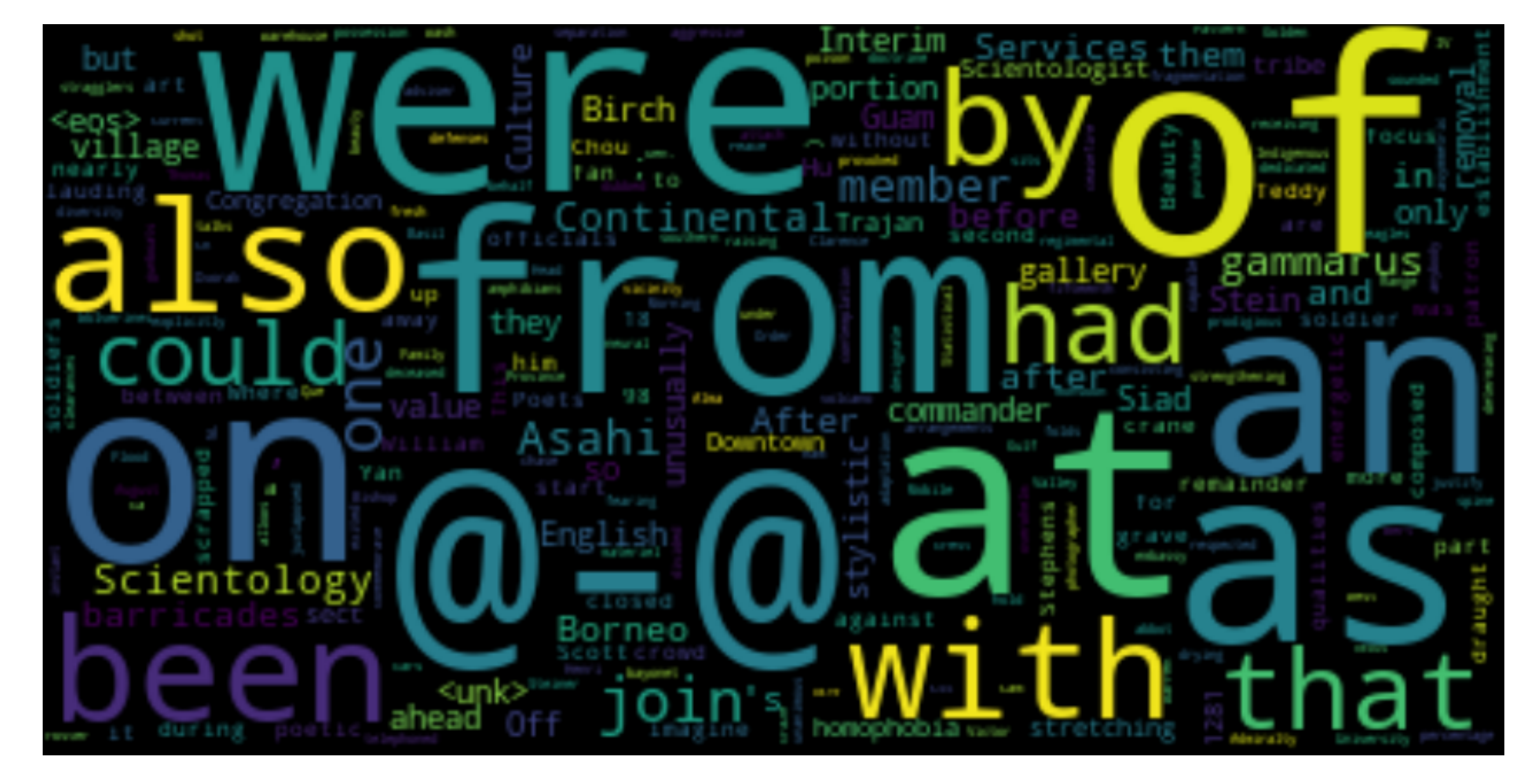}
    \caption{}
    \label{fig:from_words}
\end{subfigure}
\begin{subfigure}{0.45\columnwidth}    
\centering
    \includegraphics[width=\linewidth]{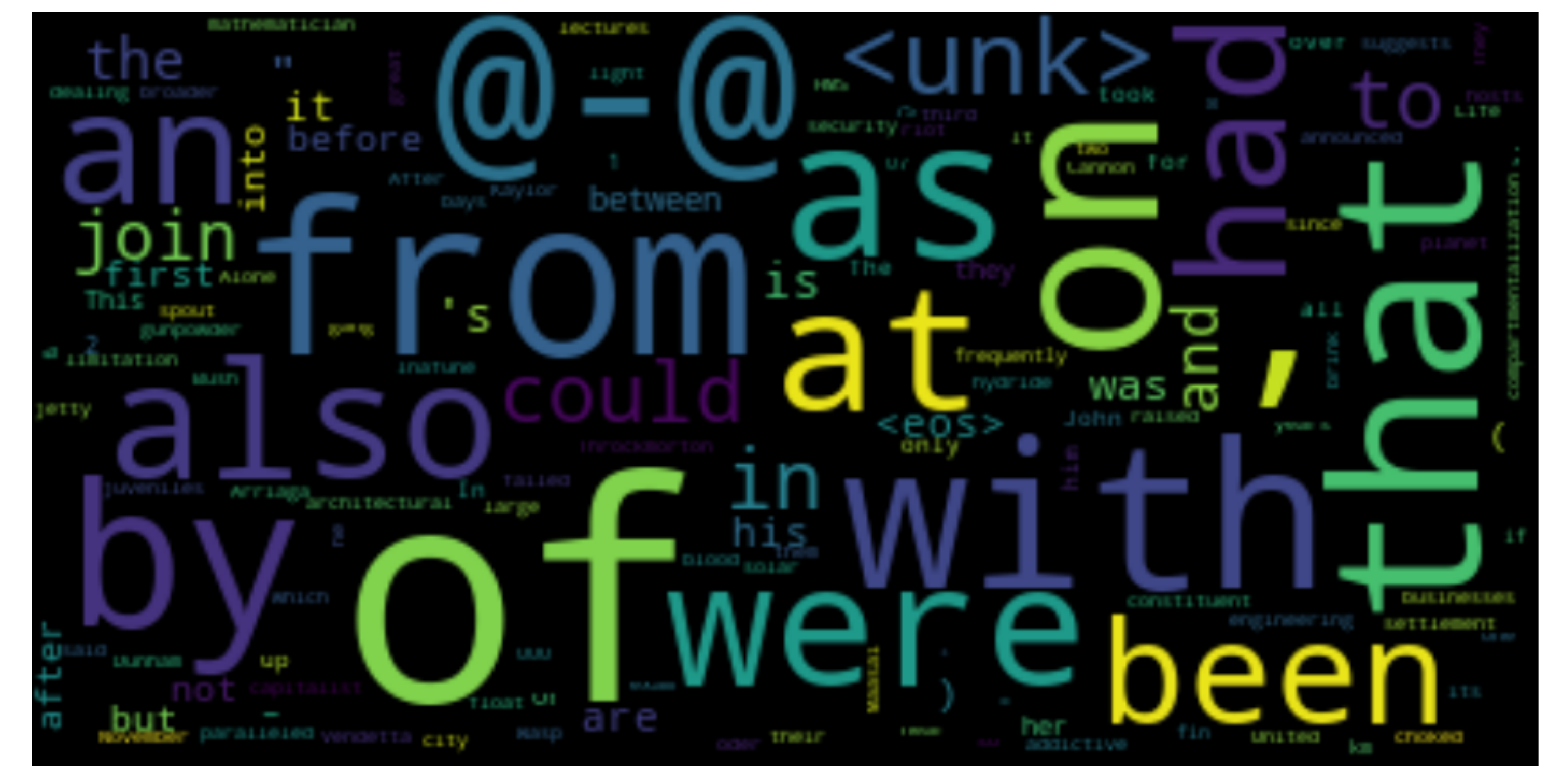}
    \caption{}
    \label{fig:to_words}
\end{subfigure}
\caption{(a) Words that were replaced in the original text. (b) Words that the model changed to in the watermarked text. Bigger fonts indicate higher frequencies.}
\vspace{-3mm}
\end{figure}

We show, in~\autoref{fig:from_words}, a word cloud for the most frequent words that were changed in the original text when watermarking, and in~\autoref{fig:to_words}, the most frequent words that were changed to in the watermarked text. As can be observed, the words in both figures are highly overlapping, therefore, we analysed the pairwise transitions between them in~\autoref{fig:matrix_words}. As we showed in~\autoref{fig:hist_words} and~\autoref{fig:matrix_words}, the model keeps the count of these top words similar, and it does not perform fixed substitutions between them. These factors support the encoding secrecy with no telltale words. Besides, there are no words that are particularly exclusive for bit holding, which has a flexibility advantage over the rule-based substitution baseline discussed in Section~\ref{baseline}. \new{For better visualization, we show in~\autoref{fig:matrix_words_no_diagonal} the words' transitions as in~\autoref{fig:matrix_words}, but without the diagonal elements where the words were not changed.}

\subsection{Different \model{} Models and Adaptive Attacks} \label{sec:append_adaptive}
\begin{figure}[!t]
\centering
\includegraphics[width=0.8\linewidth]{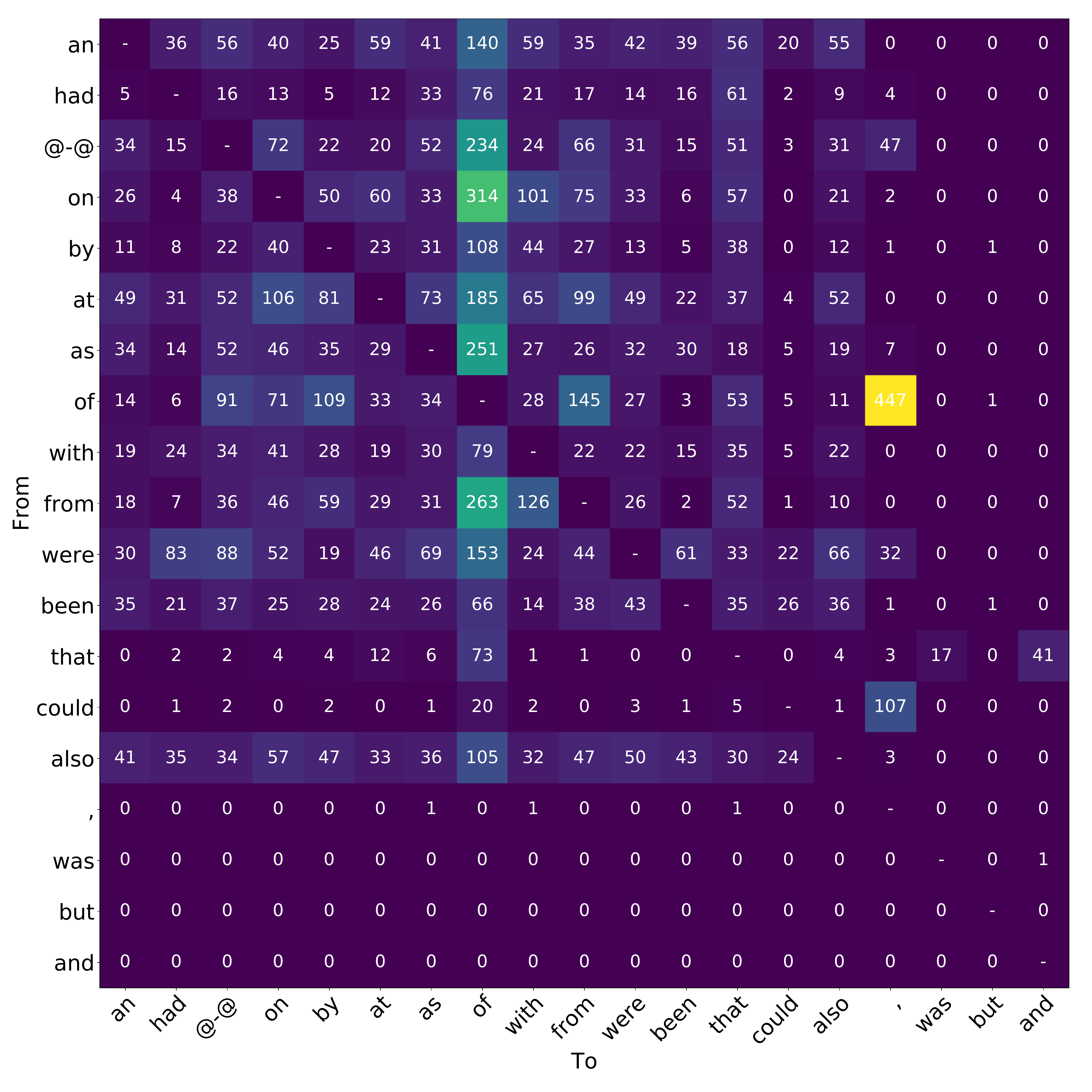}
\caption{\new{A matrix of word changes' count from the original text to modified text using \model{} (same as~\autoref{fig:matrix_words} but excluding the diagonal elements where words were not changed).}} 
\label{fig:matrix_words_no_diagonal}
\end{figure}

\begin{figure}[!t]
\centering
\includegraphics[width=0.8\linewidth]{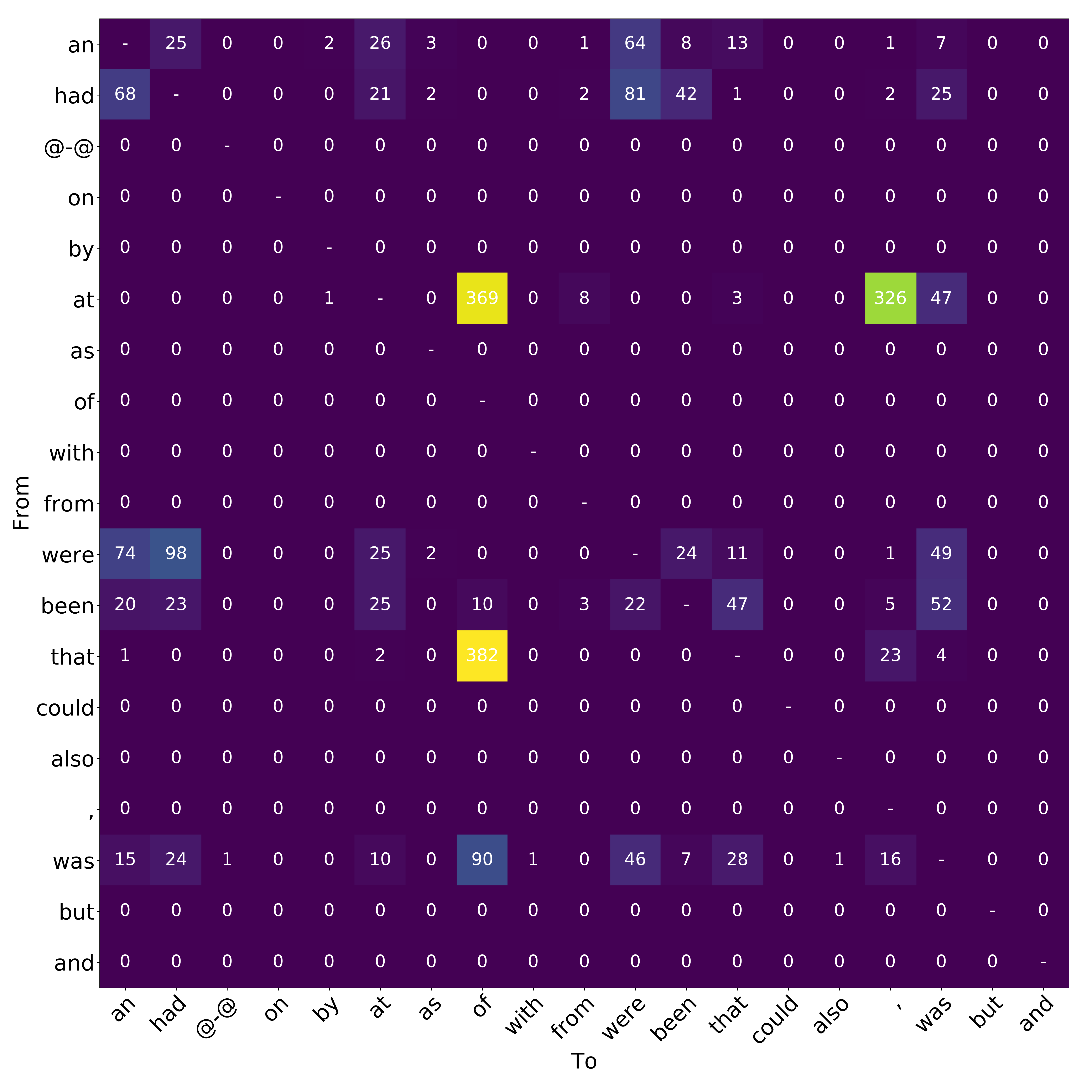}
\caption{\new{The words' transitions produced by \model{}$_{\text{adv}}$ for the most commonly changed words by \model{} (in~\autoref{fig:matrix_words_no_diagonal}).}} \label{fig:matrix_words_diff_seed}
\end{figure}

\new{In sections~\ref{sec:robust_piracy} and~\ref{sec:dewatermarking}, we discussed that attacks crafted using another trained model (\model{}$_{\text{adv}}$) are less effective in the black-box case (when applied to the first \model{} model). In this section, we first compare two independently trained models in terms of words' transitions and qualitative examples. We then show examples of adaptive attacks.}
\paragraph{Comparing different models}
\new{A message decoder of one model gives an almost random chance accuracy when used to decode another model's sentences. Thus, it is sensitive to the paired watermarking model mostly. 
A possible explanation is that different instances produce different patterns or mappings (as previously reported in data hiding studies in images~\cite{zhu2018hidden}). To investigate that, we first study whether \model{}$_{\text{adv}}$ uses the same commonly changed words to encode the information. In~\autoref{fig:matrix_words_diff_seed}, we show the transitions produced by \model{}$_{\text{adv}}$ among the commonly used words by the first \model{} model. When comparing this to~\autoref{fig:matrix_words_no_diagonal}, we notice that these words have relatively fewer transitions.} 

\new{Furthermore, we show in~\autoref{tab:model1_vs_model2}, examples of sentences that were watermarked individually (but, using the same binary message) by \model{} and \model{}$_{\text{adv}}$ producing different wording changes (for the replaced, added words, or their positions).}
\paragraph{Re-watermarking}
\new{For further investigation, we show in~\autoref{tab:rewatermarking} examples of re-watermarked sentences in the white-box and the black-box cases.} 
\newcolumntype{L}{>{\arraybackslash}m{3.8cm}}
\renewcommand{\arraystretch}{1.3}
\begin{table} [!t]
\centering
\resizebox{0.9\linewidth}{!}{%
\begin{tabular}{L|L|L}
\toprule
\textbf{Input} & \textbf{\model{}} & \textbf{\model{}$_{\text{adv}}$} \\\midrule
As is often the case with huge ancient ruins , knowledge \underline{\textit{\hlc[light_yellow]{of}}} the site was never completely lost in the region . It seems that local people never $<$unk$>$ about $<$unk$>$ and \underline{\textit{\hlc[light_yellow]{they}}} guided $<$unk$>$ expeditions to the ruins in the 1850s .& As is often the case with huge ancient ruins , knowledge \underline{\textit{\hlc[light_yellow]{by}}} the site was never completely lost in the region . It seems that local people never $<$unk$>$ about $<$unk$>$ and they guided $<$unk$>$ expeditions to the ruins in the 1850s . & As is often the case with huge ancient ruins , knowledge of the site was never completely lost in the region . It seems that local people never $<$unk$>$ about $<$unk$>$ and \underline{\textit{\hlc[light_yellow]{three}}} guided $<$unk$>$ expeditions to the ruins in the 1850s . \\ 

Jon $<$unk$>$ \underline{\textit{\hlc[light_yellow]{of}}} the professional wrestling section of the Canadian Online Explorer rated the show a 7 out of 10 , \underline{\textit{\hlc[light_yellow]{which}}} was lower than the 8 out of 10 given to the 2007 edition by Jason $<$unk$>$ . & Jon $<$unk$>$ \underline{\textit{\hlc[light_yellow]{@-@}}} the professional wrestling section of the Canadian Online Explorer rated the show a 7 out of 10 , which was lower than the 8 out of 10 given to the 2007 edition by Jason $<$unk$>$ . & Jon $<$unk$>$ of the professional wrestling section of the Canadian Online Explorer rated the show a 7 out of 10 , \underline{\textit{\hlc[light_yellow]{that}}} was lower than the 8 out of 10 given to the 2007 edition by Jason $<$unk$>$ . \\ \bottomrule
\end{tabular}}
\caption{\new{Examples of input and watermarked sentences (using the same message) by the two models.}} \label{tab:model1_vs_model2}
\vspace{-3mm}
\end{table}

\newcolumntype{L}{>{\arraybackslash}m{3.4cm}}
\renewcommand{\arraystretch}{1.5}
\begin{table} [!b]
\centering
\resizebox{0.9\linewidth}{!}{%
\begin{tabular}{c|L|L|L}
\toprule
& \textbf{Input} & \textbf{Watermarked} & \textbf{Re-watermarked} \\ \cline{2-4}

\parbox[t]{2mm}{\multirow{2}{*}{\rotatebox[origin=c]{90}{\textbf{White-box}}}} & landed a role as " Craig " in the episode " Teddy 's Story " \underline{\textit{\hlc[light_yellow]{of}}} the television series The Long Firm & landed a role as " Craig " in the episode " Teddy 's Story " \underline{\textit{\hlc[light_yellow]{from}}} the television series The Long Firm & landed a role as " Craig " in the episode " Teddy 's Story " \underline{\textit{\hlc[light_yellow]{at}}} the television series The Long Firm  \\ 

&$<$unk$>$ made a guest appearance on a two @-@ part episode arc \underline{\textit{\hlc[light_yellow]{of}}} the television series Waking the Dead & $<$unk$>$ made a guest appearance on a two @-@ part episode arc \underline{\textit{\hlc[light_yellow]{from}}} the television series Waking the Dead & $<$unk$>$ made a guest appearance on a two @-@ part episode arc \underline{\textit{\hlc[light_yellow]{with}}} the television series Waking the Dead \\ \midrule \midrule

\parbox[t]{2mm}{\multirow{2}{*}{\rotatebox[origin=c]{90}{\textbf{Black-box}}}} & Female H. gammarus reach sexual maturity when \underline{\textit{\hlc[light_yellow]{they}}} have grown to a carapace length of 80 – 85 millimetres , whereas males mature \underline{\textit{\hlc[light_yellow]{at}}} a slightly smaller size . & Female H. gammarus reach sexual maturity when they have grown to a carapace length of 80 – 85 millimetres , whereas males mature \underline{\textit{\hlc[light_yellow]{on}}} a slightly smaller size . & Female H. gammarus reach sexual maturity when \underline{\textit{\hlc[light_yellow]{to}}} have grown to a carapace length of 80 – 85 millimetres , whereas males mature \underline{\textit{\hlc[light_yellow]{on}}} a slightly smaller size . \\

 & $<$unk$>$ 's other positions \underline{\textit{\hlc[light_yellow]{at}}} the Department of Air included Air Commodore Plans from October 1957 to January 1959 , and Director General Plans and Policy from January to August 1959 . The latter assignment put \underline{\textit{\hlc[light_yellow]{him}}} in charge of the RAAF 's Directorate of Intelligence . & $<$unk$>$ 's other positions \underline{\textit{\hlc[light_yellow]{on}}} the Department of Air included Air Commodore Plans from October 1957 to January 1959 , and Director General Plans and Policy from January to August 1959 . The latter assignment put him in charge of the RAAF 's Directorate on Intelligence . & $<$unk$>$ 's other positions \underline{\textit{\hlc[light_yellow]{on}}} the Department of Air included Air Commodore Plans from October 1957 to January 1959 , and Director General Plans and Policy from January to August 1959 . The latter assignment put \underline{\textit{\hlc[light_yellow]{was}}} in charge of the RAAF 's Directorate on Intelligence . \\
\bottomrule
\end{tabular}}
\caption{\new{Examples of re-watermarking in the white-box and black-box cases.}} \label{tab:rewatermarking}
\vspace{-4mm}
\end{table}

\new{In the white-box case, we observed that the model often replaces the same word that was previously replaced in the first watermarking process. This caused the first watermark matching accuracy to drop to nearly random chance. In the black-box case, we can observe that: 1) the re-watermarking does not necessarily override the first changes (i.e., both changes can be present in the re-watermarked sentences). 2) the newly added words might not be from the most sensitive words to the first \model{} model (based on~\autoref{fig:matrix_words_no_diagonal}). These observations and the previous analysis potentially explain why re-watermarking was less effective in the black-box case.}  

\paragraph{De-watermarking}
\new{In section~\ref{sec:dewatermarking}, we evaluated an adaptive attack that tries to de-watermark the sentences rather than re-watermark them. We perform this attack by training a denoising autoencoder (DAE$_\text{paired}$, with a similar architecture to the DAE used in~\autoref{sec:robust_denoising}) on the paired training data of \model{}$_\text{adv}$ (without adding further noise). 
In~\autoref{tab:dewatermarking}, we show examples of applying this attack in the white-box and black-box cases.}

\new{In the white-box, DAE$_\text{paired}$ successfully recovered the sentences where the watermarking model caused clear syntactic flaws (such as the first example). Moreover, since DAE$_\text{paired}$ was exposed to the most frequent changes' patterns during training, it was able to reconstruct sentences with either no or less obvious artifacts (e.g., replacing `which' with `that', or `which' with `before' in the table). These changes might not be easy to detect without paired training. The second category of examples includes pairs where the watermarking changes were not reversed but were nevertheless replaced with perhaps more correct tokens. The last category shows very subtle examples that were not changed even in the white-box case.} 

\new{In the black-box, DAE$_\text{paired}$ also recovers the sentences with clear mistakes. This is similar to the DAE model that was trained on noisy data in section~\ref{sec:robust_denoising}, however, DAE$_\text{paired}$ was more successful since different models could still have some similarities (e.g., both replacing `been'). 
Since DAE$_\text{paired}$ was sensitive to the patterns that it was trained on, it often replaced words that were not changed originally by \model{} but are often changed by \model{}$_\text{adv}$ (e.g., removing `which', `three', and `they' in the third black-box category). Finally, the last black-box category shows examples where DAE$_\text{paired}$ did not perform any changes. This can be due to two reasons: 1) the changes are more subtle. 2) they were not frequently seen in the paired training data of \model{}$_\text{adv}$.}
\newcolumntype{L}{>{\arraybackslash}m{3.4cm}}
\renewcommand{\arraystretch}{1.5}
\begin{table} [!t]
\centering
\resizebox{0.9\linewidth}{!}{%
\begin{tabular}{c|L|L|L}
\toprule
& \textbf{Input} & \textbf{Watermarked} & \textbf{De-watermarked} \\ \cline{2-4}
& with a body length up to 60 centimetres \underline{\textit{\hlc[light_yellow]{(}}} 24 in ) & with a body length up to 60 centimetres \underline{\textit{\hlc[light_yellow]{of}}} 24 in ) & with a body length up to 60 centimetres \underline{\textit{\hlc[light_yellow]{(}}} 24 in )  \\

& which \underline{\textit{\hlc[light_yellow]{they}}} must shed in order to grow & which \underline{\textit{\hlc[light_yellow]{three}}} must shed in order to grow & which \underline{\textit{\hlc[light_yellow]{they}}} must shed in order to grow  \\ 

\parbox[t]{2mm}{\multirow{9}{*}{\rotatebox[origin=c]{90}{\textbf{White-box}}}} & $<$unk$>$ \underline{\textit{\hlc[light_yellow]{is}}} remembered for ... & $<$unk$>$ \underline{\textit{\hlc[light_yellow]{been}}} remembered for ... & $<$unk$>$ \underline{\textit{\hlc[light_yellow]{is}}} remembered for ... \\

& , \underline{\textit{\hlc[light_yellow]{which}}} was lower than the 8 out of 10 given to the 2007 edition by Jason $<$unk$>$ . & , \underline{\textit{\hlc[light_yellow]{that}}} was lower than the 8 out of 10 given to the 2007 edition by Jason $<$unk$>$ . & , \underline{\textit{\hlc[light_yellow]{which}}} was lower than the 8 out of 10 given to the 2007 edition by Jason $<$unk$>$ . \\

& ,  \underline{\textit{\hlc[light_yellow]{which}}} he was granted by  $<$unk$>$  on the May 29 episode of Impact !  & ,  \underline{\textit{\hlc[light_yellow]{before}}} he was granted by  $<$unk$>$ on the May 29 episode of Impact !  & ,  \underline{\textit{\hlc[light_yellow]{which}}} he was granted by  $<$unk$>$  on the May 29 episode of Impact ! \\  \cline{2-4}

& Today the fort \underline{\textit{\hlc[light_yellow]{is}}} open throughout the year & Today the fort \underline{\textit{\hlc[light_yellow]{not}}} open throughout the year & Today the fort \underline{\textit{\hlc[light_yellow]{was}}} open throughout the year \\ \cline{2-4}

& On the night \underline{\textit{\hlc[light_yellow]{before}}} such an event neither $<$unk$>$ or $<$unk$>$ Gale could get those minutes & On the night \underline{\textit{\hlc[light_yellow]{of}}} such an event neither $<$unk$>$ or $<$unk$>$ Gale could get those minutes & On the night \underline{\textit{\hlc[light_yellow]{of}}} such an event neither $<$unk$>$ or $<$unk$>$ Gale could get those minutes \\ \midrule \midrule

& which have \underline{\textit{\hlc[light_yellow]{been}}} referred to as the " midnight @-@ sun lobster " . & which have \underline{\textit{\hlc[light_yellow]{from}}} referred to as the " midnight @-@ sun lobster " .& which have \underline{\textit{\hlc[light_yellow]{been}}} referred to as the " midnight @-@ sun lobster " . \\
& several research @-@ $<$unk$>$ allegations that \underline{\textit{\hlc[light_yellow]{were}}} brought against him & several research @-@ $<$unk$>$ allegations that \underline{\textit{\hlc[light_yellow]{from}}} brought against him & several research @-@ $<$unk$>$ allegations that \underline{\textit{\hlc[light_yellow]{were}}} brought against him \\ \cline{2-4} 

\parbox[t]{2mm}{\multirow{3}{*}{\rotatebox[origin=c]{90}{\textbf{Black-box}}}} & the United $<$unk$>$ Band \underline{\textit{\hlc[light_yellow]{had}}} voted to stop $<$unk$>$ associate $<$unk$>$ & the United $<$unk$>$ Band \underline{\textit{\hlc[light_yellow]{that}}} voted to stop $<$unk$>$ associate $<$unk$>$ & the United $<$unk$>$ Band \underline{\textit{\hlc[light_yellow]{was}}} voted to stop $<$unk$>$ associate $<$unk$>$ \\ \cline{2-4} 
 
 & This stage involves \underline{\textit{\hlc[light_yellow]{three}}} $<$unk$>$ and lasts for 15 – 35 days . & This stage involves \underline{\textit{\hlc[light_yellow]{three}}} $<$unk$>$ and lasts for 15 – 35 days .& This stage involves \underline{\textit{\hlc[light_yellow]{an}}} $<$unk$>$ and lasts for 15 – 35 days . \\ 
& and \underline{\textit{\hlc[light_yellow]{three which}}} have diverged due to small effective population sizes & and \underline{\textit{\hlc[light_yellow]{three which}}} have diverged due to small effective population sizes & and \underline{\textit{\hlc[light_yellow]{which they}}} have diverged due to small effective population sizes \\ \cline{2-4} 

& The first pair of $<$unk$>$ is armed \underline{\textit{\hlc[light_yellow]{with}}} a large , asymmetrical pair of claws . & The first pair of $<$unk$>$ is armed \underline{\textit{\hlc[light_yellow]{by}}} a large , asymmetrical pair of claws . & The first pair of $<$unk$>$ is armed \underline{\textit{\hlc[light_yellow]{by}}} a large , asymmetrical pair of claws . \\
& Churchill has argued that blood quantum laws have \underline{\textit{\hlc[light_yellow]{an}}} inherent $<$unk$>$ purpose . & Churchill has argued that blood quantum laws have \underline{\textit{\hlc[light_yellow]{been}}} inherent $<$unk$>$ purpose . & Churchill has argued that blood quantum laws have \underline{\textit{\hlc[light_yellow]{been}}} inherent $<$unk$>$ purpose . \\
& Homarus gammarus is found across the north \underline{\textit{\hlc[light_yellow]{@-@}}} eastern Atlantic Ocean & Homarus gammarus is found across the north \underline{\textit{\hlc[light_yellow]{of}}} eastern Atlantic Ocean & Homarus gammarus is found across the north \underline{\textit{\hlc[light_yellow]{of}}} eastern Atlantic Ocean \\

\bottomrule
\end{tabular}}
\caption{\new{Examples of de-watermarking in the white-box and black-box cases.}} \label{tab:dewatermarking}
\vspace{-6mm}
\end{table}

\subsection{Generation-based hiding} \label{apendix_gen}
We present more details about the baseline of generation-based hiding in Section~\ref{lstm}.

\subsubsection{Architecture}
We add a `data hiding' component to the AWD-LSTM~\cite{merity2017regularizing} by feeding the message to the language model LSTM and simultaneously train a message decoder that is optimized to reconstruct the message from the output sequence. The input message is passed to a linear layer to match the embeddings' dimension, it is then repeated and added to the word embeddings at each time step. The language model is then trained with the cross-entropy loss: $ L_\textit{1} = \mathbb{E}_{p_{\textit{data}}(S)}[-\log p_{\textit{model}}(S)]$.

To allow end-to-end training, we use Gumbel-Softmax. The message decoder has a similar architecture to the AWD-LSTM and it takes the one-hot samples projected back into the embedding space. To reconstruct the message, the hidden states from the last layer are average-pooled and fed to a linear layer. We tie the embeddings and the pre-Softmax weights. The message reconstruction loss is the binary cross-entropy: $L_\textit{2} = -\sum_{i=1}^{q} b_i \log(b^{'}_i) + (1-b_i) \log(1-b^{'}_i)$.

The model is trained with a weighted average of both losses: $L= w_\textit{1}*L_\textit{1} + w_\textit{2}*L_\textit{2}$.

\subsubsection{Training details}
We mainly used the same hyperparameters and setup of~\cite{merity2017regularizing}, however, we found it essential to decrease the learning rate of ASGD than the one used; we use an initial learning rate of 2.5 instead of 30 for the language modelling LSTM and a smaller learning rate of 0.5 for the message decoding LSTM. We also found it helpful for a successful message encoding to pre-train the AWD-LSTM of the message decoder as a language model. Following the original implementation, we fine-tune the model after the initial training by restarting the training, to allow the ASGD optimizer to restart the averaging. Similar to \model{}, we use a message length of 4 bits. To allow multiple operating points of text utility vs. bit accuracy, we fine-tune the model again by assigning lower weight to the message loss. We start the training by $w_1 = 1, w_2=2$, and decrease $w_2$ for each fine-tuning step to reach a new operating point.
\newcolumntype{L}{>{\arraybackslash}m{8cm}}
\renewcommand{\arraystretch}{1.3}
\begin{table} [!b]
\centering
\resizebox{0.85\linewidth}{!}{%
\begin{tabular}{l|L}
\toprule
\textbf{Rating} & \textbf{Description} \\ \midrule
5 & The text is understandable, natural, and grammatically and structurally correct. \\
4 & The text is understandable, but it contains minor mistakes. \\
3 & The text is generally understandable, but some parts are ambiguous. \\
2 & The text is roughly understandable, but most parts are ambiguous. \\
1 & The text is mainly not understandable, but you can get the main ideas. \\ 
0 & The text is completely not understandable, unnatural, and you cannot get the main ideas. \\
 \bottomrule
\end{tabular}}
\caption{Ratings explanations given in the user study.} \label{tab:study_description}
\end{table}
\begin{figure} [!b]
    \centering
    \includegraphics[width=0.6\linewidth]{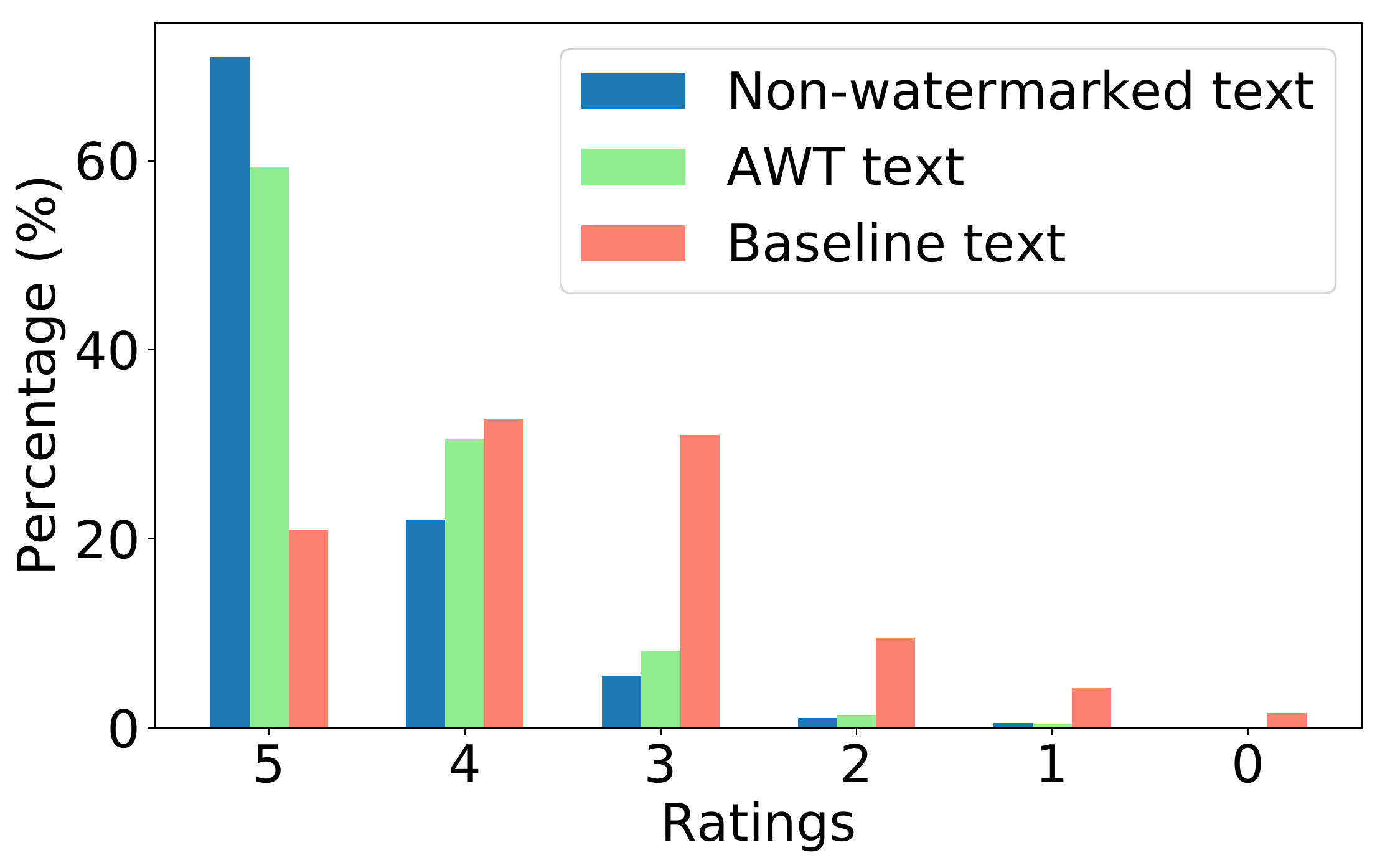}
    \caption{Histograms of ratings given to the three types of sentences in the user study.}
    \label{fig:study_ratings}
\end{figure}
\begin{table} [!b]
\centering
\resizebox{0.9\linewidth}{!}{%
\begin{tabular}{l|llllll}
\toprule
& \textbf{Judge 1} & \textbf{Judge 2} & \textbf{Judge 3} & \textbf{Judge 4} & \textbf{Judge 5}  & \textbf{Judge 6} \\  \midrule
\textbf{Non-wm} & 4.86\rpm0.4 & 3.98\rpm0.96 & 4.47\rpm0.62 & 4.77\rpm0.48 & 4.84\rpm0.44 & 4.8\rpm0.52\\
\textbf{Wm} & 4.76\rpm0.47 & 3.98\rpm1.09 & 4.13\rpm0.64 & 4.58\rpm0.61 & 4.71\rpm0.49 & 4.63\rpm0.6 \\
\textbf{Baseline} & 3.4\rpm1.28 & 3.57\rpm1.21 & 3.37\rpm0.81 & 3.32\rpm1.02 & 3.4\rpm1.09 & 4.03\rpm1.19 \\ \bottomrule
\end{tabular}}
\caption{Per-judge averaged ratings for the three types of sentences.} \label{tab:study_judges}
\end{table}

\newcolumntype{L}{>{\arraybackslash}m{5.7cm}}
\renewcommand{\arraystretch}{1.4}
\begin{table} [!b]
\centering
\resizebox{0.9\linewidth}{!}{%
\begin{tabular}{L|L}
\toprule
\textbf{Input} & \textbf{Synonym-baseline} \\ \midrule
Caldwell said it was \underline{\textit{\hlc[light_yellow]{easy}}} to obtain \underline{\textit{\hlc[light_yellow]{guns in}}} New Mexico : " we \underline{\textit{\hlc[light_yellow]{found}}} it was pretty \underline{\textit{\hlc[light_yellow]{easy}}} to \underline{\textit{\hlc[light_yellow]{buy guns}}} . & Caldwell said it was \underline{\textit{\hlc[light_yellow]{soft}}} to obtain \underline{\textit{\hlc[light_yellow]{artillery In}}} New Mexico : " we \underline{\textit{\hlc[light_yellow]{rule}}} it was pretty \underline{\textit{\hlc[light_yellow]{soft}}} to \underline{\textit{\hlc[light_yellow]{purchase accelerator}}} . \\

Caldwell said she and $<$unk$>$ went to a university library to \underline{\textit{\hlc[light_yellow]{find}}} the identity " of someone dying \underline{\textit{\hlc[light_yellow]{very young}}} " , \underline{\textit{\hlc[light_yellow]{next}}} went to public records and asked for a \underline{\textit{\hlc[light_yellow]{copy}}} of a \underline{\textit{\hlc[light_yellow]{birth certificate}}} & Caldwell said she and $<$unk$>$ went to a university library to \underline{\textit{\hlc[light_yellow]{found}}} the identity " of someone dying \underline{\textit{\hlc[light_yellow]{real new}}} " , \underline{\textit{\hlc[light_yellow]{adjacent}}} went to public records and asked for a \underline{\textit{\hlc[light_yellow]{replicate}}} of a \underline{\textit{\hlc[light_yellow]{parentage certification}}} \\

 \bottomrule
\end{tabular}}
\caption{Examples of the synonym substitution baseline sentences that were included in the user study.} \label{tab:baseline_examples}
\vspace{-3mm}
\end{table}
\subsection{User Study} \label{app_study}
We demonstrate in~\autoref{tab:study_description} the ratings' descriptions given in the instructions of the user study. In~\autoref{fig:study_ratings}, we show a histogram of ratings given to the three types of sentences included. We show in~\autoref{tab:study_judges}, the per-judge averaged ratings where we can observe that all judges gave \model{} higher ratings than the baseline. We show examples of the baseline sentences in~\autoref{tab:baseline_examples} along with the corresponding original sentences (paired sentences were not included in the study). 



%



\end{document}